\documentclass[fleqn,usenatbib]{mnras}

\usepackage{newtxtext,newtxmath}
\usepackage{multirow}

\usepackage[T1]{fontenc}

\DeclareRobustCommand{\VAN}[3]{#2}
\let\VANthebibliography\thebibliography
\def\thebibliography{\DeclareRobustCommand{\VAN}[3]{##3}\VANthebibliography}

\usepackage{graphicx}	
\usepackage{amsmath}	
\usepackage{tabularray}
\usepackage[dvipsnames]{xcolor}

\newcommand{\farc}{$^{\prime\prime}$}

\title[The 4MOST IR AGN survey]{An obscured quasar census with the 4MOST IR AGN survey: design, predicted properties, and scientific goals}

\author[C. Andonie et al.]{Carolina Andonie,$^{1,2}$\thanks{E-mail: cpandonie@mpe.mpg.de}
David M. Alexander,$^{1}$
Claire Greenwell,$^{1}$
Sotiria Fotopoulou,$^{3}$
Ryan Hickox,$^{4}$
\newauthor
David J Rosario,$^{5}$
Carolin Villforth,$^{6}$
Johannes Buchner,$^{2}$
Jens-Kristian Krogager,$^{7,8}$
Brivael Laloux,$^{9,2}$
\newauthor
Andrea Merloni,$^{2}$
Mara Salvato,$^{2}$
Ole Streicher,$^{10}$
and Wei Yan$^{11}$
\\
$^{1}$ Centre for Extragalactic Astronomy, Department of Physics, Durham University, Durham, DH1 3LE, UK\\
$^{2}$ Max-Planck-Institut für Extraterrestrische Physik, Gießenbachstraße, D-85748 Garching, Germany\\
$^{3}$ School of Physics, HH Wills Physics Laboratory, University of Bristol, Tyndall Avenue, Bristol, BS8 1TL, UK\\
$^{4}$ Department of Physics and Astronomy, Dartmouth College, 6127 Wilder Laboratory, Hanover, NH 03755, USA\\
$^{5}$ School of Mathematics, Statistics and Physics, Newcastle University, Newcastle upon Tyne, NE1 7RU, UK \\
$^{6}$ Department of Physics, University of Bath, Claverton Down, Bath BA2 7AY, UK\\
$^{7}$ French-Chilean Laboratory for Astronomy, IRL 3386, CNRS and U. de Chile, Casilla 36-D, Santiago, Chile \\
$^{8}$ Centre de Recherche Astrophysique de Lyon, Universit{\'e} de Lyon 1, UMR5574, 9 Av Charles Andr{\'e}, 69230 Saint-Genis-Laval, France \\
$^{9}$ INAF – Osservatorio Astronomico di Capodimonte, Via Moiariello 16, 80131 Napoli, Italy\\
$^{10}$ Leibniz Institute for Astrophysics Potsdam (AIP), An der Sternwarte 16 14482 Potsdam, Germany \\
$^{11}$ Wartburg College, 100 Wartburg Blvd. Waverly, IA 50677\\
}

\date{Accepted XXX. Received YYY; in original form ZZZ}

\pubyear{2023}

\defcitealias{2018Assef}{A18}

\begin{document}
\label{firstpage}
\pagerange{\pageref{firstpage}--\pageref{lastpage}}
\maketitle

\begin{abstract}
We present the 4MOST IR AGN survey, the first large-scale optical spectroscopic survey characterizing mid-infrared (MIR) selected obscured active galactic nuclei (AGN). The survey targets $\approx 212,000$ obscured infrared (IR) AGN candidates over $\approx 10,000 \rm \: deg^2$ down to a magnitude limit of $r_{\rm AB}=22.8 \, \rm mag$ and will be $\approx 100 \times$ larger than any existing obscured IR AGN spectroscopic sample. We select the targets using a MIR colour criterion applied to the unWISE catalogue from the {\it WISE} all-sky survey, and then apply a $r-W2\geq 5.9 \rm \: mag$ cut; we demonstrate that this selection will mostly identify sources obscured by $N_{\rm H}>10^{22} \rm \: cm^{-2}$. The survey complements the 4MOST X-ray survey, which will follow up $\sim 1\rm M$ {\it eROSITA}-selected (typically unobscured) AGN. We perform simulations to predict the quality of the spectra that we will obtain and validate our MIR-optical colour-selection method using X-ray spectral constraints and UV-to-far IR spectral energy distribution (SED) modelling in four well-observed deep-sky fields. We find that: (1) $\approx 80-87\%$ of the {\it WISE}-selected targets are AGN down to $r_{\rm AB}=22.1-22.8 \: \rm mag$ of which  $\approx 70\%$ are obscured by $N_{\rm H}>10^{22} \: \rm cm^{-2}$, and (2) $\approx 80\%$ of the 4MOST IR AGN sample  will remain undetected by the deepest {\it eROSITA} observations due to extreme absorption. Our SED fitting results show that the 4MOST IR AGN survey will primarily identify obscured AGN and quasars ($\approx 55\%$ of the sample is expected to have $L_{\rm AGN,IR}>10^{45} \rm \: erg \: s^{-1}$) residing in massive galaxies ($M_{\star}\approx 10^{10}-10^{12} \rm \: M_{\odot}$) at $z\approx 0.5-3.5$ with $ \approx 33\%$ expected to be hosted by starburst galaxies.

\end{abstract}




\section{Introduction}

Supermassive black holes (SMBHs) are found at the centres of most massive galaxies \citep[e.g.,][]{1998Magorrian, 2003Marconi}, and are thought to play an important role in galaxy evolution \citep[e.g.,][]{2012Alexander,2013Kormendy}. Most SMBH growth occurs in a phase of efficient accretion, where SMBHs are observed as active galactic nuclei (AGN), or in extremely luminous cases, as quasars\footnote{In this work, we refer to quasars as AGNs with a $8-1000\rm \: \mu m$ (IR) AGN luminosity $L_{\rm AGN, IR}>10^{45} \rm \: erg \: s^{-1}$, which is approximately equivalent to an AGN bolometric luminosity of $L_{\rm AGN, bol}>3\times 10^{45} \rm \: erg \: s^{-1}$, regardless of whether they are optically bright and unobscured or optically faint and obscured. }. The tight relationships between the SMBH mass and galaxy properties such as the stellar velocity dispersion and the mass of the bulge provide compelling observational evidence supporting a connection between SMBH accretion and the evolution of their host galaxies \citep[e.g.,][]{1998Magorrian,2000Ferrarese,2000Gebhardt,2003Marconi,2013Kormendy,2014Madau}. The key early growth phase of galaxies and their SMBHs is expected to be predominantly obscured \citep{2008aHopkins,2012Alexander,2018Blecha, 2018HickoxyAlexander}, with large reservoirs of gas and dust feeding both the SMBH and the galaxy. Hence, obscured quasars provide key information on the parallel evolution between SMBHs and galaxies. However, due to observational limitations, the reliable identification and characterization of obscured quasars is challenging, presenting a significant hurdle when investigating SMBH and galaxy growth.

So far, the majority of the sizeable, large-scale AGN optical spectroscopic surveys target AGN and quasars that are luminous in the optical band \citep[e.g.,][]{2020Rakshit, 2023Alexander, 2024StoreyFisher}, such as surveys from the Sloan Digital Sky Survey (SDSS, \citealp[][]{2000York}) and the Dark Energy Spectroscopic Instrument (DESI, \citealp{2016DESI1}). SDSS quasar surveys have made a remarkable contribution to our AGN knowledge characterizing accretion disk (AD) physics \citep[e.g.,][]{2015LaMassa,2016Lusso}, the broad emission line region \citep[e.g.,][]{2006Nagao,2021Temple,2023Temple}, the broad absorption line region \citep[e.g.,][]{2006Trump,2020Rankine,2022Petley}, the clustering properties of quasars \citep[e.g.,][]{2017Laurent,2020Powell}, and optically bright Type 2 AGN \citep[e.g.,][]{2003Zakamska,2004Zakamska,2008Zakamska}, among other key results. However, due to sensitivity limitations, SDSS only targets optically bright AGN; hence, it misses the wide majority of the faint, obscured systems, as the optical band is strongly affected by absorption. This presents a particular challenge for understanding the full AGN population as it is now known that the majority of the AGN population is obscured \citep[e.g.,][]{2006MartinezSansigre,2008Fiore, 2014Ueda, 2015Aird, 2015Ricci, 2024Laloux}.

X-ray observations are arguably one of the most efficient methods to identify AGN, including the obscured AGN population, for several reasons \citep[see][]{2015ABrandt}: (1) X-ray emission appears to be a near-universal property of AGN, (2) X-ray photons can penetrate relatively high column densities of gas with $N_{\rm H}\approx 10^{21}-10^{24} \rm \: cm^{-2}$, identifying obscured AGN that optical selections would miss, and (3) the contamination from the host galaxy emission in the X-ray band is negligible in most cases. Furthermore, the X-ray emission from the entire AGN population is imprinted on the cosmic X-ray background \citep[e.g,.][]{1995Comastri, 2007Gilli, 2014Ueda, 2019Ananna}, providing a key probe of SMBH accretion in the universe. In the local universe, we have an excellent census of the obscured AGN population with the BAT AGN Spectroscopic Survey (BASS, \citealp{2017Koss}), based on the 70- and 105-months Swift-BAT All-sky Hard X-Ray survey \citep{2013Baumgartner,2018Oh}. BASS is an all-sky survey in the ultra-hard X-ray band ($\rm 14-195 \: keV$) able to penetrate very large levels of absorption, even in the Compton-thick (CT, $N_{\rm H}>10^{24} \rm \: cm^{-2}$) regime \citep[e.g.,][]{2015Ricci}. BASS provides optical spectroscopy for an almost obscuration-unbiased AGN sample at least up to $N_{\rm H}\approx 10^{24}\rm \: cm^{-2}$ in the local universe ($z<0.2$) or for relatively bright X-ray sources (14-195\,keV limiting depth of $8.40\times 10^{-12}\rm \, {erg}\:s^{-1}\:cm^{-2}$, \citealp[][]{2018Oh}). However, at larger redshifts, due to sensitivity limitations, when the level of obscuration is CT, even the deepest X-ray surveys will miss most heavily obscured and CT AGN \citep[e.g.,][]{2004Treister, 2008Alexander, 2017Lansbury}.

An alternative method for identifying obscured AGN involves using the mid-infrared (MIR) band. Bright MIR emission is a key signature of SMBH accretion, and arises from the reprocessing of optical/UV radiation produced within the accretion disk by circumnuclear dust (the "dusty torus"). Hence, MIR emission is not strongly affected by absorption and is able to identify obscured AGN that may escape optical and even X-ray selection approaches \citep[e.g.,][]{2012Donley,2016DelMoro,2022Andonie}. However, a challenge with the MIR band is that it can be highly contaminated by star formation (SF) emission from the host galaxy, making AGN identification challenging, in particular when the host galaxy dominates the MIR emission. To address this problem, a number of MIR AGN identification techniques have been developed. One of the most popular criteria is based on MIR colours, which are sensitive to the dust temperature and, hence, can be successfully used to distinguish AGN from star-forming galaxies \citep[e.g.,][]{2005Stern,2007Lacy,2012Donley,2012Mateos,2018Assef}. Currently, AGN selection approaches using MIR colours are typically defined using photometry from the {\it Spitzer} \citep{2004Werner} and {\it Wide-field Infrared Survey Explorer} ({\it WISE}, \citealp[]{2010Wright}) telescopes. {\it WISE} has made a significant contribution to the discovery of luminous MIR quasars, which are typically characterized by red MIR colours. There are various AGN selection techniques based on $W1$ and $W2$ {\it WISE} magnitudes that reliably identify AGN \citep[e.g.,][]{2012Stern,2018Assef}. Given that {\it WISE} is an all-sky survey, these selections have the potential to find a very large number of AGN across the entire sky. 

Multiple studies have published spectroscopic catalogues based on MIR-selected AGN. For example, there are several dedicated spectroscopic surveys following up MIR-selected AGN in small fields of the sky, containing several hundred up to a few thousand objects \citep[e.g.,][]{2013Lacy, 2012Kochanek,2024Li}. A few works have also studied samples of a few thousand {\it WISE} selected AGN across the entire sky, using archival optical spectroscopic data \citep[e.g.,][]{2018DiPompeo,2023Carroll}; hence, those samples do not provide a uniform and homogeneous spectroscopically identified MIR AGN sample. Consequently, large-scale spectroscopic surveys targeting MIR-selected AGN are, so far, nonexistent.

Recently, several studies have explored the properties of MIR-selected AGNs finding interesting results that, as a whole, provide strong evidence for an evolutionary connection between SMBHs and galaxies. First, obscured and unobscured quasars seem to reside in different large-scale environments, where (on average) obscured sources are hosted in more massive halos and are more clustered than their unobscured counterparts \citep[e.g.,][]{2011Hickox, 2014DiPompeo, 2018Powell,2023Petter}. Second, the host galaxies of obscured quasars have larger SFRs than unobscured quasars \citep[e.g.,][]{2015Chen,2022Andonie}, and are more likely to reside in compact starbursts which can produce an obscuration comparable with that of the dusty torus \citep[e.g.,][]{2019Circosta,2024Andonie}. Third, IR AGN are more likely to be found in close galaxy pairs compared to optically or X-ray selected systems which are more typically unobscured AGN \citep[e.g.,][]{2014Satyapal, 2024Dougherty,2024LaMarca}. These results indicate environmental differences between obscured and unobscured AGN, suggesting that at least some fraction of the obscured MIR AGN population represents a distinct phase in the overall evolution of galaxies, a hypothesis proposed by several observational and theoretical studies \citep[e.g.,][]{1988Sanders,2005DiMatteo,2008aHopkins}. However, this scenario has been difficult to directly test since (1) many of the aforementioned studies have been performed in deep but small fields (and hence they are limited in source statistics), or (2) they lack spectroscopic confirmation, relying on more uncertain photometric redshifts. 

Here, we present the 4MOST\footnote{4MOST (4-metre Multi-Object Spectroscopic Telescope); \citealp{2019deJong} is a wide-field spectroscopic survey facility which will be mounted at the VISTA telescope, and is expected to obtain optical spectra for more than 80 million sources over an area of $>17000 \rm \: deg^{-2}$ with a 5-year survey starting late 2025/early 2026. 4MOST will complement large multi-wavelength surveys such as those undertaken by {\it eROSITA}, {\it Gaia}, and {\it Euclid}, as well as ground-based optical imaging and spectroscopic facilities such as LSST, SDSS, and DESI.} IR AGN survey (PI: D.M. Alexander), the first large-scale optical spectroscopic survey characterizing the MIR-selected obscured AGN population. The objective of the 4MOST IR AGN survey is to provide the first large and complete obscured MIR AGN census up to redshift $\sim 1$ with a spectroscopically identified IR AGN sample $\sim 100\times$ larger than existing studies. This will allow for the most statistically robust characterisation of the obscured IR AGN population, including identifying new AGN subpopulations and testing different AGN-host galaxy evolutionary scenarios. The 4MOST IR AGN survey is designed to complement the 4MOST X-ray AGN survey (PI: A. Merloni, \citealp[][]{2019Merloni}), which targets {\it eROSITA} \citep[][]{2012Merloni,2021Predehl} selected AGNs. {\it eROSITA} has a predominantly soft
X-ray (0.5--2\,keV) response, and largely identifies luminous unobscured AGN \citep[][]{2022Liu}, while the 4MOST IR AGN survey will predominantly identify the obscured AGN missed by {\it eROSITA}. Aside from the 4MOST IR AGN survey, there are other surveys that have the potential to investigate obscured AGNs such as DESI \citep[][]{2023Alexander,2023Fawcett}, the Optical, Radio Continuum and HI Deep Spectroscopic Survey \citep[ORCHIDSS;][]{2023Duncan}, the 4MOST Gaia Purely Astrometric Quasar Survey (PAQS; \citealp{2023MsngrKrogager}) and the Chilean AGN/Galaxy Extragalactic Survey \citep[ChANGES;][]{2023MsngrBauer}. All these surveys combined will open up new grounds for the exploration of the optical spectroscopic properties of obscured IR AGNs. For more information on the 4MOST capabilities and instrumental specifications, see \citet{2019deJong}.

This paper presents the design of the 4MOST IR AGN survey, our target selection approach, and predictions for the AGN and host galaxy properties of the identified sources. The structure of the paper is as follows. In Section\,\ref{sec:data}, we describe the datasets used to construct our catalogue and to test our AGN selection. In Section\,\ref{sec:AGNident}, we describe the 4MOST IR AGN survey design, describing the obscured IR AGN selection function and the methods used to test their performance. In Section\,\ref{sec:4MOST}, we then present our final catalogue, the predictions for the properties of the survey, and the scientific goals of our survey. Finally, Section\,\ref{sec:conc} summarizes this work.

Throughout this work, we adopt a concordance cosmology \citep{2013Hinshaw}, and a Chabrier initial mass function \citep[][]{2003Chabrier}. The magnitudes are in Vega units unless we state otherwise.

\section{Datasets and methods used to define the 4MOST IR AGN survey} \label{sec:data}

This section describes the different datasets used to define the 4MOST IR AGN survey. In brief, to identify obscured IR AGN candidates, we first select AGN using the \citet{2018Assef} 90\% AGN reliability colour selection technique (hereafter \citetalias{2018Assef} R90), which requires a $3\sigma$ detection in $W1$ and a $5\sigma$ detection in $W2$, and selects the sources that meet the following criteria

\begin{equation} \label{eq:A18R90}
W1-W2>
    \begin{cases}
      \alpha_{R} \exp \{ \beta_R (W2-\gamma_R)^2 \},\, W2>\gamma_R,\\
      \alpha_{R}, \, W2\leq \gamma_R,
    \end{cases}    
\end{equation}

\noindent where $\alpha_R=0.65$, $\beta_R=0.153$, and $\gamma_R=13.86$ (see also Figure\,\ref{fig:A90_cat} of Appendix\,\ref{ap:reliability}). We then select the obscured sources using an optical-to-MIR colour cut, following \citet{2007Hickox}. While the entire survey covers $\sim 10,000\: \rm  deg^2$ of the sky, we test and optimize the selection in small but well-observed fields using UV-to-FIR spectral energy distribution template fitting and X-ray spectral constraints. Subsections\,\ref{sec:data:LS} and\,\ref{subsec:wise} describe the optical and IR datasets used to perform the target selection. Subsections\,\ref{sec:data:sed} and\,\ref{subsec:Xraydata} describe the datasets and methods used to test the AGN reliability of our selection approach, and predict the AGN and host galaxy properties of the sample.

\subsection{Parent sample: DESI Legacy Surveys DR10} \label{sec:data:LS}

The parent sample of the 4MOST IR AGN survey is the Legacy Surveys (LS) DR10 South\footnote{See the LS DR10 description here: \url{https://www.legacysurvey.org/dr10/description/}.}. The original LS map 14,000 $\rm deg^2$ of the sky in the optical bands $g$, $r$, and $z$ \citep{2019Dey}. LS is primarily constructed using three imaging surveys. The observations in the South Galactic Cap are conducted by The Dark Energy Camera Legacy Survey ({\it DECaLS}) and are carried out by the Dark Energy Camera \citep[DECam;][]{2015Flaugher} at the Blanco Telescope at the Cerro Tololo Inter-American Observatory. In the North Galactic Cap, the observations are conducted by the Beijing-Arizona Sky Survey ({\it BASS}) and the Mayall z-band Legacy Survey ({\it MzLS}). The {\it BASS} observations were taken at the University of Arizona Steward Observatory 2.3 m (90 inch) Bart Bok Telescope at the Kitt Peak National Observatory, and the {\it MzLS} observations are taken at the Mayall Telescope at the Kitt Peak National Observatory. LS DR10 expanded the original LS to $\rm > 20,000 \:deg^2$, adding optical $i$-band DECam observations from NOIRLab in the Southern Galactic Cap. LS DR10 also provides deblended {\it WISE} photometry in the $W1$, $W2$, $W3$, and $W4$ bands, which is PSF-fitted and measured at the position of the $griz$ detected sources \citep{2019Dey}. Overall, the LS DR10 contains $\sim2.8$ billion unique sources.

The depth of the LS is not uniform and depends on the number of exposures in the area. The point-source $5\sigma$ depths for two observations in AB magnitudes\footnote{See \url{https://www.legacysurvey.org/dr10/description/\#depths} for more details} (90\% of the survey has at least two observations in $grz$; \citealp[][]{2019Dey}) are $g=24.7$, $r=23.9$, $i=23.5$, and $z=23$.

\subsection{{\it WISE} data and matching with LS DR10 catalogue} \label{subsec:wise}

As previously mentioned, we identify the IR AGN sample using the \citetalias{2018Assef} R90 selection, which was calibrated using AllWISE photometry. Since the publication of \citetalias{2018Assef}, newer {\it WISE} catalogues provide deeper {\it WISE} photometry and larger numbers of {\it WISE} sources, which could potentially provide more complete AGN samples. Hence, to optimize our IR AGN selection, we test the efficiency of the \citetalias{2018Assef} R90 selection for three different {\it WISE} datasets: the AllWISE\footnote{\url{http://wise2.ipac.caltech.edu/docs/release/AllWISE/}} catalogue, the unWISE catalogue \citep{2019Schlafly}, and the LS DR10-derived {\it WISE} catalogue. After performing extensive tests, we decided to adopt the unWISE fluxes to identify the IR AGN in the 4MOST IR AGN survey. See Appendix\,\ref{ap:reliability} for our tests and justifications of this choice, in addition to the details of each catalogue and their efficiency in identifying AGN. Since the LS DR10 is our parent sample and provides the optical positions of the 4MOST IR AGN survey, we need to find the unWISE counterparts of the LS sources. The FWHM of the $W1$ and $W2$ PSFs are ${\approx}6$\farc{} \citep{2010Wright} with typical positional uncertainties ranging $0.4{-}1$\farc{}\footnote{The {\it WISE} positional uncertainties depend on the $\rm S/N$ of the sources. Before matching the catalogues, we pre-select the {\it WISE} sources with $W1\:\rm S/N{>}3$ and $W2\:\rm S/N{>}5$ (see Section\,\ref{sec:AGNident}). We find that 95\% of the {\it WISE} sources have a $W1\:\rm S/N{>}20$. Hence, we expect the positional uncertainties of our sources to be typically $<0.5$\farc{}. See the WISE All-Sky Data Release Explanatory Supplement for details (\url{https://wise2.ipac.caltech.edu/docs/release/allsky/expsup/sec6_4.html}).}, while the LS optical PSF is typically $\sim 1$\farc{}, with a typical positional uncertainty of 0.1\farc{} for AGN-like sources \citep[e.g.,][]{2022Salvato}. Hence, we must be careful when matching the catalogues to avoid source misidentification. To do this, we find all the matches between the {\it WISE} and the optical LS positions within 2\farc{} separation. When one {\it WISE} source has multiple optical counterparts, we pick the optical source with the largest W2 flux in the LS DR10 catalogue, which is associated with the central galaxy instead of an extended sub-structure. We chose a  2\farc{} matching radius after visualizing the unWISE and LS images of a random sample of our sources. We find that the typical distance between the MIR and optical positions for the same source is always $<2$\farc{}; hence, a larger matching radius would significantly increase the number of source misidentifications between two point-like sources. In addition, we calculate the probability of having a spurious match within a 2\farc{} matching radius based on the source density of \citetalias{2018Assef}-selected AGNs ($\sim 140 \rm \: deg^{-2} $; see Table\,\ref{t:reliability} in Appendix\,\ref{ap:reliability}), and we find that it is only $\sim 0.05\%$. Furthermore, we find that the probability of a true match between an optical and {\it WISE} position at $>$2\farc separation, given the {\it WISE} and LS positional uncertainties, is $<3\%$, which is consistent with that found in previous works \citep[e.g.,][]{2024Greenwell}. A 2\farc{} matching radius between {\it WISE} and optical catalogues has also been found by previous work to be a good compromise to minimize the spurious matches \citep[e.g.,][]{2012Mateos, 2013Assef, 2018Assef,2024Greenwell}. We apply the same method to find the LS counterparts of AllWISE sources in our comparisons.

\subsection{UV-to-FIR SED fitting} \label{sec:data:sed}

To test the AGN reliability of our obscured IR AGN colour selection approach and also predict the properties of the 4MOST IR AGN survey, we perform detailed UV-to-FIR spectral energy distribution (SED) template fitting in four deep and well-observed fields in the sky: COSMOS, XMM-LSS, Bo\"{o}tes, and ELAIS-S1 fields (see Table\,\ref{t:data}). These fields were selected because they have UV-to-far-IR (FIR) photometric coverage and X-ray spectral observations. This subsection describes the photometric datasets and the SED modelling adopted in this work.

\subsubsection{Multiwavelength photometric datasets} \label{sec:data:photo}

The photometry used to perform SED fitting was taken from the Herschel Extragalactic Legacy Project (HELP, \citealp[][]{2019Shirley, 2021Shirley}) for the  Bo\"{o}tes, XMM-LSS, and ELAIS-S1 fields. The HELP survey provides homogeneous optical, near-IR (NIR), MIR, and FIR photometry for the {\it Herschel} SPIRE extragalactic survey fields: the Herschel Multi-tiered Extragalactic Survey \citep{2012Oliver} and the Herschel Atlas survey \citep{2010Eales}. In the case of COSMOS, we take the UV-to-NIR photometry from the COSMOS2015 catalogue \citep{2016Laigle}, and the MIR and FIR photometry from the \citet{2018Jin} catalogue. For all sources, we take the spectroscopic redshift when available, otherwise, we use the photometric redshift (see \citealp[][]{2021Shirley} for details on the redshifts of the HELP fields, and Section\,2.1 of \citealp[][]{2022Andonie} for details on the redshift assignment for the COSMOS field).

The available UV-to-near-IR (NIR) photometry varies for each field since they have been observed by different instruments, but the MIR and FIR observations come from the same telescopes: {\it Spitzer} and {\it Herschel}. The MIR and FIR photometry includes the fluxes from the following bands: {\it Spitzer} IRAC fluxes at 3.6, 4.5, 5.8, and 8 $\mu \rm m$; {\it Spitzer} MIPS at 24 $\mu \rm m$; {\it Herschel} PACS at 100 and 160 $\mu \rm m$; and {\it Herschel} SPIRE at 250, 350, and 500 $\mu \rm m$. Table\,\ref{t:data} summarizes the different UV-to-NIR datasets used in each field.

\begin{table*}
\centering

 \begin{tabular}{lp{3cm}p{5cm}p{5cm}} 
 \hline
 \hline
\noalign{\smallskip}
 Field &  Photometric catalogue & UV-to-NIR photometry &X-ray catalogue   \\
\noalign{\smallskip}
 \hline
\noalign{\smallskip}
COSMOS & COSMOS2015 \citep{2016Laigle} and \citet{2018Jin} IR catalogue &  - CFHT: $u*$\newline - Subaru Suprime-Cam: $V$, $B$, $r+$, $i+$, $z++$\newline - UltraVISTA-DR2: $Y$, $J$, $H$, and $Ks$ & {\it Chandra} X-ray spectral constraints from \citet{2023Laloux}   \\
 \noalign{\smallskip}
XMM-LSS & HELP \citep{2021Shirley} &  - CFHT/Megacam:$U$\newline - Subaru/Suprime: $g$, $r$, $i$, $z$ \newline - VISTA: $Y$, $J$, $H$, and $Ks$   & {\it XMM-Newton} point-source catalogue from \citet{2018Chiappetti}  \\
\noalign{\smallskip}
 Bo\"{o}tes &HELP \citep{2021Shirley}& - Large Binocular Telescope Observatory: $U$\newline - KPNO/Mosaic: $B$, $r$, $i$, $z$\newline - NOAO/NEWFIRM: $J$, $H$, $K$ & {\it Chandra} point-source catalogue from \citet{2020Masini}\\
\noalign{\smallskip}
ELAIS-S1 &HELP \citep{2021Shirley}& - CTIO/DECam: $g$, $r$, $i$, $z$\newline - VISTA: $Y$, $J$, $H$, and $Ks$. & {\it XMM-Newton} point source catalogue from \citet{2021Ni} and spectral constraints from \citet{2023Yan} \\
\noalign{\smallskip}
\hline
 \noalign{\smallskip}
 \end{tabular}
 \caption{Multiwavelength datasets used in each field to perform the UV-to-FIR SED fitting and the X-ray analysis. } \label{t:data}
\end{table*} 

\subsubsection{SED modelling} \label{sec:sedfitting}

To perform SED fitting, we use the multicomponent Bayesian SED fitting code \textsc{fortesfit}\footnote{\url{https://github.com/vikalibrate/FortesFit}} \citep[][]{2019Rosario}. \textsc{fortesfit} is a publicly available \textsc{python}-based hierarchical Bayesian modelling framework for multi-wavelength SED fitting. It allows the incorporation of user-defined SED models and the combination of these in a fit with flexible specifications of the prior distribution on modelling parameters. \textsc{fortesfit} also allows the construction of full posterior distributions of all free parameters, providing accurate parameter uncertainties. It has been successfully used in several works to measure the physical properties of AGNs and galaxies from fitting their broad-band UV-to-radio SEDs \citep[e.g.,][]{2019Rosario, 2021Scholtz, 2022Andonie, 2023Laloux, 2024Andonie}. 

In our implementation of the SED fitting for our sample, we use \textsc{fortesfit} along with a set of models that feature emission from the accretion disc and torus of the AGN, as well as stellar and dust components of the host galaxy (see below). \textsc{fortesfit} does not explicitly assume energy balance in the galaxy emission, as performed in some other SED fitting codes (e.g., CIGALE). As accurate galaxy star-formation histories are not the main focus of this analysis, strict energy balance is not a critical requirement for our fits. 

In our \textsc{fortesfit} implementation, we use the \textsc{python} adaptation \citep[][]{2014Buchner}  of the Nested Sampling algorithm MultiNest \citep[][]{2008Feroz, 2009Feroz, 2019Feroz} as a fitting engine. \textsc{fortesfit} uses the likelihood function outlined by \citet{2012Sawicki}, which accounts for photometric detections and upper limits self-consistently by treating them as a combination of independent Gaussian likelihoods (i.e, a blend of $\chi^2$ and error functions).

We adopt a similar SED fitting approach to that taken in \citet{2022Andonie} to identify the AGN and to model the AGN and host galaxy properties of our sample. Here, we fit the observed-frame $0.3-500 \rm \: \mu m$ SED of the objects. In modelling the SEDs, we include the four following components: (1) the unabsorbed stellar emission characterised by the \citet{2003BC} stellar-population models, attenuated according to the Milky Way reddening law from \citet{2000Calzetti}; (2) the UV emission from the accretion disk following the empirical model of \citet{2006Richards} and the \citet{1992Pei} reddening law for the Small Magellanic Clouds; (3) the IR emission from the AGN torus, based on the empirical DECOMPIR AGN model of \citet{2011Mullaney}; and (4) a semi-empirical star-forming galaxy model which reproduces the full range of dust temperatures observed in galaxies \citep{2014Dale}. 

\begin{table*}
\centering

 \begin{tabular}{lcccc} 
 \hline
 \hline
\noalign{\smallskip}

\bf Component &\bf  Parameter & \bf Description &{\bf Range} & {\bf Reference} \\
\noalign{\smallskip}
 \hline
\noalign{\smallskip}
\noalign{\smallskip}
Stellar population emission (SP)  & $\log M_{\star} \rm \: [M_{\odot}]$ & stellar mass & $\rm [7,13]$ & \citet[][]{2003BC}\\
 \noalign{\smallskip}
 & $\log t'\rm \: [Gyr]$  & age of the stellar population & $\rm [8,10.1]$&\\
 \noalign{\smallskip}
 & $\log \tau\rm \: [log\:Gyr]$  & star-formation timescale  & $\rm [0.01,15]$ &\\
 \noalign{\smallskip}
& $\rm E(B-V)_{SP} \: [mag]$  & galaxy reddening & $\rm [0,0.5]$ & \citet[][]{2000Calzetti}\\
\noalign{\medskip}

Accretion disk (AD) & ${\rm log \:} L_{2500}\rm \: [erg \: s^{-1} \: Hz^{-1}]$ & luminosity at 2500{\AA} & $[26,36]$ & \citet[][]{2006Richards} \\
 \noalign{\smallskip}
 & $\rm E(B-V)_{AD} \: [mag]$  & accretion disk reddening & $\rm [0,1]$ & \citet[][]{1984Prevot}\\
\noalign{\medskip}

Torus & ${\rm log \:} L_{\rm AGN, IR} \rm \: [erg \: s^{-1}]$ & $8-1000 \mu m$ AGN luminosity & $[38,48]$ & \citet[][]{2011Mullaney} \\
 \noalign{\smallskip}
& $\Gamma_{\rm s}$  & short-wavelength slope & $[-0.3,1.4]$ & \\
 \noalign{\smallskip}
& $\Gamma_{\rm L}$  & long-wavelength slope & $[-1,0.5]$ &\\
 \noalign{\smallskip}
& $\lambda_{\rm Brk} \rm \: [\mu m]$  & turnover wavelength & $17.5^*$\\
 \noalign{\smallskip}
 & $\lambda_{\rm BB}\rm \: [\mu m]$  & black body peak wavelength & $40^*$\\
 \noalign{\smallskip}
 & $T_{\rm BB, s} \rm \: [K]$  & hot dust temperature & $1500^*$ \\
 \noalign{\medskip} 

Star formation (SF) & ${\rm log \:} L_{\rm SF, IR} \rm \: [erg \: s^{-1}]$ & $8-1000 \rm \: \mu m$ star-formation luminosity & $[42,48]$ &\citet[][]{2014Dale}\\
 \noalign{\smallskip}
& $\alpha_{\rm SF}$  & shape parameter & $[0.06,4]$ &\\
 \noalign{\smallskip}
\hline
 \noalign{\smallskip}
 \end{tabular}
 \caption{Description and allowed ranges of the parameters of each component identified in our SED fitting models. For more details on each model, please refer to the references. Parameters with a * symbol are fixed.  } \label{t:params}
\end{table*}

To break the degeneracy between fitting the accretion disk and the stellar emission in the rest-frame optical-near-IR waveband, in the case of X-ray detected sources, we use the well-established $L_{2500}$--$L_{\rm 2keV}$ relation for quasars reported by \citet{2016Lusso}. We apply a Gaussian prior centred at the estimated value for $L_{2500}$, with a standard deviation of $\sigma = 0.3 \rm \: dex $. In the case of X-ray undetected sources, we do not include the AD emission in our SED modelling. As demonstrated below in Section\,\ref{subsec:predictionsAGN}, X-ray undetected AGN are typically obscured by $N_{\rm H}>10^{22} \rm \: cm^{-2}$. Hence, most or all of the optical emission coming from the AD will be extinguished, and it is a reasonable assumption to associate all of the observed optical emission with the stellar population. We note that $\lesssim 15\%$ of the AGN with $N_{\rm H}>10^{21.5} \rm \: cm^{-2}$ are expected to show signs of broad lines or a reddened accretion disc in their optical spectra \citep[e.g.,][]{2014Merloni}; hence, not including the accretion disk component in the SED modelling could cause a small fraction of the X-ray undetected sources to have overestimated stellar masses. However, we do not consider this a major issue as the main objective of the SED modelling is to identify the IR AGNs. We summarize the different components and parameters of our SED modelling in Table\,\ref{t:params}. For more details on each model and SED fitting implementation, we refer the interested reader to Section 3 of \citet{2022Andonie}. Figure\,\ref{fig:SEDcomp_ex} in Appendix\,\ref{ap:sed} shows examples of the best-fitting SED models for two 4MOST IR AGN sources in the XMM-LSS field.

We classify a source as an AGN only if the $8-1000 \rm \: \mu m $ AGN luminosity ($L_{\rm AGN, IR}$) posterior distribution is well constrained, following the approach of \citet{2022Andonie} (see Section\,3.3 of that paper). We consider $L_{\rm AGN, IR}$ to be well constrained (i.e., the source hosts an AGN) when the 1st percentile of $\log L_{\rm AGN, IR}/(\rm erg\:s^{-1})$ is larger than 38.5 (i.e., $>0.5 \rm \: dex$ from the lower limit of the prior; see Table\,\ref{t:params}). Similarly, we consider $L_{\rm SF, IR}$ to be well constrained when the 1st percentile of $\log L_{\rm SF, IR}/(\rm erg\:s^{-1})$ is higher than 42.5. We note that the thresholds $\log L_{\rm AGN, IR, 1st\: percentile}/(\rm erg\:s^{-1})=38.5$ and $\log L_{\rm SF, IR, 1st\: percentile}/(\rm erg\:s^{-1})=42.5$ do not have a redshift dependence and they are only an indication on whether the AGN or SF luminosity posterior distributions have a non-zero probability of taking the lowest possible value (i.e., having a value $<0.5 \rm \: dex$ from the lower limit of the prior).

Throughout the text, to compute the median values and standard deviation of $L_{\rm AGN, IR}$, $ L_{\rm SF, IR}$, and $ M_{\star}$ of different samples, we use \textsc{posterior stacker}\footnote{See \url{https://github.com/JohannesBuchner/PosteriorStacker/tree/main}}. This code takes as input samples of posterior distributions and fits them using a Gaussian model \citep[][]{2020Baronchelli}, taking into account all of the information contained in the posterior distribution of each parameter to analyze the samples.

\subsection{X-ray data, spectral constraints, and X-ray stacking} \label{subsec:Xraydata}

In this work, we use different X-ray datasets ({\it Chandra}; {\it XMM-Newton}) to perform a variety of analyses; see Table~\ref{t:data}. Our primary X-ray dataset is the {\it Chandra} data in the COSMOS field, which we use to demonstrate our obscured AGN selection approach. The XMM-LSS,  Bo\"{o}tes, and ELAIS-S1 X-ray datasets are used solely to make predictions of the basic obscuration properties of the 4MOST IR AGN survey. Finally, the X-ray data from the {\it eROSITA} Final Equatorial-Depth Survey are utilised as an equivalent of the 4MOST X-ray AGN survey to compare with the properties of the 4MOST IR AGN survey.

\subsubsection{COSMOS survey}

The deepest and most complete X-ray observations in COSMOS were carried out using {\it Chandra}, as part of the \textit{Chandra} COSMOS-Legacy survey \citep[][]{2016Civano}. This is a 4.6 Ms \textit{Chandra} program that has observed the 2.2 deg$^2$ area of the COSMOS field, with a homogeneous exposure time of $\sim$ 160 ks, yielding a 2-10 keV flux limit of $\rm F_{2-10keV} \sim 1.9\times 10^{-15}  \: erg \: cm^{-2} \: s^{-1}$. 

The properties of the X-ray sources in the COSMOS field have been extensively studied in previous works. In this work, we adopt the X-ray spectral fitting measurements from \citet{2023Laloux}, who used a Bayesian approach and adopted a physically motivated torus model (UXCLUMPY, \citealp[][]{2019Buchner}) to robustly constrain the physical parameters of all the X-ray detected sources in COSMOS. They provide the $2-10\rm \: keV$ X-ray luminosity ($L_{\rm 2-10 \: keV}$), the line of sight column density ($N_{\rm H}$), and the obscured fraction of each source ($f_{\rm obs}$; e.g., the probability of each source of being obscured by $N_{\rm H}>10^{22}\rm \: cm^{-2}$).

\subsubsection{Bo\"{o}tes, XMM-LSS, and ELAIS-S1 surveys}

The Bo\"{o}tes, XMM-LSS, and ELAIS-S1 fields do not have X-ray observations as deep as COSMOS or the same level of published X-ray spectral constraints. However, they have available X-ray point source catalogues which are suitable for the analyses we want to perform; i.e., to provide an estimate of the X-ray luminosity and make predictions of the X-ray properties of the 4MOST IR AGN sample such as the fraction of X-ray detected sources and the fraction of obscured sources with $N_{\rm H}>10^{22} \rm \: cm^{-2}$. The X-ray luminosities are only used in our SED modelling to place a loose prior on the accretion disk luminosity (see Section\,\ref{sec:sedfitting}). Hence, the variety in the X-ray quality and information that we have for the various fields is not a significant limitation.

The  Bo\"{o}tes field has been extensively observed by {\it Chandra} over more than 15 years. Here, we use the catalogue of the Chandra Deep Wide-Field Survey from \citet{2020Masini}, which combines all available {\it Chandra} observations, reaching a  $2-7\rm \: keV$ limiting depth of $9 \times 10^{-16} \rm \: erg \: cm^{-2} \: s^{-1}$. Their catalogue provides estimates for the intrinsic X-ray luminosity and $N_{\rm H}$ measurements derived from hardness ratio analyses. The hardness ratio (HR) is defined as $\rm HR=(H-S)/(H+S)$, where H and S are the counts in the hard ($2-7 \rm \: keV$) and soft ($0.5-2 \rm \: keV$) bands, respectively. This method can provide an estimation of the $N_{\rm H}$, as obscuration mostly affects the soft X-ray band, while hard X-ray photons are able to penetrate larger column densities (for more details, see Section\,6.2 of \citealp{2020Masini}). 

The XMM-LSS field has been observed by {\it XMM-Newton} ($2-10\rm \: keV$ limiting depth of $3 \times 10^{-15} \rm \, erg \, cm^{-2} \, s^{-1}$) and has a public X-ray source catalogue \citep{2018Chiappetti}, which contains information on the observed flux and count rates of each source at different bands. We estimate the observed X-ray luminosity of the sources using the provided observed flux in the $2-10 \rm \: keV$ band. For this, we follow \citet{2003Alexander} and calculate $L_{2-10 \rm \: keV}=4\pi d^2_{\rm L}f_{2-10 \rm \: keV} (1+z)^{\Gamma-2}$, where $d^2_{\rm L}$ is the luminosity distance, $f_{2-10 \rm \: keV}$ is the observed flux in the $2-10 \rm \: keV$ band, and $\Gamma$ is the photon index which we set to $\Gamma=1.8$. We note that the observed $2-10 \rm \: keV$ X-ray luminosity provides a good estimate of the intrinsic X-ray luminosity as for the majority of the sources the absorption correction is small in the observed $2-10 \rm \: keV$ band\footnote{For example, assuming an absorbed power-law model, for an AGN with $N_{\rm H}=10^{23} \rm \: cm^{-2}$ at $z\sim 1$, the observed X-ray luminosity would correspond to 90\% of the absorption-corrected X-ray luminosity.}.

The ELAIS-S1 field has also been observed by {\it XMM-Newton} as part of the XMM-SERVS Survey (\citealp{2021Ni}, $0.5-10 \rm \: keV$ limiting depth of $1.3 \times 10^{-14} \rm \, erg \, cm^{-2} \, s^{-1}$), for which the X-ray point source catalogue contains the information on the observed flux of each source at different bands. \citet{2023Yan} studied the most obscured AGN in ELAIS-S1 and published the X-ray spectral constraints only for the heavily obscured AGN with $N_{\rm H}>5\times 10^{23} \rm cm^{-2}$. The $N_{\rm H}$ values for the rest of the sources were acquired by private communication. To estimate the X-ray luminosity, we follow the same approach as for the XMM-LSS field.

\subsubsection{{\it eROSITA} Final Equatorial-Depth Survey}

The {\it eROSITA} Final Equatorial-Depth Survey (eFEDS, \citealp{2022Brunner}) is a 140\,deg$^2$ area observed by {\it eROSITA} as part of the performance verification phase ahead of the planned four years of {\it eROSITA} all-sky scanning operations. eFEDS contains $27,910$ X-ray sources (out of which $\sim 22,000$ are AGN) detected down to $6.5 \times 10^{-15} \rm \: erg \: cm^{-2} \: s^{-1}$ in the $0.5-2 \rm \: keV$ band. The objective of eFEDs was to observe a small region to the full depth of the all-sky {\it eROSITA} survey. However, due to unforeseen circumstances, {\it eROSITA} only performed about half of the planned all-sky observations (only 4-5 scans were completed, instead of the planned 8 scans); hence, the eFEDS observations are about two times deeper than the final observations across the majority of the sky, and $\sim 4$ times deeper than the publicly available 1st-pass all-sky survey \citep[][]{2024Merloni}. 

In this work, we use the multi-wavelength counterparts catalogue from \citet{2022Salvato} to obtain the optical positions and the photometric information of the eFEDS sources, which were originally taken from the Legacy Surveys DR 8 (LS DR8)\footnote{We note that in Appendix\,\ref{ap:reliability}, we mention that the LS handles optically extended sources in a way that is not useful for identifying IR AGN. However, all the eFEDS sources are X-ray AGN by selection, and most of them have an optically point-like morphology. Hence, using {\it WISE} photometry from the LS DR 8 is not an issue for the analysis performed in the eFEDS (see Section\,\ref{sec:comparison_ero}).}. In addition, we also use the AGN catalogue from \citet{2022Liu}, which published the X-ray spectral fitting results of all the sources, such as the intrinsic $L_{2-10 \rm \: keV}$ and $N_{\rm H}$.

\subsubsection{X-ray stacking} \label{sub:xstacking}

To analyze the obscuration properties of the obscured IR AGN candidates that are undetected in the X-ray band, we performed X-ray stacking using the publicly available \textsc{python} code \textsc{StackFast}\footnote{The \textsc{StackFast} code is available in the following websites: \url{https://www.tonima-ananna.com/research} and \url{https://gitfront.io/r/user-5016120/9154498a888cfd94976fc57b51b447460c69b268/StackFast/}.} \citep[see also][]{2024Ananna}. In brief, \textsc{StackFast} is an X-ray stacking code that finds all of the {\it Chandra} observations of a given sample of objects and extracts the photons within the 90\% energy encircling radius ($r_{ee,90}$) around the position of each object. To calculate the net photons of each object, \textsc{StackFast} adopts a background with annular shape of radii (1.3-6) $r_{ee,90}$. It then calculates fluxes for each individual object and the mean stacked flux for the entire sample in the soft (0.5-2 keV) and hard (2-7 keV) X-ray band.

\section{4MOST IR AGN survey design} \label{sec:AGNident}

This section describes the steps performed to define the 4MOST IR AGN survey (subsection\,\ref{sec:selectionfunction}) and demonstrates two key factors: (1) the effectiveness of the optical-to-MIR colour cut to distinguish between obscured and unobscured AGN (subsection\,\ref{subsec:rW2}), and (2) the reliability of our AGN identification approach (subsection\,\ref{sec:finalselection}).

\subsection{The 4MOST IR AGN survey selection function} \label{sec:selectionfunction}

\begin{figure}
\centering
\includegraphics[scale=0.6]{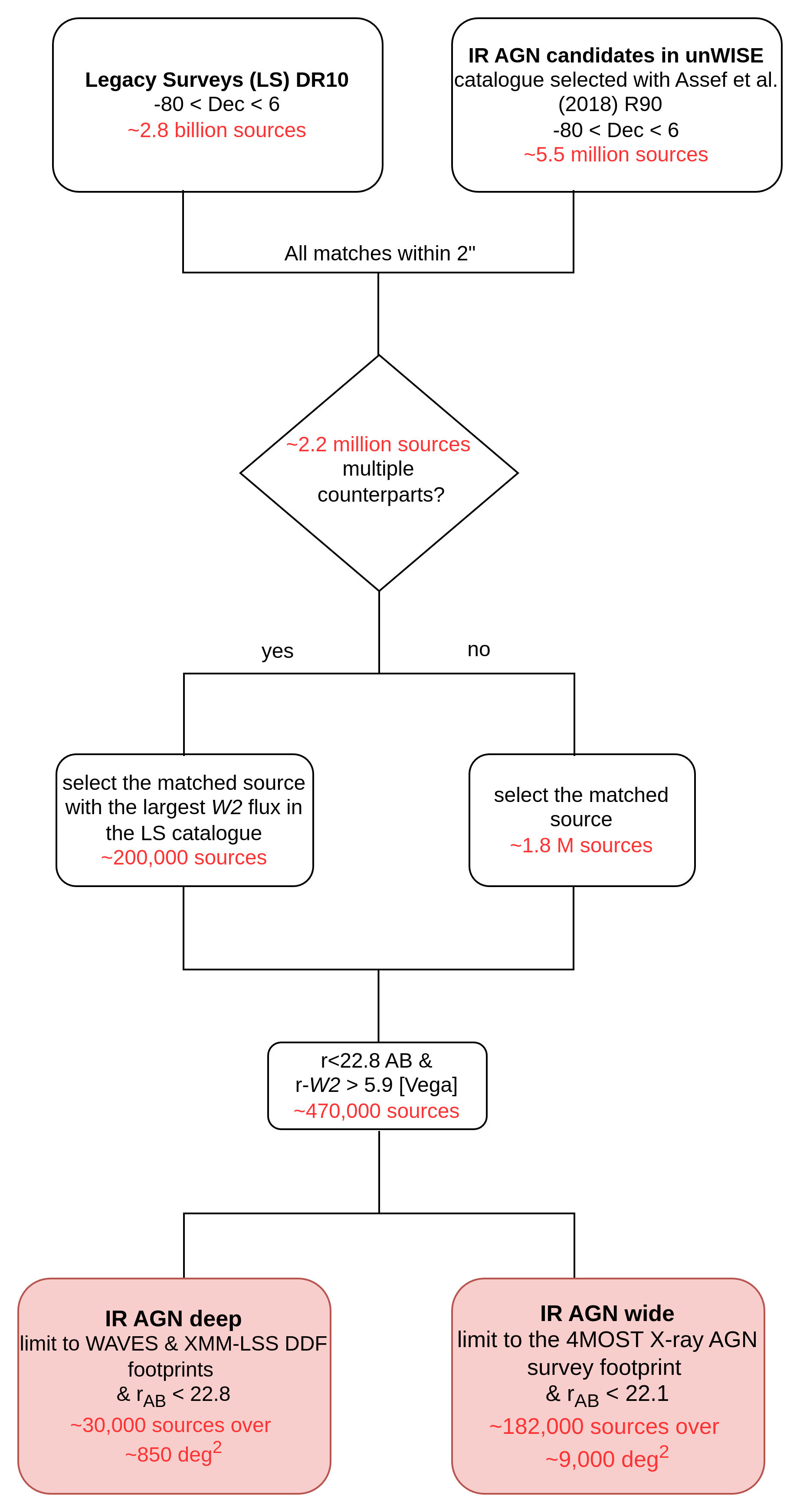}
\caption{4MOST IR AGN survey selection function. We first select all the IR AGN that meet the \citet{2018Assef} criteria using the unWISE catalogue with $\rm -80<Dec<6$, and find all the possible matches within 2\farc{} of targets in the Legacy Survey DR10 (LS10). When one unWISE source has multiple optical counterparts in the LS10, we pick the source with the largest {\it WISE W2} flux in the LS10 catalogue. To select the obscured IR AGN, we apply the colour cut $r-W2_{\rm unWISE}\geq 5.9 \rm \: mag$, and then restrict our sources to $r_{\rm AB}<22.8$, which corresponds to the optical limit of the 4MOST IR AGN survey. Given that our survey is designed to complement the 4MOST X-ray survey, we limit the catalogue to the 4MOST {\it eROSITA} footprint. Finally, we define the two survey components within this footprint: IR AGN deep, which will target sources down to $r_{\rm AB}=22.8$ in the WAVES and XMM-LSS (Deep Drilling Fields area) regions ($\sim 850\rm \: deg^2$), and IR AGN wide, which will target sources down to $r_{\rm AB}=22.1$ across the rest of the survey footprint ($\sim 9000\rm \: deg^2$; see Figure\,\ref{fig:coveragemap}). } 
\label{fig:selection}
\end{figure}

The 4MOST IR AGN survey selection function has several steps, which are illustrated in Figure\,\ref{fig:selection}. Every decision made during the selection has been previously justified in Section\,\ref{subsec:wise}, or will be well justified and validated later in this section using detailed UV-to-FIR SED template fitting and X-ray spectral constraints. We first find all the matches between the LS DR10 and the IR AGN catalogue, following Section\,\ref{subsec:wise} using the unWISE catalogue (see Subsection\,\ref{sec:finalselection}). Given the aims of our survey and the properties of the 4MOST instrument, we choose a magnitude limit of $r_{\rm AB}<22.8$ in order to achieve at least our minimum goals (i.e., a measured spectroscopic redshift; see Section\,\ref{subsec:spec_predictions}), for the majority of sources. Then, in order to select the most obscured AGN, we apply the optical to MIR colour cut of $r-W2\geq 5.9 \rm \: mag$ \citep[][]{2007Hickox}, following Section\,\ref{subsec:rW2}. This optical to MIR colour cut also delivers an obscured sample that complements the properties of the 4MOST X-ray AGN survey (see Section\,\ref{sec:comparison_ero} and Figure\,\ref{fig:CompeRO}). At this point, we have $\sim 470,000$ IR AGN candidates. As we need the IR AGN survey to spatially overlap with the 4MOST X-ray survey, we limit our catalogue to the 4MOST X-ray survey footprint, reaching $\sim 212,000$ sources, applying a $r_{\rm AB}<22.1$ cut for those that lie in the ``IR AGN wide" footprint (see below).

\begin{figure*}
\centering
\includegraphics[trim={0cm 0 0cm 0cm},clip,scale=0.35]{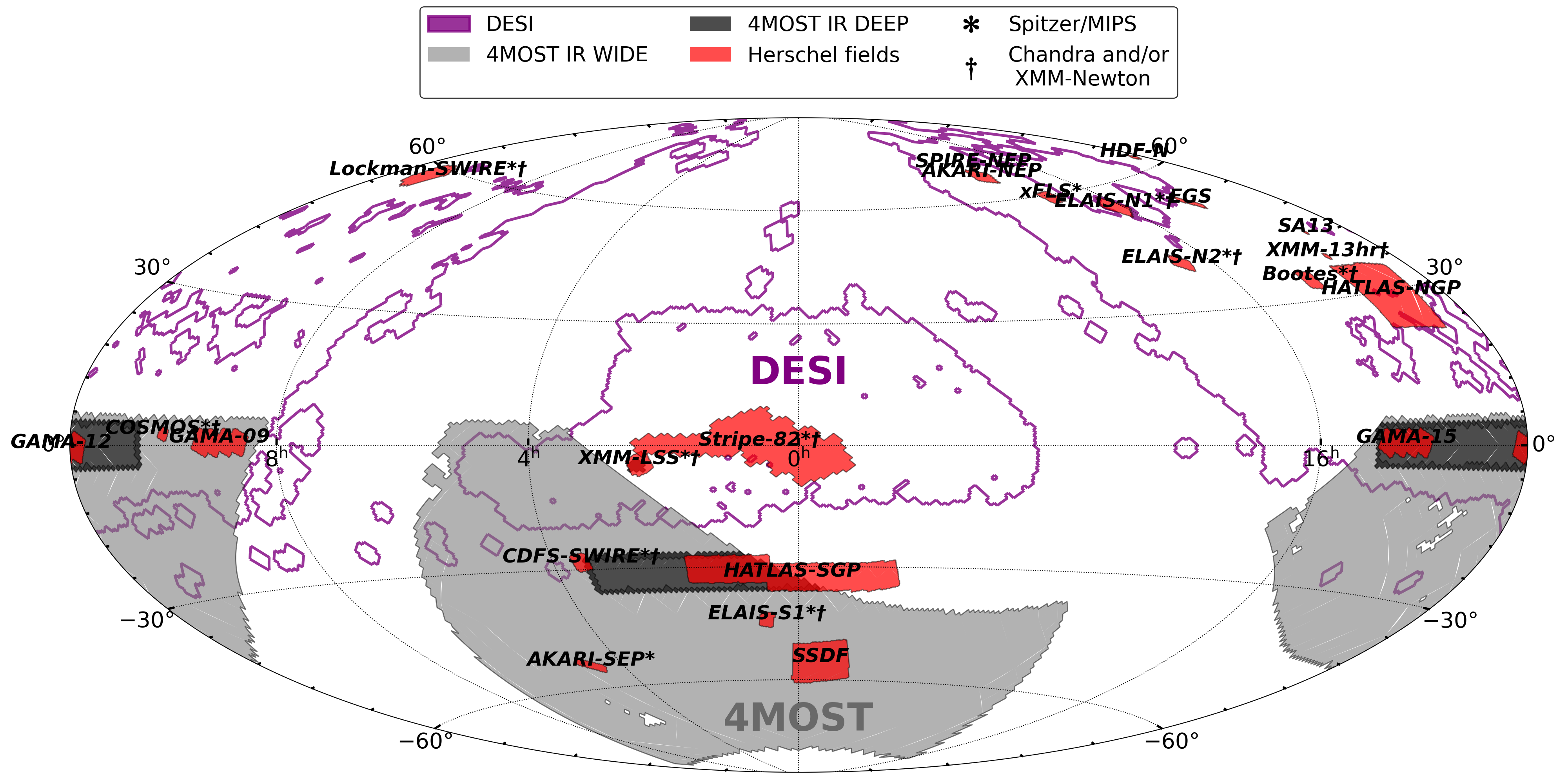}
\caption{Aitoff projection of the 4MOST IR AGN survey sources in Equatorial coordinates (J2000). The light grey highlighted region indicates the IR AGN wide component ($r_{\rm AB}<22.1$), and the dark grey highlighted region indicates the IR AGN deep component ($r_{\rm AB}<22.8$). The full 4MOST IR AGN survey footprint has been observed by {\it eROSITA} and follows that of the 4MOST X-ray AGN survey. We also show some multi-wavelength surveys overlapping with the 4MOST footprint. Red regions show fields with {\it Herschel} coverage, black asterisks show fields with {\it Spitzer} MIPS $24\rm \mu m$ coverage, black crosses show fields with deeper X-ray coverage from {\it Chandra} or {\it XMM-Newton}, and the purple open region indicates the coverage map of the DESI survey which partially overlaps ($\sim 2500\,\rm deg^2$) with our 4MOST IR AGN survey, including all of the equatorial IR AGN deep field.} 
\label{fig:coveragemap}
\end{figure*}

The 4MOST IR AGN survey has two components; see the spatial footprint of the survey in Figure\,\ref{fig:coveragemap}. A deep component, which will target $\approx 30,000$ sources down to $r_{\rm AB}<22.8$ over $\sim 850 \rm \: deg^2$ with more extensive deeper multi-wavelength photometry, and a wide component, which will target $\approx 182,000$ sources down to $r_{\rm AB}<22.1$ over $\sim 9,000 \rm \: deg^2$. The deep component follows the footprints of two extragalactic surveys: the Wide Area Vista Extragalactic Survey (WAVES, \citealp[][]{2019Driver}), which is confined to the footprint of the Kilo-Degree Survey (KiDS, \citealp{2013deJong}) and the VISTA Kilo-Degree Infrared Galaxy Survey (VIKING, \citealp{2013Edge}); and the XMM-LSS field as defined by the LSST Deep Drilling Fields \citep[DDF,][]{2018Brandt}. This ensures that our deep survey will have complementary NIR photometry over the entire footprint, and additional MIR and FIR photometry in the XMM-LSS area. Figure\,\ref{fig:coveragemap} also shows the overlap of the 4MOST IR AGN survey with other key public spectroscopic and photometric surveys. Notably, we highlight the overlap of $\sim 2,500 \,\rm deg^2$ between the 4MOST IR AGN survey and DESI, including the entire equatorial IR AGN deep field, which will result in exceptionally deep and extensive optical spectroscopic data across this region.

We note that full UV-FIR SED modelling would be the most effective way to select obscured IR AGN; however, we cannot use this approach to reliably construct the 4MOST IR AGN survey because the required deep multiwavelength observations out to far-IR wavelengths only exist over a small fraction of the sky (see Figure\,\ref{fig:coveragemap} for the {\it Herschel} coverage). Instead we adopt {\it WISE}, an all-sky MIR survey, which offers us the widest possible areal coverage and a reliable obscured AGN selection strategy.

\subsection{Quantifying the $ r-W2$ colour as a proxy for AGN obscuration} \label{subsec:rW2}

To select obscured AGN, the key objective of the 4MOST IR AGN survey, we require a diagnostic sensitive to obscuration. \citet{2007Hickox} presented a simple optical to MIR colour cut to select obscured AGN. Unlike the MIR emission, the optical emission is very sensitive to obscuration and gets extinguished at relatively low obscuration levels. Hence, the optical to MIR colour is expected to vary with different amounts of obscuration. \citet{2007Hickox} proposed to use the colour $r-\rm [4.5]$ as a proxy for AGN obscuration, where $[4.5]$ correspond to the {\it Spitzer} IRAC 2 channel at $4.5 \rm \: \mu m$. Based on the observed bimodal distribution of $r-\rm [4.5]$ colour as well as the results of X-ray stacking analyses, they demonstrated that the boundary $r-\rm [4.5]\gtrsim \: 6.1 \rm \: mag$ selects obscured AGN with an average $N_{\rm H} \approx 3\times 10^{22} \rm \: cm^{-2}$. A decade later, \citet{2017Hickox} extended these analyses by studying the SED and spectral properties of SDSS quasars as a function of the optical to MIR colour, finding that obscured and unobscured quasars have remarkably similar UV-MIR SEDs with the major difference being clear UV-optical extinction in the obscured quasars. However, they did not explore the dependency of $r-\rm [4.5]$ with the absorbing column density. 

In this paper, we extend the \citet{2007Hickox,2017Hickox} analyses to explore the relationship between $r-\rm [4.5]$ and AGN obscuration using the X-ray spectral constraints in the COSMOS field, by analyzing the line-of-sight column density ($N_{\rm H}$), the obscured fraction ($f_{\rm obs}$, see Section\,\ref{subsec:Xraydata} for details), and the optical extinction towards the accretion disc. For this analysis, we performed UV-to-FIR SED fitting of all the X-ray AGN in COSMOS to constrain the accretion disk reddening $\rm E(B-V)_{AD}$ and the IR AGN luminosity, following the procedure described in Section\,\ref{sec:sedfitting}. Since we use {\it WISE} $W2 \: \rm 4.6 \mu m$ to construct the 4MOST IR AGN survey instead of {\it Spitzer}/IRAC 2 (since there is {\it Spitzer}/IRAC coverage over only a small fraction of the sky), we also extend these analyses using $r-W2$ \citep[e.g.,][]{2015DiPompeo,2023Petter}. However, {\it Spitzer}/IRAC is much deeper than {\it WISE} in the COSMOS field, providing better source statistics. We note that in our analyses here, we adopt $r-[4.5]\geq 5.9\: \rm  mag$ as the (predominantly) obscured AGN threshold to align with the definition of our 4MOST IR AGN survey. This $r-[4.5]$ cut is slightly less conservative than the $r- [4.5]\geq 6.1\: \rm mag$ threshold adopted in \citet{2007Hickox}. However, we note that this change does not quantitatively change the results presented in this paper and selects more sources in the interesting obscured--unobscured threshold where we may identify rare systems such as red and extremely red quasars \citep[][]{2015Ross,2017Hamann,2023Fawcett}.

\begin{figure}
\centering
\includegraphics[trim={2cm 0cm 4cm 0cm},clip, scale=0.38]{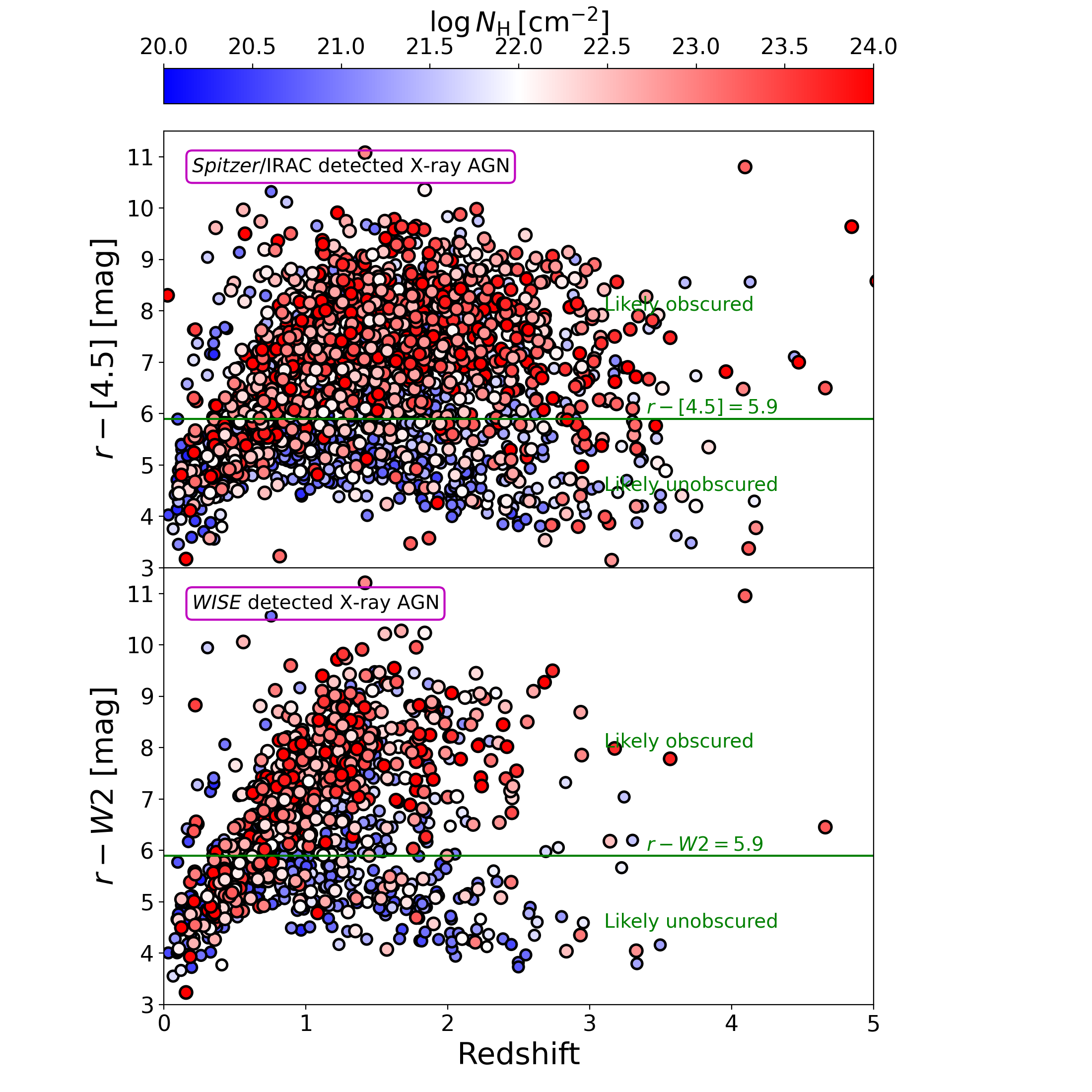}

\caption{Optical to MIR colour against redshift, colour-coded by the line of sight logarithmic column density ($\log N_{\rm H}$) for all the X-ray AGN with X-ray spectral constraints in the {\it Chandra} COSMOS field. The top panel plots the $r-[4.5]$ colour for all the sources detected by {\it Spitzer}/IRAC 2 at $4.5 \rm \:\mu m$, and the bottom panel plots the $\rm r-W2$ colour for all the sources with a W2 detection in the unWISE catalogue. The green line shows $r-[4.5]/W2=5.9 \: \rm mag$, the colour cut adopted in this work to separate obscured from unobscured AGN, based on \citet{2007Hickox}. The majority of the AGN with $N_{\rm H}\geq 10^{22}\rm \:cm^{-2}$ have $r-[4.5]/W2\geq 5.9$ colour. }

\label{fig:rW2_cosmos}
\end{figure}

\begin{figure}
\centering
\includegraphics[trim={2cm 0cm 0cm 0cm},clip, scale=0.38]{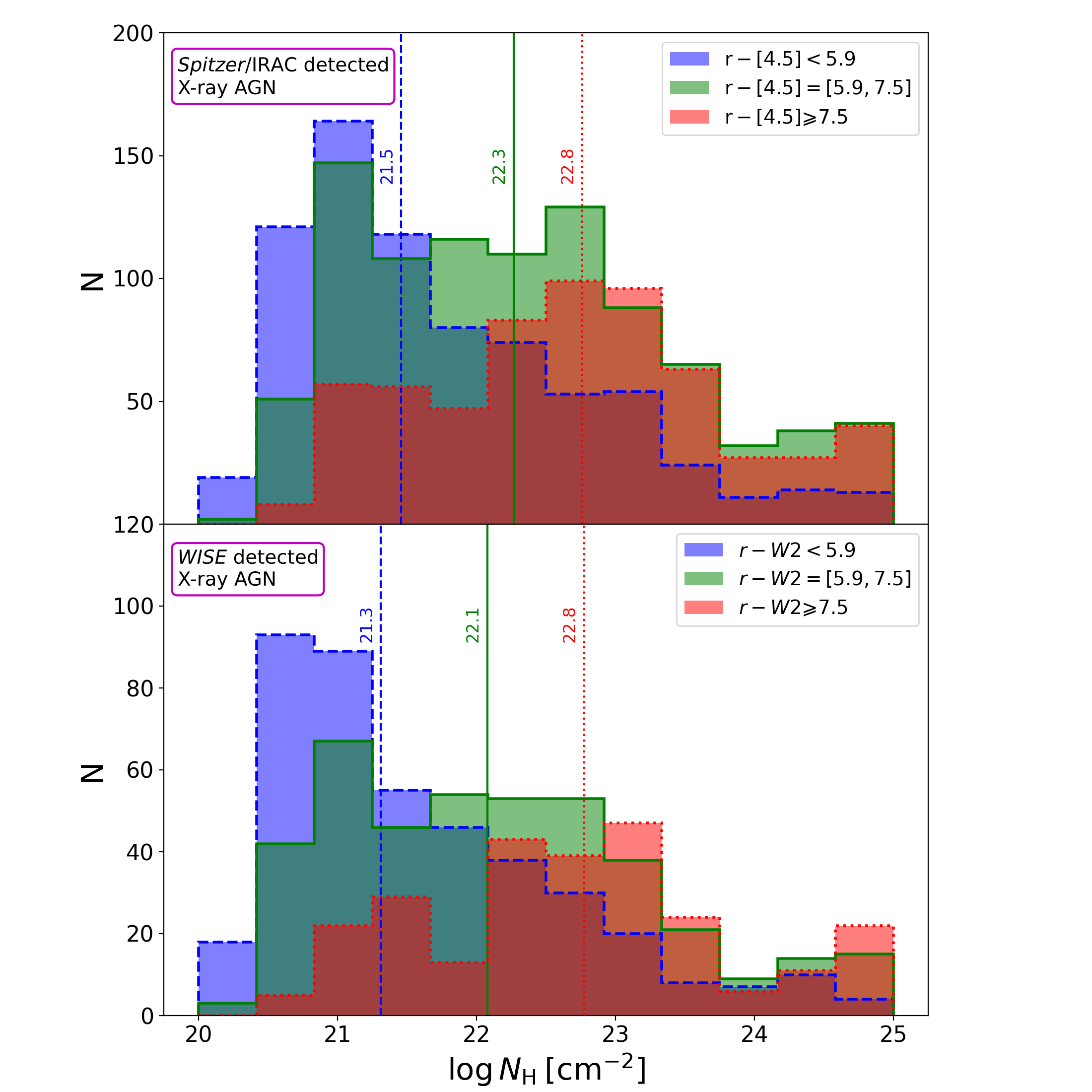}

\caption{Distributions of the logarithmic column density, $\log N_{\rm H} \: \rm (cm^{-2})$, for the X-ray AGN in COSMOS with $r-[4.5]/W2 <5.9 \rm \: mag$ (dashed blue), $5.9\: {\rm mag}\leq r-[4.5]/W2 < 7.5 \rm \: mag$ (solid green), and  $r-[4.5]/W2 \geq 7.5 \rm \: mag$ (dotted red). The vertical lines show the median $\log N_{\rm H} \: \rm (cm^{-2})$ of each colour distribution, following the same colour code and line style as the histograms. The top panel plots the sources detected by {\it Spitzer}/IRAC, and the bottom panel plots the sources detected in the unWISE catalogue. The $r-[4.5]/W2 \geqslant 5.9 \rm \: mag$ colour cut predominantly selects obscured AGN.   } 
\label{fig:NHdist_rW2}
\end{figure}

\begin{figure}

\centering

\includegraphics[trim={0.cm 0cm 0 0
cm},clip, scale=0.39]{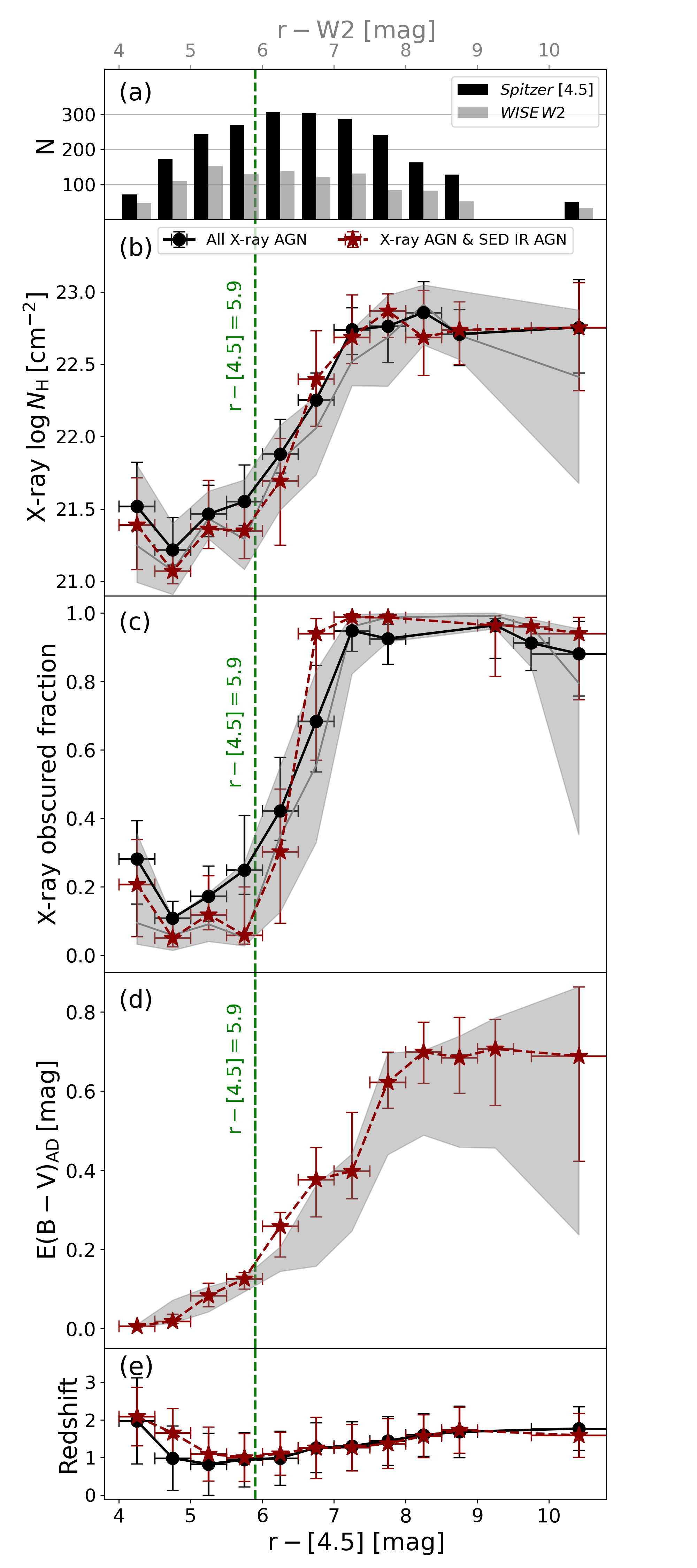}

\caption{Different obscuration parameters and redshift as a function of the $r-[4.5]$ (and $r-W2$) colour. The black circles represent all the X-ray AGN in the COSMOS field, and the red stars are the X-ray AGN that are also identified as IR AGN by the UV-to-FIR SED fitting. The error bars in the x-axis represent the size of the bin in all the panels. The y-axis values are the median value of the given obscuration parameter in each $r-[4.5]$ colour bin, and the error bars represent the 95\% confidence interval of the median of each parameter, calculated using bootstrap resampling within each colour bin. The grey shaded area in each panel illustrates the relationship between each obscuration parameter as a function of $r-W2$ instead (top y-axis). {\it Panel (a)} shows the number of sources per $r-[4.5]$ (black bar) and $r-W2$ (grey bar) colour bin; {\it panel (b)} shows the line-of-sight X-ray column density ($N_{\rm H}$) in logarithmic scale (the median was also calculated in logarithmic space); {\it panel (c)} shows the obscured AGN fraction $f_{\rm obs}$ calculated from the $N_{\rm H}$ posterior distribution (see Section\,\ref{subsec:Xraydata} for details); {\it panel (d)} shows the accretion disk reddening $\rm E(B-V)_{AD}$ calculated from the UV-to-FIR SED fitting; and {\it panel (e)} shows the median and standard deviation of the redshift distribution of each  $r-[4.5]$ colour bin.  }

\label{fig:NH_obsfrac_rW2}
\end{figure}

Figure\,\ref{fig:rW2_cosmos} shows the optical to MIR colour against redshift for the X-ray AGN in the COSMOS field. The top panel plots $r-\rm [4.5]$ for the sources detected by {\it Spitzer}/IRAC 2 and the bottom panel plots $r-W2$ for the sources with $W2$ detections in the unWISE catalogue. We find consistent results when using {\it Spitzer}/IRAC 2 and $W2$\footnote{From now on, we use the expression $r- [4.5]/W2$ to refer to $r-[4.5]$ and $r-W2$ at the same time.}. In both cases, $\sim 65\%$ of the X-ray AGN with \mbox{$r-[4.5]/W2\geqslant 5.9 \rm \: mag$} colours are obscured by \mbox{$N_{\rm H} >10^{22} \rm \: cm^{-2}$}, while only $\sim 30-35\%$ of the X-ray AGN with \mbox{$r-[4.5]/W2< 5.9 \rm \: mag$} colours have \mbox{$N_{\rm H} >10^{22} \rm \: cm^{-2}$}. This is better illustrated in Figure\,\ref{fig:NHdist_rW2}, which shows the distribution of $\log N_{\rm H}$ for sources with \mbox{$r-[4.5]/W2<5.9 \rm \: mag$}, \mbox{$5.9\:{\rm mag}\leq r-[4.5]/W2<7.5 \rm \: mag$} and \mbox{$r-[4.5]/W2\geqslant 7.5 \rm \: mag$}, and also plots vertical lines at the median $\log N_{\rm H}$ of each colour distribution. A progression can clearly be seen where AGNs with redder $r-[4.5]/W2$ colours have larger $N_{\rm H}$. We find that the \mbox{$r-[4.5]/W2< 5.9 \rm \: mag$} colour cut is more efficient in identifying X-ray unobscured AGN with \mbox{$N_{\rm H} <10^{22} \rm \: cm^{-2}$} than the \mbox{$5.9\:{\rm mag}\leq r-[4.5]/W2<7.5 \rm \: mag$} colour cut. Consequently, the \mbox{$5.9\:{\rm mag}\leq r-[4.5]/W2<7.5 \rm \: mag$} cut is much more effective at identifying X-ray obscured AGN with \mbox{$N_{\rm H} >10^{22} \rm \: cm^{-2}$}. Notably, the majority ($\sim 75\%$) of the AGNs with \mbox{$r-[4.5]/W2\geqslant 7.5 \rm \: mag$} are obscured by \mbox{$N_{\rm H} >10^{22} \rm \: cm^{-2}$}. We find that the median obscuration of sources with \mbox{$r-[4.5]/W2<5.9 \: \rm mag$} is \mbox{$\log N_{\rm H, r-4.5/W2<5.9}/(\rm cm^{-2}) \approx 21.5$}, while the median obscuration of sources with $r-[4.5]/W2\geq 5.9 \rm \: mag$ is \mbox{$\log N_{\rm H, r-4.5/W2 \geqslant 5.9}/(\rm cm^{-2}) \approx 22.5$}, consistent with \citet{2007Hickox}. Similar median values are found when we only consider the X-ray AGNs that are also identified as IR AGNs from our broadband SED fitting (\mbox{$\log N_{\rm H, r-4.5/W2<5.9}/(\rm cm^{-2}) \approx 21.3$} and \mbox{$\log N_{\rm H, r-4.5/W2\geqslant 5.9}/(\rm cm^{-2}) \approx 22.6)$}.

Figure\,\ref{fig:NH_obsfrac_rW2} depicts the variation of the X-ray column density $N_{\rm H}$ (panel b), the obscured fraction $f_{\rm obs}$ (panel c), and the accretion disk reddening $\rm E(B-V)_{AB}$ (panel d) as a function of $r-[4.5]/W2$ colour. We perform this analysis for all X-ray AGNs, and separately examine the subset that also has an IR AGN component identified by our SED fitting. We can see a clear positive trend between the two X-ray obscuration features and $\rm E(B-V)_{AB}$ with the optical-to-MIR colour. {Panels (b) and (c) indicate that the wide majority of the objects with $r-[4.5]/W2 \gtrsim 6.5 \rm \: mag$ are obscured by $N_{\rm H} > 10^{22} \rm \: cm^{-2}$.  For $r-[4.5]/W2 \gtrsim 5 \: \rm mag$, $N_{\rm H}$ and $f_{\rm obs}$ systematically increase with the optical-to-MIR colour reaching a constant value at $r-[4.5]/W2 \approx 8 \rm \: mag$. These trends suggest that when $N_{\rm H}$ increases, the amount of dust, and hence, dust reddening in the system also increases, progressively extinguishing the optical emission, which is also observed in panel (d). At $r-[4.5]/W2 \gtrsim 7.5 \: \rm mag$, the $N_{\rm H}$, $f_{\rm obs}$, and $\rm E(B-V)_{AB}$ reach a constant value probably due to the AGN emission in the optical waveband being completely extinguished at those high obscuration levels. Indeed, the optical wavelength is expected to be fully extinguished at $A_V\gtrsim 5$ \citep[e.g.,][]{2016SchnorrMuller,2016Burtscher}, which is equivalent to \mbox{$N_{\rm H} \gtrsim 10^{22} \rm \: cm^{-2}$} (assuming the gas-to-dust ratio observed in our galaxy \citealp{1995Predehl})\footnote{We note that we would need high S/N optical spectroscopy, as the one that will be provided by 4MOST, to constrain the accretion disk reddening beyond $\rm E(B-V)_{AD}\gtrsim 0.7$).}. Hence, at $r-[4.5]/W2 \gtrsim 7.5 \: \rm mag$, where the obscured fraction is $>80-90\%$ ($N_{\rm H} > 10^{22} \rm \: cm^{-2}$)\footnote{We note that, at $r-W2>7.5 \rm \, mag$, the obscured fraction calculated using the $f_{\rm obs}$ parameter is larger than the value obtained when just calculating the fraction of sources with $N_{\rm H}>10^{22} \, \rm cm^{-2}$ ($\sim 75\%$), which is not surprising as these quantities were calculated using different approaches.}, we expect the optical emission to be dominated by host galaxy light, and hence, the $r-[4.5]/W2$ colour will saturate. In addition, at those very high levels of obscuration, the MIR emission at $\sim 4.5 \rm \mu m$ can also be absorbed, reaching a regime where the value of the $r- [4.5]/W2$ colour will only indicate whether the source is obscured or not, but will not give an estimate of the exact level of obscuration. Notably, we find no strong dependence between redshift and $r-[4.5]/W2$ colour (see panel e of Figure\,\ref{fig:NH_obsfrac_rW2}); hence, the positive trend between obscuration and $r [4.5]/W2$ is not driven by redshift effects. 

This subsection demonstrated that the $r-[4.5]$ and $r-W2$ colours are good proxies of AGN obscuration. Particularly, the correspondence between the X-ray obscuration features and $r-[4.5]/W2$ is notably impressive, given that they are fully independent quantities. In addition, as will be demonstrated in the following section, the $r-W2>5.9$ cut also provides a good balance between obscuration properties and sample size. We note that the results of this section are also consistent with the X-ray properties of extremely red objects, which are typically selected using optical to near-IR colours (e.g., $r-K$), and have also been shown to be X-ray obscured \citep[e.g.,][]{2002Alexander,2002Brusa,2005Brusa,2005Mainieri,2010Brusa}. 

\subsection{AGN reliability of our obscured IR AGN selection} \label{sec:finalselection}

This subsection demonstrates the effectiveness of the \citetalias{2018Assef} R90 colour selection (see Equation\,\ref{eq:A18R90} and Figure\,\ref{fig:A90_cat} of Appendix\,\ref{ap:reliability}) to reliably identify AGN. The \citetalias{2018Assef} R90 AGN selection is based on $W1-W2$ colours (see Figure\,\ref{fig:A90_cat}) with $3\sigma$ and $5\sigma$ detections in $W1$ and $W2$, respectively. As demonstrated later in this work (see Section\,\ref{subsec:predictions} and Appendix\,\ref{ap:reliability}) and also in previous studies \citep[e.g.,][]{2021Barrows}, the \citetalias{2018Assef} R90 selection provides a robust AGN sample, spanning across a wide AGN luminosity and redshift range. 

Here, we use detailed UV-to-FIR SED fitting (see Section\,\ref{sec:sedfitting}) to investigate the performance of the \citetalias{2018Assef} R90 selection for the unWISE catalogue. For this, we identify \citetalias{2018Assef} R90 IR AGN candidates in four deep, well-observed fields in the sky with UV-to-FIR photometric coverage described in Section\,\ref{sec:data:photo}: the COSMOS,  Bo\"{o}tes, ELAIS-S1, and XMM-LSS fields. We then calculate the AGN reliability of the \citetalias{2018Assef} R90 selection by assuming that the reliable AGN are those identified by our SED template fitting approach described in Section\,\ref{sec:sedfitting}.

Figure\,\ref{fig:rel_unwise} shows the AGN reliability of the \citetalias{2018Assef} R90 selection as a function of $r-W2$ for the unWISE catalogue\footnote{We note that \citetalias{2018Assef} used UV-to-MIR SED fitting to arrive at their 90\% AGN reliability threshold using AllWISE photometry (see \citet[][]{2013Assef} for details on their SED fitting approach). In our work, we also independently assess the reliability of the approach using other deeper {\it WISE} photometry, such as unWISE and the LSDR10, and we use a more sophisticated SED modelling which also includes the FIR band.}, for $r_{\rm AB}<22.1$ and $r_{\rm AB}<22.8$, which are the depths of our wide and deep IR AGN survey components, respectively. The AGN reliability as a function of the $r-W2$ colour is important as we apply the colour cut $r-W2\geq 5.9 \rm \: mag$ to identify the obscured IR AGN (see Section\,\ref{subsec:rW2}). We find that at $r-W2<4 \rm \: mag$ colours, the sample is highly contaminated by quiescent, massive galaxies (see Figure\,\ref{fig:ap:contamination} in Appendix\,\ref{ap:a:wise_rel}) with an AGN reliability of only $\approx 20\%$. However, at $r-W2\geq 4 \rm \: mag$, the AGN reliability drastically improves to $>80\%$. At $r-W2\geq 5.9 \rm \: mag$, which is our obscured AGN selection, the \citetalias{2018Assef} R90 selection provides a clean AGN sample with an AGN reliability of $87\%$ for $r_{\rm AB}<22.1$ and $80\%$ for $r_{\rm AB}<22.8$. Table\,\ref{t:reliability_unwise} provides a summary of the AGN reliability and number density of the \citetalias{2018Assef} R90 selection applied to unWISE photometry.

\begin{figure}
\centering
\includegraphics[scale=0.45]{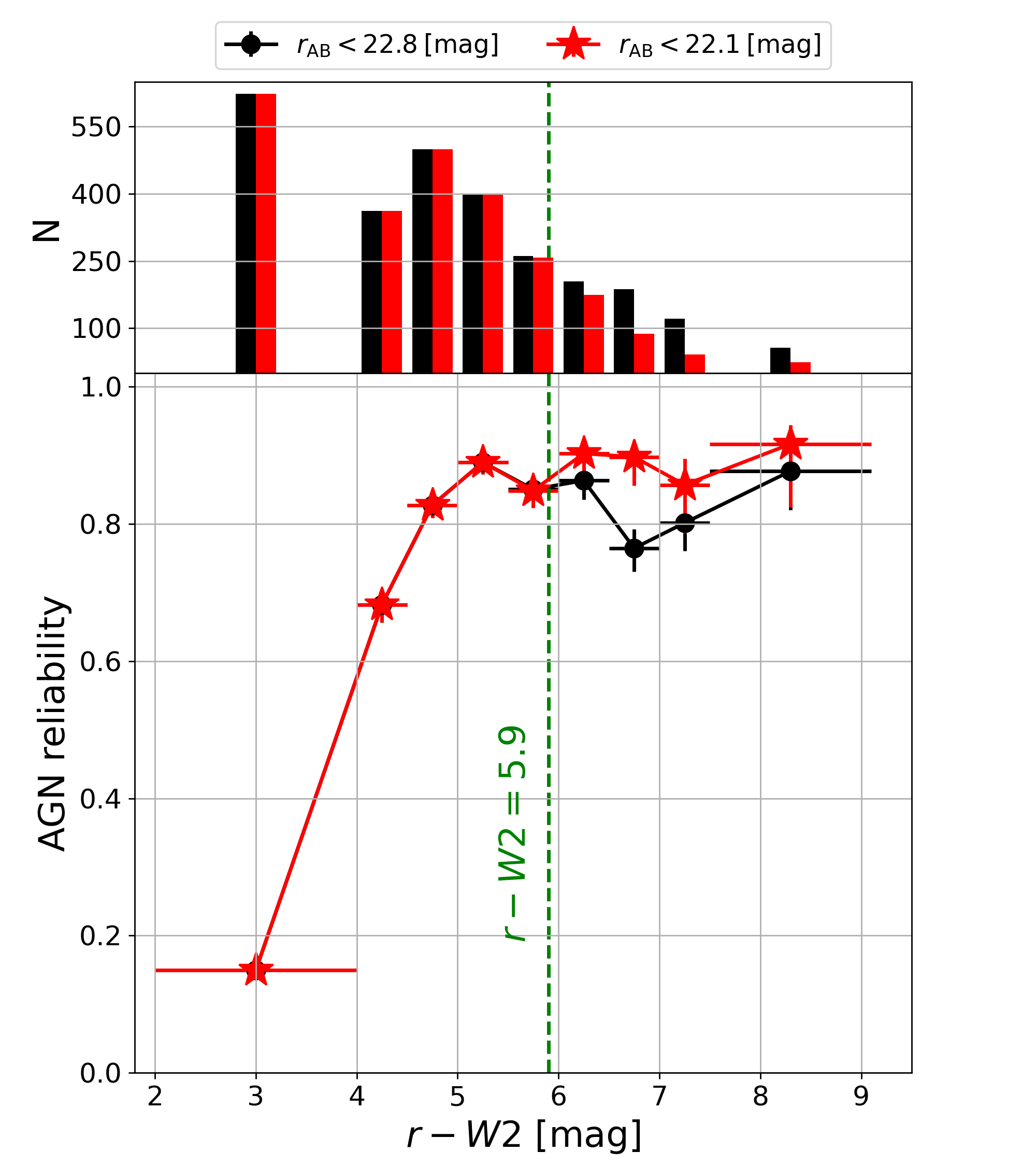}

\caption{AGN reliability as a function of the $r-W2$ colour for the AGN candidates selected by the \citet{2018Assef} R90 AGN colour selection and using unWISE photometry down to $r_{\rm AB}<22.1 \, \rm mag$ (black circles) and $r_{\rm AB}<22.8 \, \rm mag$ (red stars). The AGN reliability corresponds to the fraction of colour-selected AGN that are also identified as AGN by our SED fitting. The errorbar sizes in the x-axis represent the half width of the $r-W2$ bins while the y-axis errorbars correspond to the $1\sigma$ binomial error. The widths of the $r-W2$ bins are 0.5 mags for all bins except the lowest ($r-W2=2-4\rm \, mag$) and highest ($r-W2=7.5-9\rm \, mag$) values, which have bin widths of 2 mags and 1.6 mags, respectively, since there are not enough sources with such extremes $r-W2$ colours. The green dotted vertical line marks $r-W2=5.9 \, \rm mag$, which is our threshold to separate obscured from unobscured IR AGN candidates. The upper panel shows the number of sources per $r-W2$ bin down to $r_{\rm AB}<22.1 \, \rm mag$ (red bar) and $r_{\rm AB}<22.8 \, \rm mag$ (black bar) }.\label{fig:rel_unwise}
\end{figure}

We conclude that our obscured IR AGN selection (i.e., \citetalias{2018Assef} R90 selection plus $r-W2\geq 5.9 \rm \: mag$) applied to unWISE photometry provides a robust and clean AGN sample to define the 4MOST IR AGN survey. For the interested reader, Appendix\,\ref{ap:reliability} presents a more exhaustive analysis of the efficiency of the \citetalias{2018Assef} R90 selection applied to different {\it WISE} catalogues, its source of contamination, and also demonstrates that the unWISE photometry provides the best option to construct the 4MOST IR AGN sample.

\begin{table}
\centering

\begin{tabular}{c|c|l|l|l|l|l|}
 \hline
 \hline
\noalign{\smallskip}
Selection & & R (\%) & N (deg$^{-2}$) \\
\noalign{\smallskip}

\hline
\hline
\noalign{\smallskip}
\citetalias{2018Assef} R90 AGN & $r_{\rm AB}<22.8$ & 68 &140  \\
\noalign{\smallskip}
 & $r_{\rm AB}<22.1$ & 69  &125  \\

\noalign{\medskip}
\citetalias{2018Assef} R90 AGN + $r-W2\geq 5.9$ & $r_{\rm AB}<22.8$ & 80 & 32 \\
\noalign{\smallskip}

& $r_{\rm AB}<22.1$ &87 & 18 \\
\noalign{\smallskip}

\hline
\hline

\end{tabular}
\caption{AGN reliability (R) and AGN number density per deg$^2$ (N) of the AGN candidates identified by the \citetalias{2018Assef} R90 selection using unWISE photometry. The AGN reliability is calculated based on the UV-to-FIR SED fitting results. We also report the values for the obscured \citet{2018Assef} AGN with $r-W2\geq 5.9 \rm \: mag$ colours (bottom rows). }
\label{t:reliability_unwise}
\end{table}

\section{4MOST IR AGN survey: predictions and objectives} \label{sec:4MOST}

This section presents our predictions for the properties of the targets in the 4MOST IR AGN survey and our future research plans. Our final survey consists of $\approx 212,000$ obscured MIR-selected AGN candidates and has two components: (1) the IR AGN wide component, which will observe $\approx 182,000$ mostly obscured IR AGN over an area of $\sim 9,000 \rm \: deg^2$ down to $r_{\rm AB}=22.1$, and (2) the IR AGN deep component, which will observe $\approx 30,000$ obscured IR AGN over a smaller but extensively observed area of $\sim 850 \rm \: deg^2$ down to $r_{\rm AB}=22.8$. Subsection\,\ref{subsec:spec_predictions} presents predictions for the optical spectral properties of the 4MOST IR AGN survey;  subsection\,\ref{subsec:predictions} presents predictions for the AGN and host galaxy properties of 4MOST IR AGN sample calculated in deep extensively observed regions described in Section\,\ref{sec:data:sed}; subsection\,\ref{sec:comparison_ero} demonstrates that the 4MOST IR AGN survey complements the 4MOST X-ray AGN survey by targeting AGN with different properties; and in subsection\,\ref{sec:goals}, we present the scientific goals and future research plans we expect to execute with the 4MOST IR AGN survey.

\subsection{Spectroscopic predictions for the 4MOST IR AGN survey} \label{subsec:spec_predictions}

To inform our target selection (particularly our optical magnitude limits) and to test that the 4MOST spectra will lead to scientifically useful spectra, we simulated targets using the 4MOST PAQS spectral simulator\footnote{\url{https://github.com/jkrogager/simpaqs}}. This tool simulates the optical spectra of individual targets and calculates the expected exposure times using the 4MOST exposure time calculator (ETC). We take the following steps to produce simulated spectra with appropriate properties:

\begin{enumerate}
    \item Randomly select targets for simulation from our target catalog\footnote{10,000 for each of the two sub-surveys.}. These objects have $r$-band magnitude measurements, and we choose targets so that the magnitude distributions match those found in detailed studies of the deep fields (see Sections\,\ref{sec:data:photo} and \ref{subsec:predictions}). We also assign redshifts to these targets so that the distribution across the redshift-magnitude plane of simulated targets will be similar to the likely observed distributions.
    \item Based on the distribution of the obscured fraction as a function of $r$-band magnitude calculated using the predictions in the deep fields (see Section\,\ref{subsec:predictions}), we randomly assign a Type 1 or Type 2 quasar template to each target. The Type 1 template was produced from SDSS sources observed as part of the SPIDERS survey \citep{2020Comparat}. The Type 2 template is a composite of narrow line AGNs meeting our target-selection approach found in the DESI early data release \citep{DESICollab2024}.
    \item Simulate an observed spectrum with the PAQS spectral simulator for each target based on its redshift, magnitude, and spectral template (see Figure\,\ref{fig:opt_spec}). This tool also calculates an associated exposure time which is adjusted automatically to produce a S/N according to rules defined specifically for the IR AGN sub-surveys. 
    \item Apply the redshift and classification algorithm of 4MOST, which uses principal component analysis (PCA) template fitting (xPCA; Krogager et al., in prep.), to determine a best fitting redshift and object type to each simulated spectrum.
\end{enumerate}

\begin{figure*}
\centering
\includegraphics[width=0.8\textwidth]{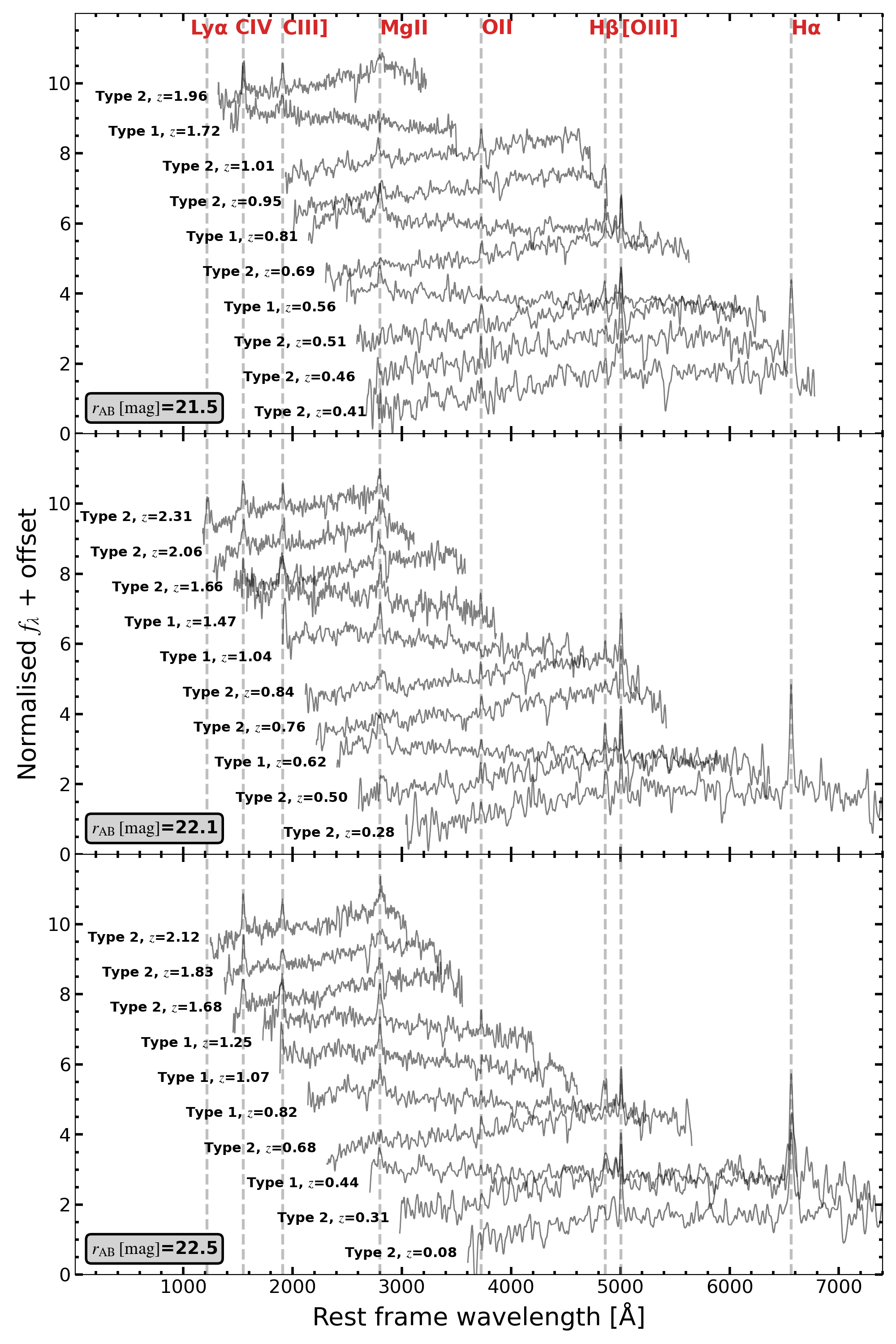}
\caption{Randomly selected simulated optical spectra with smoothing applied at $r$-band magnitudes of 21.5 (top panel), 22.1 (middle panel), and 22.5 (bottom panel). Major emission lines are indicated by vertical dashed grey lines. Spectra at a range of redshifts and with both template types are shown.}
\label{fig:opt_spec}
\end{figure*}

\begin{figure}
\centering
\includegraphics[width=\columnwidth]{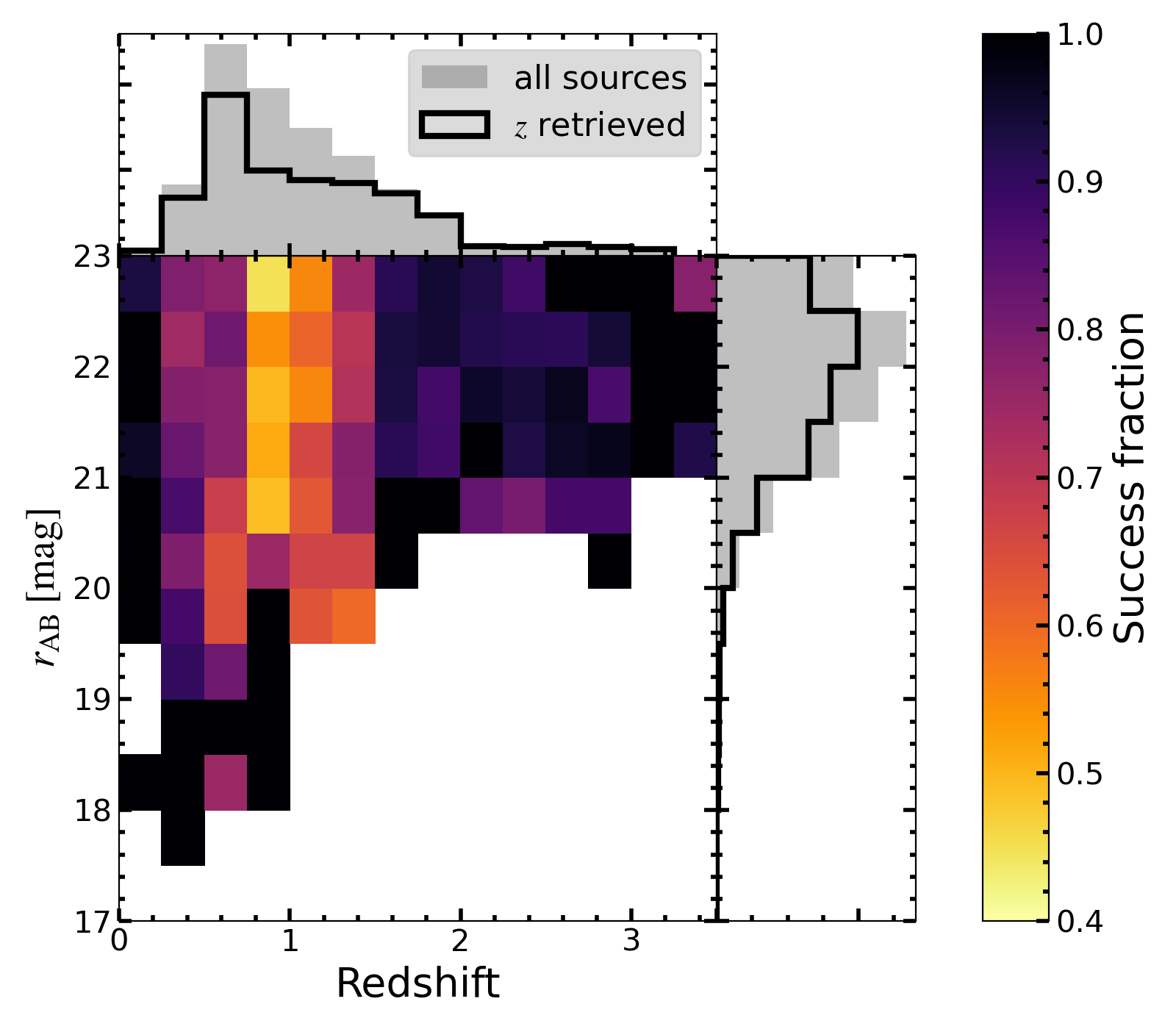}
\caption{Redshifts recovered from simulated spectra as a function of $r$-band magnitude and redshift. The central panel shows the fraction of targets with a fitted redshift that matches the simulation input redshift compared to total targets within each bin. Darker colours represent a higher fraction of successful results and empty bins contain no simulated targets. Histograms show these results as a function of each parameter separately: total distribution of targets (filled grey bins) and distribution of targets with redshifts recovered (solid black lines). This figure shows results for the deep survey - results for the wide survey are very similar below the $r$-band magnitude cut-off.}
\label{fig:opt_spec_recov}
\end{figure}

Based on these simulated optical spectra and template fitting results, we can assess the full results. For the wide survey simulations (mean optical magnitude of $r_{\rm AB} \approx 21.4$), the mean and maximum exposure times are 14.5 and 37.2 minutes, respectively, yielding a total exposure time for all targets of 44,000 hours. For the deep survey (mean optical magnitude of $r_{\rm AB} \approx 21.8$), the mean and maximum exposure times of the simulated targets are 29.4 and 96.4 minutes, respectively, which results in a total exposure time of 15,000 hours.


The minimum science goal of the IR AGN survey is to measure optical spectroscopic redshifts (see Section\,\ref{sec:goals}). We define a redshift as successfully recovered if the redshift returned by xPCA is close to the redshift input to the simulation, i.e. $\frac{|z_{sim} - z_{fit}|}{1 + z_{sim}}<0.05$\footnote{The specific choice in threshold makes little difference below $\sim0.1$.}. Figure\,\ref{fig:opt_spec_recov} shows that the redshift recovery rate varies with redshift but not significantly with magnitude, as also found for the 4MOST-ORCHIDSS survey \citep{2023Duncan}. This can be due to the fact that the optical spectra of AGN typically have high equivalent width emission lines, which can be detected even in optically faint sources, as shown in Figure\,\ref{fig:opt_spec}. Additionally, in our survey, the exposure time varies with magnitude in order to obtain a consistent S/N (i.e., fainter sources are observed for longer, as described in Section~\ref{subsec:spec_predictions}, point (iii)). With all redshifts included the total recovery rate is $\approx$~73\% ($\approx$~72\%) for the wide (deep) sub-survey. However, after removing the ``redshift desert'' of $0.75<z<1.25$, where the observed optical spectra often contain only a single emission line ($\rm [OII] \lambda 3727\AA$) that we can use to measure redshifts, the recovery rate increases to $\approx$~80\% ($\approx$~82\%). We expect the total redshift recovery rate to improve with developments in the 4MOST pipelines (particularly the inclusion of photometric redshift priors, which have been shown to increase the performance significantly; Krogager et al., in prep.) and visual inspection (VI) analysis as the project moves forward. Quasars in particular can be difficult for general pipelines, and large improvements can be made using VI and combining the pipeline with additional ``afterburner" identification approaches \citep[e.g.,][]{2023Alexander,2023Chaussidon}. A simple check of simulated sources finds that in cases where the pipeline assigns the wrong redshift, the correct redshift can be easily identified with VI in many cases. For example, common mis-identifications such as [O\,{\sc ii}] $\rightarrow$ Ly$\alpha$, Mg\,{\sc ii} $\rightarrow$ Ly$\alpha$, [O\,{\sc ii}] $\rightarrow$ [O\,{\sc iii}] produce a significant fraction of wrongly assigned redshifts, depending on the strength of these emission lines.

Expanded analyses will involve detection and fitting of specific emission lines, with the aim of exploring the accretion properties of the targets. Figure\,\ref{fig:opt_spec} shows simulated spectra, in which several important AGN emission lines are visible such as C\,{\sc iv}, C\,{\sc iii}], Mg\,{\sc ii}, [O\,{\sc ii}] $\lambda 3727\AA$, $\rm H\beta$, [O\,{\sc iii}] $\lambda 5008$\AA, and $\rm H\alpha$. Continuum emission is also clearly detectable. Full fitting of these simulated emission lines would not be informative because their specific properties depend on the input template and would therefore not provide an idea of expected results beyond the original assumptions adopted in the simulations.

Based on the results of this section, we are confident that, when strong emission lines are present (which is likely to be the case except for very extreme and unusual AGN), we will measure the redshifts and emission line fluxes for the majority of 4MOST IR AGN survey targets. Furthermore, with sufficient S/N, we will also be able to perform more detailed optical spectral fitting and analyse further properties, such as the examples described in Section\,\ref{sec:goals}. With this unprecedentedly large number of optical spectra we will perform, for the first time, large statistical population analyses of obscured IR AGNs and quasars.

\subsection{Photometric predictions for the 4MOST IR AGN target properties } \label{subsec:predictions}

This subsection explores the AGN and host galaxy properties of the obscured IR AGN candidates (i.e., the sources identified by our obscured IR AGN selection based on their {\it WISE} and $r-W2$ colours; see Section\,\ref{sec:finalselection}) in the deep fields. These properties will be used to predict the AGN and host galaxy properties of both components of the 4MOST IR AGN survey: IR AGN deep ($\rm r_{\rm AB}<22.8$) and IR AGN wide ($\rm r_{\rm AB}<22.1$). For this analysis, we apply our obscured IR AGN selection to the targets in the COSMOS,  Bo\"{o}tes, XMM-LSS, and ELAIS-S1 fields, where we have the UV-to-FIR SED fitting (see Section\,\ref{sec:sedfitting}) and X-ray observations from {\it Chandra} or {\it XMM-Newton} to investigate the X-ray obscuration properties. Our 4MOST IR AGN selection identifies 762 obscured IR AGN candidates with $r_{\rm AB}<22.8$ in those fields, from which 653 sources ($85\%$) have a high-quality spectroscopic or photometric redshift estimate and sufficient UV-to-FIR photometry to perform SED fitting; i.e., they are detected by at $24 \rm \: \mu m$ by {\it Spitzer} MIPS and have at least a FIR flux upper-limit from {\it Herschel}, which is crucial to constrain the dust-obscured AGN and star-formation components.

\subsubsection{AGN properties: luminosity and obscuration}  \label{subsec:predictionsAGN}

Figure\,\ref{fig:LAGN_z} shows the $8-1000\rm \: \mu m$ AGN luminosity ($L_{\rm AGN, IR}$) versus the redshift of the sample. As presented in Section\,\ref{sec:finalselection}, the SED fitting identifies an AGN in $\sim 80\%$ and $\sim 87\%$ of the obscured IR AGN candidates down to $r_{\rm AB}<22.8$ and $r_{\rm AB}<22.1$, respectively. For the sources which are not identified as AGN, the SED fitting could only provide an upper limit on their AGN luminosity. We note that the SED fitting can fail to identify an AGN in the case of low luminosity AGN, in high star-formation rate systems where the host-galaxy emission dilutes the emission from the AGN, or when there is poor photometric coverage. 

We find that the redshift of the obscured IR AGN candidates ranges $z\approx 0-3.5$, with median values of $z\approx0.8$ and $z\approx1$ for $r_{\rm AB}<22.1$ and $r_{\rm AB}<22.8$, respectively. In addition, we observe that both samples have very similar AGN luminosities: the IR AGN luminosity ranges $\log L_{\rm AGN,IR}/(\rm erg\:s^{-1})=43.5-47$, with a median value of $\log L_{\rm AGN,IR}/(\rm erg\:s^{-1})=45.1\pm 0.8 \: (45 \pm 0.9)$ down to $r_{\rm AB}<22.1 \: (22.8)$, and $\approx 55\%$ of the sources are IR quasars with an AGN luminosity $L_{\rm AGN,IR}>10^{45} \: \rm erg\:s^{-1}$. Based on these results, we predict that the 4MOST IR AGN deep and wide surveys will observe $\sim 16,500$ and $\sim 100,000$ IR quasars, respectively. To statistically test whether the IR AGN deep and wide samples have different redshift and AGN luminosity properties, we measure the p-value of the distributions using a log rank test, which also accounts for censored data (i.e., upper limits). For this, we use the \textsc{python} package \textsc{lifelines}\footnote{See \url{https://pypi.org/project/lifelines/}}. We find that the p-value of the $L_{\rm AGN,IR}$ distributions is $\approx 0.6$, indicating that there are not significant differences between both samples. In the case of the redshift distributions, we find that the p-value between the wide and deep samples is $\approx 0.04$, suggesting stronger statistical differences, which is expected given that the sources in the IR AGN deep sample generally have larger redshifts. 

\begin{figure}

\centering

\includegraphics[scale=0.45]{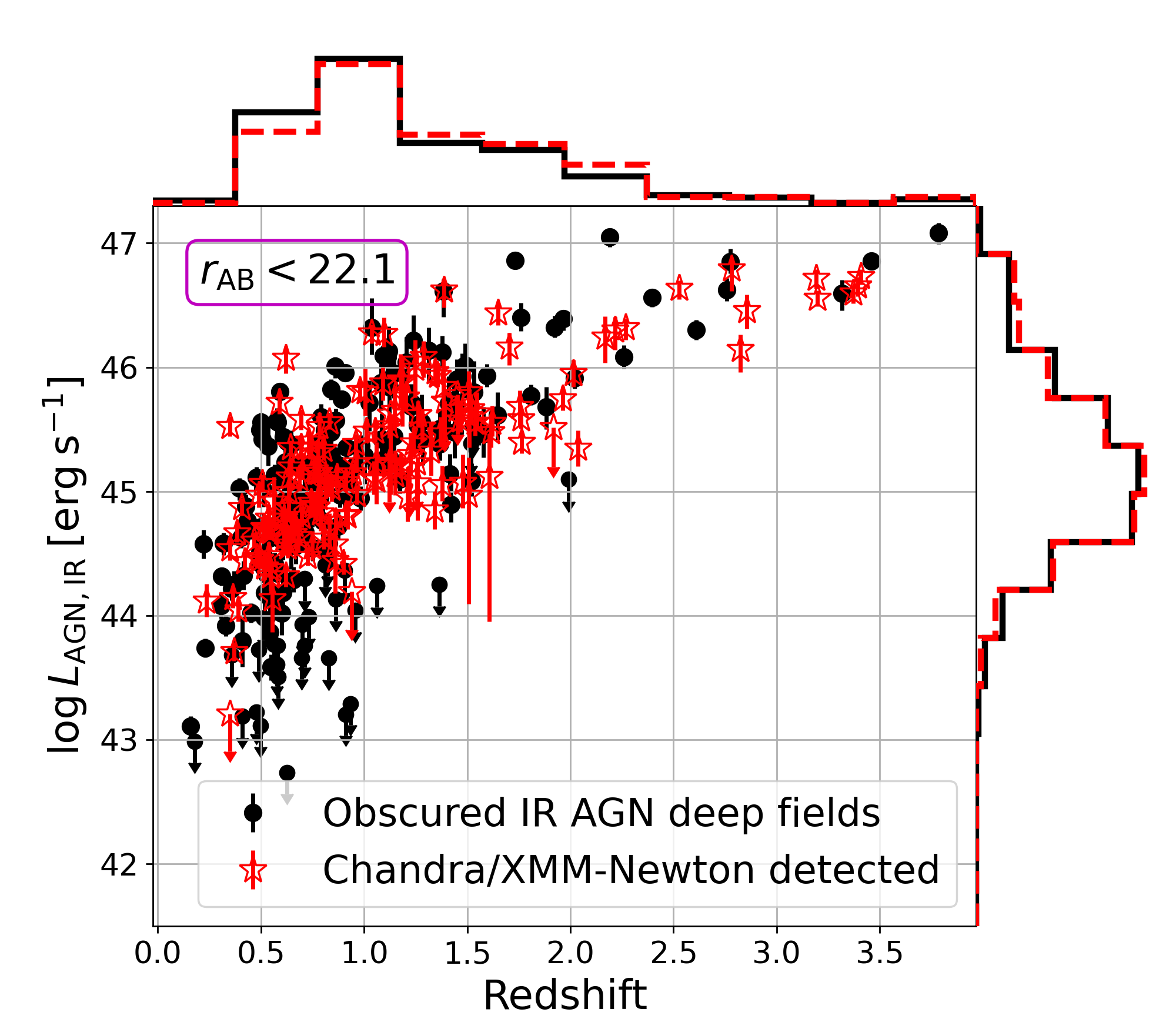}
\includegraphics[scale=0.45]{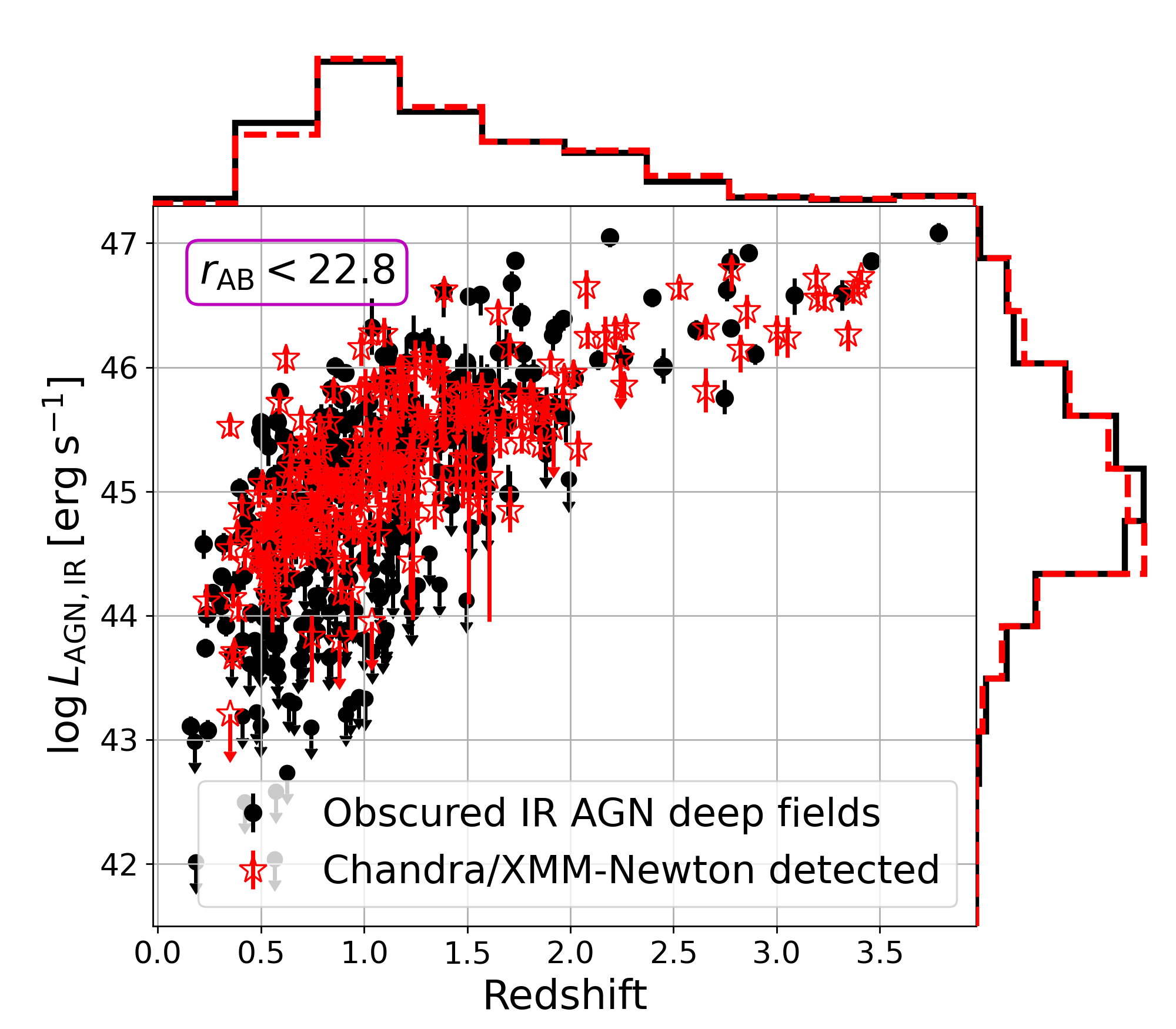}

\caption{The logarithmic $8-1000 \rm \: \mu m$ AGN luminosity ($L_{\rm AGN, IR}$) against the redshift for sources that meet the 4MOST IR AGN selection for the wide survey (top panel) and deep survey (bottom panel) in the deep fields (black circles). The red stars are the IR AGN that are also X-ray AGN detected by {\it Chandra} or {\it XMM-Newton}. The $L_{\rm AGN, IR}$ upper limits correspond to the 99-th percentile of the $L_{\rm AGN, IR}$ posterior distribution for the AGN candidates without a well-constrained AGN component in their SED. The histograms on the top and right side of each figure show the redshift and $L_{\rm AGN, IR}$ distributions of the IR AGNs (black solid) and the IR AGNs detected in the X-ray band (red dotted). } 
\label{fig:LAGN_z}
\end{figure}

Figure\,\ref{fig:LAGN_z} also shows the obscured IR AGN candidates that are detected in the X-ray band. We find that $\sim 45-50\%$ of the sample are detected by {\it Chandra} or {\it XMM-Newton} down to \mbox{$f_{0.5-10\rm keV}\approx (0.09-1) \times 10^{-14}  \rm \: erg\:cm^{-2} \: s^{-1}$} , while the other $\sim 50-55\%$ are undetected. If we apply a simple X-ray flux cut using the eFEDS limiting depth, we find that only $\sim 20\%$ of the sample would be detected by {\it eROSITA} down to \mbox{$f_{0.5-2 \rm \: keV}=1.5\times 10^{-15} \rm \: erg\:cm^{-2} \: s^{-1}$}, consistent with the results found in Section\,\ref{sec:comparison_ero}. We find that the X-ray detected IR AGN have consistent redshift and $L_{\rm AGN, IR}$ distributions with the obscured IR AGN candidates: the p-values of the redshift distributions and $L_{\rm AGN, IR}$ distributions are always $>0.8$. In addition, the COSMOS,  Bo\"{o}tes and ELAIS-S1 fields have measurements of the line-of-sight column density $N_{\rm H}$. We find that 83 out of 120 obscured IR AGN candidates that are detected in the X-ray band and also have $N_{\rm H}$ constraints are obscured by $N_{\rm H}>10^{22}\rm \: cm^{-2}$, corresponding to $\sim 70\%$ of the X-ray detected sample. This value is in quantitative agreement with Section\,\ref{subsec:rW2}, where it was found that $\sim 66\%$ of the X-ray AGN in the COSMOS field with $r-W2\geq 5.9 \rm \: mag$ are obscured by $N_{\rm H}>10^{22}\rm \: cm^{-2}$.

To assess the level of obscuration for the X-ray undetected obscured IR AGN candidates, we follow the same approach as \citet{2022Andonie} and stack all the X-ray undetected systems in the Bo\"{o}tes and COSMOS fields using \textsc{StackFast} (see Section\,\ref{sub:xstacking} for details). We only perform this analysis in fields with {\it Chandra} observations since the background emission of the observations is fairly low, which makes it possible to obtain reliable background-subtracted count rates. We note that in the stacking, we also include the X-ray undetected systems that do not have enough photometry to perform SED fitting. 

Table\,\ref{t:HR} shows the stacking results in the soft and hard X-ray bands for the X-ray undetected obscured IR AGN candidates in COSMOS,  Bo\"{o}tes, and COMOS+ Bo\"{o}tes combined. In all three cases, we detect a strong signal in both X-ray bands. We also calculate the HR of the samples to estimate the level of obscuration for the X-ray undetected sources. Figure\,\ref{fig:HR} shows the HR against the median redshift of each sample, which is calculated only for the sources with either spectroscopic or photometric redshift estimates (most sources without photometry do not have redshift estimates). The plot also shows the expected change in the HR with redshift for a fixed column density and intrinsic X-ray spectral slope ($\Gamma = 1.8$) calculated using PIMMS v4.12d\footnote{See \url{https://cxc.harvard.edu/toolkit/pimms.jsp}. We choose to use the Chandra-Cycle 14 response files as that corresponds to the period when the majority of the Chandra observations for the \textit{Chandra} COSMOS Legacy survey were taken. The majority of the  \textit{Chandra} observations in the  Bo\"{o}tes field were taken in Chandra-Cycle 18, but we found that the \textit{Chandra} response did not change significantly between Cycle 14 and Cycle 18.}. The HR suggests that the majority of the X-ray undetected obscured IR AGN candidates are obscured by $N_{\rm H}>10^{22}\rm \: cm^{-2}$, which is consistent with that expected given the results in Section\,\ref{subsec:rW2}. In addition, the clear X-ray signal detected in the hard band strongly suggests that the majority of the X-ray undetected systems are real AGN, as we would not expect to detect a strong hard X-ray signal coming from inactive star-forming galaxies. Indeed, adopting the relationship between the SFR and $2-10\rm \: keV$ luminosity from \citet[][]{2003Ranalli}, the X-ray undetected obscured IR AGN candidates would need to have, on average, $\rm SFR\approx 750\rm \: M_{\odot} \: yr^{-1}$ to produce the stacked X-ray counts observed in the hard X-ray band. However, our SED fitting finds that in the COSMOS and Bootes fields, the mean SFR of the X-ray undetected obscured IR AGN candidates is $\rm SFR\approx 25^{+181}_{-22} \rm \: M_{\odot} \: yr^{-1}$ (which is consistent with the SFR of all obscured IR AGN candidates; see next subsection), and that only $\sim 4\% (4/110 )$ of the X-ray undetected sources have $\rm SFR> 750 \rm \: M_{\odot} \: yr^{-1}$. Therefore, we are confident that the strong stacked hard X-ray signal is dominated by AGN processes, and only has a small contribution from SF. Moreover, the population could include many Compton-thick AGN with intrinsically hard but faint observed soft X-ray emission due to scattering \citep[e.g.,][]{2021Carroll,2023Carroll}.

\begin{figure}
\centering
\includegraphics[trim={0.cm 0cm 1.5cm 1.5cm},clip, scale=0.4]{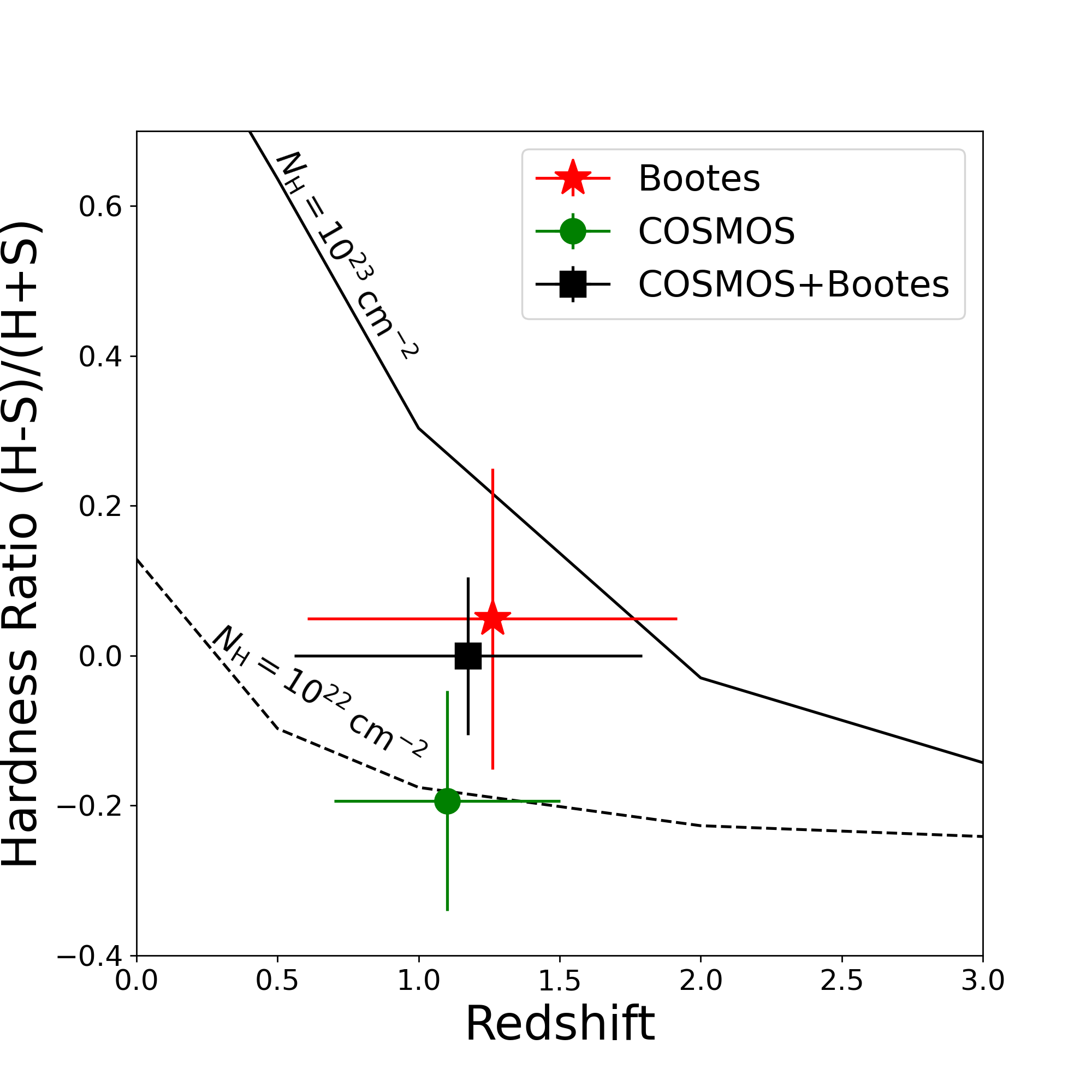}
\caption{ Hardness ratio (HR) against redshift for the X-ray undetected IR AGN candidates in the Bo\"{o}tes field (red), the COSMOS field (green), and the Bo\"{o}tes and COSMOS field combined (black). The X-ray stacking results are reported in Table\,\ref{t:stacking} . The dotted black and
continuous lines represent the change of the HR with redshift for $N_{\rm H} = 10^{22} \rm \: cm^{-2}$ and $N_{\rm H} = 10^{23} \rm \: cm^{-2}$, respectively. The X-ray stacking results suggest that, on average, the X-ray undetected source are obscured by $N_{\rm H} \gtrsim 10^{22} \rm \: cm^{-2}$. } 
\label{fig:HR}
\end{figure}

Assuming the same fraction of X-ray undetected systems are obscured by $N_{\rm H}>10^{22}\rm \: cm^{-2}$ as found for the X-ray detected obscured IR AGN candidates (i.e., $\sim 70\%$), we predict that the 4MOST IR AGN survey will select $\sim 148,000$ X-ray obscured IR AGN with $N_{\rm H}>10^{22}\rm \: cm^{-2}$. Table\,\ref{t:predictions} summarizes the predictions made in this section for the AGN luminosity and obscuration properties of the final 4MOST IR AGN surveys.

\begin{table}
\centering
\scalebox{0.69}{
\begin{tabular}{l c c c c c} 
 \hline
 \hline
\noalign{\smallskip}
Sample & N & S & H & HR & $z$ \\ 
&  & ($\rm counts \: s^{-1}$) & ($\rm counts \: s^{-1}$) &   \\ 
 (1) & (2)  & (3) & (4) & (5)  \\ 
\hline
\noalign{\smallskip}
COSMOS  &30&$(1.4\pm 0.3)\times 10^{-5}$&$(2.2\pm 0.4)\times 10^{-5}$& $-0.22\pm 0.1$&$1.2\pm 0.6$\\ 
\noalign{\smallskip}

\noalign{\smallskip}
 Bo\"{o}tes   &90&$(2.3\pm 0.6)\times 10^{-5}$&$(2.1\pm 0.6)\times 10^{-5}$& $0.05\pm 0.2$&$1.3\pm 0.7$\\ 
\noalign{\smallskip}

COSMOS+ Bo\"{o}tes  &120&$(2.1\pm 0.3)\times 10^{-5}$&$(2.2\pm 0.3)\times 10^{-5}$& $-0.02\pm 0.1$ & $1.2\pm 0.6$\\ 
\noalign{\smallskip}
\hline
\hline

\end{tabular}}
\caption{X-ray stacking results for the {\it Chandra} undetected obscured IR AGN candidates in the COSMOS and  Bo\"{o}tes fields. Column description: (1) sample size; (2) mean count rate in the soft (0.5-2\,keV) X-ray band; (3) mean count rate in the hard (2-7\,keV) X-ray band; (4) the hardness radio $\rm H=(H-S)/(H+S)$ of the sample; and (5) median and standard deviation of the redshift distribution of the sample.}  \label{t:stacking}
\label{t:HR}
\end{table}

\begin{table*}
\centering
\scalebox{0.86}{
 \begin{tabular}{lcccccccl} 
 \hline
 \hline
\noalign{\smallskip}
 4MOST Sample & N &$z$ & $\rm N_{IR\: quasar}$  &${\rm N}_{N_{\rm H}>10^{22}}$ &$\log L_{\rm AGN, IR}$   & $\log 
 M_{\star}$ & $\log L_{\rm SF, IR}$ & SFR \\
\noalign{\smallskip}

& & & & & ($\rm \: erg \: s^{-1}$) &($\rm M_{\odot}$)&($\rm \: erg \: s^{-1}$)&($\rm M_{\odot}  \: yr^{-1}$)\\
\noalign{\smallskip}
(1) & (2)&(3)&(4)&(5)&(6)&(7)&(8)&(9)\\
\noalign{\smallskip}
 \hline
\noalign{\smallskip}

IR AGN wide & $\sim 182,000$ & $0.8\pm 0.6$& $\sim  100,000$ & $\sim 
 127,000$ &$45.1\pm0.8 \: (45.6\pm0.5)$  & $10.8\pm 0.6\:(11.1\pm0.5)$ & $44.8\pm 0.8 \: (45\pm 0.8) $ &  $27^{+172}_{-22} \:(50_{-42}^{+270}$) \\
\noalign{\smallskip}
IR AGN deep & $\sim 30,000$ & $1\pm 0.6$ & $\sim 16,500$ & $\sim 21,000$ &$45\pm0.9 \: (45.6\pm0.5)$    & $10.8\pm 0.6\:(11.1\pm0.5)$ &  $44.8\pm 0.9 \: (45\pm 0.8) $ & $27^{+194}_{-23} \: (50_{-42}^{+270}$) \\

 \noalign{\smallskip}
\hline
 \noalign{\smallskip}
 \end{tabular}}
 \caption{Predictions for some AGN and host galaxy properties of all the obscured IR AGN candidates in the deep ($r_{\rm AB}<22.8$) and wide ($r_{\rm AB}<22.1$) components of the 4MOST IR AGN survey, based on the UV-to-FIR SED fitting and X-ray analyses in the deep fields. Column description: (1) the sample; (2) sample size; (3) median and standard deviation of the redshift; (4) predicted number of IR quasars with $L_{\rm AGN, IR}>10^{45} \rm \: erg \: s^{-1}$; (5) predicted number of X-ray obscured AGN with $N_{\rm H}>10^{22} \:\rm cm^{-2}$ (number based on both X-ray detected AGN and stacking analyses of X-ray undetected AGN); (6) median and standard deviation of the $8-1000 \: \rm \mu m $ AGN luminosity ($L_{\rm AGN, IR}$) of the obscured IR AGN candidates and the IR quasars (in parenthesis) in the deep fields; (7) median and standard deviation of the stellar masses for the obscured IR AGN candidates and the IR quasars (in parenthesis) in the deep fields; (8) median and standard deviation of the $8-1000 \: \rm \mu m $ SF luminosity ($L_{\rm SF, IR}$) of the obscured IR AGN candidates and the IR quasars (in parenthesis) in the deep fields; (9) median and standard deviation of the star-formation rates (SFR) calculated from $L_{\rm SF, IR}$ for the obscured IR AGN candidates and the IR quasars (in parenthesis) in the deep fields.} 
 
 \label{t:predictions}
\end{table*}

\subsubsection{Host galaxy properties: SFR and stellar masses} \label{subsec:hostprop}

Our SED fitting also provides constraints on the $8-1000\rm \: \mu m$ star formation luminosity $L_{\rm SF,IR}$ and stellar masses of the sample, allowing us to make predictions for the host-galaxy properties of the 4MOST IR AGN sample. We estimate the star-formation rates (SFRs) using $L_{\rm SF,IR}$ and assuming the \citet{1998Kennicutt} relation. Figure\,\ref{fig:Hostprop} shows the SFR against the stellar masses for the obscured IR AGN candidates down to $r_{\rm AB}<22.1$ and $r_{\rm AB}<22.8$. Overall, we find that the host galaxy properties of the wide and deep samples do not significantly differ. Our SED fitting could reliably measure the SFR for $\sim 45\%$ of the sample, while for the other $\sim 55\%$, we can only provide an upper limit, either because of a lack of sufficient photometric coverage or because the AGN emission dominates the IR SED. For sources with $r_{\rm AB}<22.1$, the SFR ranges $\approx 4-1300\rm \: M_{\odot} \: yr^{-1}$, with a median value of $\approx 27^{+172}_{-22} \rm \: M_{\odot} \: yr^{-1}$, and for sources with $r_{\rm AB}<22.8$, the SFR ranges $\approx 2-3300\rm \: M_{\odot} \: yr^{-1}$, with a median value of $\approx 27^{+194}_{-23} \rm \: M_{\odot} \: yr^{-1}$. Although the median SFR values of the wide ($r_{\rm AB}<22.1$) and deep ($r_{\rm AB}<22.8$) samples are consistent, we find that the p-value of the SFR distributions is $\approx 0.1$, which provides marginal evidence for a statistical difference between both distributions. This is probably due to the fact that the deep sample has more sources with large SFRs (which is reflected in its modestly larger SFR standard deviation), which is expected given that the deep sample also has higher redshifts. These values remain almost the same when we only consider the sources that are also identified as IR AGN by the SED fitting. 

We also find that the obscured IR AGN candidate host galaxies are generally massive, with stellar masses ranging $\log M_{\star}/\rm M_{\odot} \approx 9-12.5$, and with a median value of $M_{\star}/\rm M_{\odot} \approx 10.8\pm 0.6$ for both the wide and deep samples, respectively. In addition, we find that the p-value between the deep and wide samples is $\approx 0.45$, indicating that there are no significant differences in the stellar masses of both samples.

We also calculate the host galaxy properties of the IR quasars with $L_{\rm AGN,IR}>10^{45} \: \rm erg\:s^{-1}$ only. We find that the IR quasars are hosted in more massive ($M_{\star}/\rm M_{\odot} \approx 11.1\pm 0.5$) and more powerful star-forming ($\rm SFR \approx 50^{+270}_{-43} \rm \: M_{\odot} \: yr^{-1}$) host galaxies than the overall IR AGN sample, consistent with these more luminous AGN hosting more massive SMBHs. The SFR and stellar masses results are reported in Table\,\ref{t:predictions}.

\begin{figure*}
\centering
\includegraphics[scale=0.55]{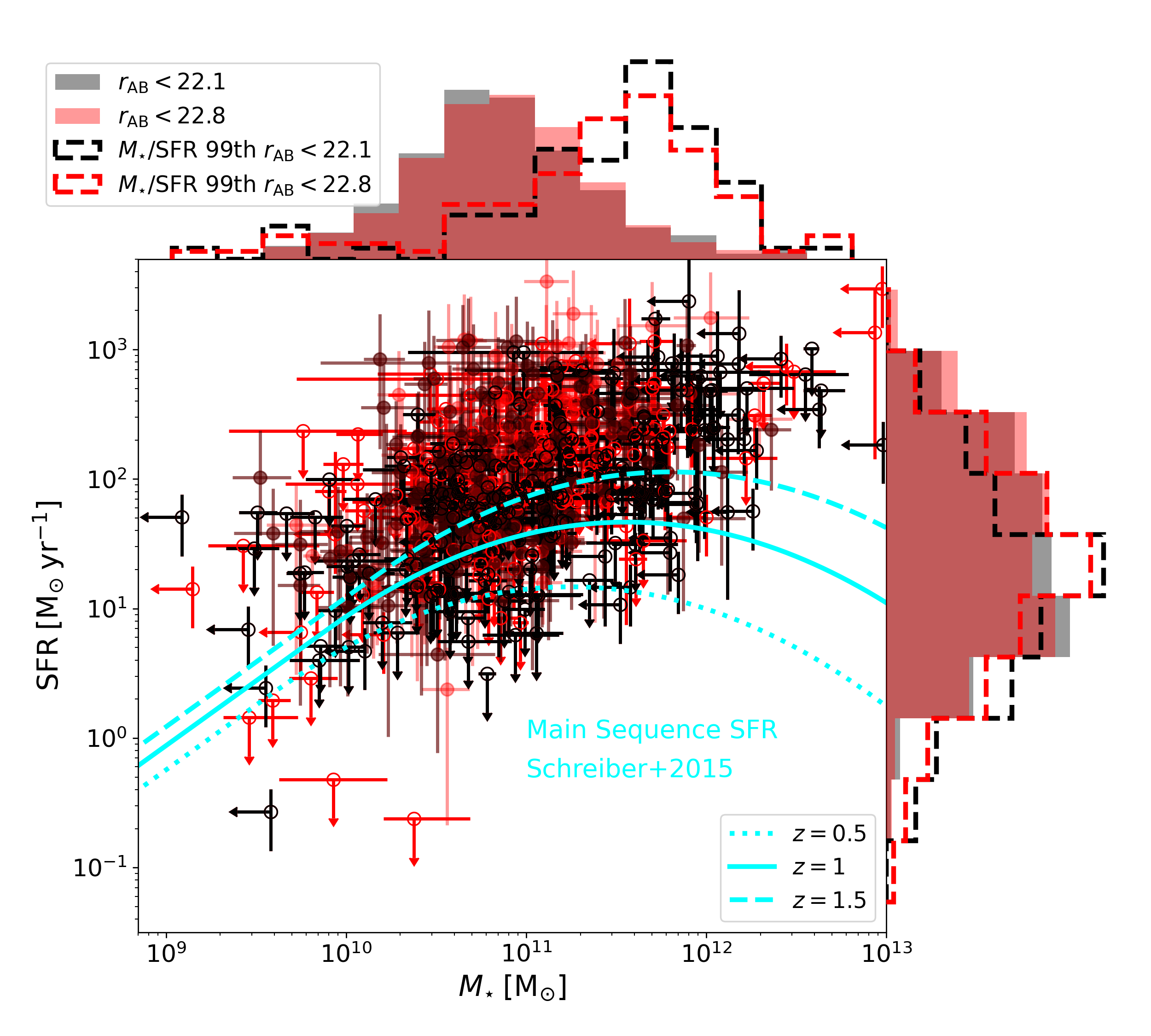}

\caption{ SFR against the stellar mass $M_{\star}$ of the obscured IR AGN candidates in the deep fields down to $r_{\rm AB}<22.1 \, \rm mag$ (black/grey) and  $r_{\rm AB}<22.8 \, \rm mag$ (red/light red) as a representation of the 4MOST IR AGN wide and deep surveys, respectively. The SFRs are calculated from the $8-1000 \rm \: \mu m$ star-formation luminosity. The cyan curves represent the main sequence of star-forming galaxies relation from \citet{2015Schreiber} for $z=0.5$ (dotted curve), $z=1$ (solid curve), and $z=1.5$ (dashed curve), converted from the Salpeter to the Chabrier IMF (see Footnote~\ref{IMFchange}). The top histograms show the distribution of $M_{\star}$ while the right histograms show distributions of SFR. In both cases, the filled light-colour histograms represent the well-constrained SFR and $M_{\star}$, while the dotted, empty histograms represent the 99th percentile of the posterior distribution of unconstrained SFR and $M_{\star}$. }  
\label{fig:Hostprop}
\end{figure*}

Motivated by the results of \citet{2022Andonie, 2024Andonie}, who found a large fraction of IR quasars residing in starburst host galaxies, we calculated the amount of obscured IR AGN candidates in the starburst phase adopting the main sequence of star-forming galaxies relation from \citet{2015Schreiber}\footnote{We rescale the \citet{2015Schreiber}'s SFR$-M_{\star}$ relationship from the Salpeter \citep{1955Salpeter} to the Chabrier IMF adopting the following commonly used relations reported in \citet{2014Madau}: $\rm SFR_{Chabrier}=0.64 \times SFR_{Salpeter} $ and $\rm M_{\star, Chabrier}=0.61 \times M_{\star, Salpeter} $.}\label{IMFchange}} (see the cyan curves in Figure,\ref{fig:Hostprop}). We classify a source as in the starburst phase when it is $>0.3$~dex above the SFR$-M_{\star}$ main sequence relationship. We find 123 and 198 IR AGN candidates residing in starburst host galaxies down to $r_{\rm AB}<22.1$ and $r_{\rm AB}<22.8$, respectively, which corresponds to $\approx 33\%$ of each sample. This fraction is in full agreement with \citet{2022Andonie}, who also found that $\sim 29\%$ (170/578) of their SED-selected IR quasar sample in the starburst phase. Extrapolating to the rest of 4MOST IR AGN survey footprint, we predict that the 4MOST IR AGN survey will identify  $\sim 70,000$ IR AGN residing in starburst host galaxies, where the star formation is likely to provide a significant (potentially dominant) contribution to the AGN obscuration \citep[e.g.,][]{2019Circosta, Gilli2022, 2024Andonie}.

\subsection{Comparison to the 4MOST X-ray AGN survey with \textbf{\textit{eROSITA}}} \label{sec:comparison_ero}

The 4MOST X-ray AGN survey will target $\sim 1\rm M$ AGN selected by {\it eROSITA}. The 4MOST IR AGN survey is designed to complement the 4MOST X-ray AGN survey by identifying the obscured AGN that {\it eROSITA} will miss due to its predominantly soft X-ray sensitivity response. In this subsection, we use the {\it eROSITA} observations in eFEDS (hereafter eFEDS X-ray AGN) as an equivalent of the 4MOST X-ray AGN survey to demonstrate that our 4MOST IR AGN survey will target AGN with different properties to the 4MOST X-ray AGN sample. The eFEDs region is the ideal field for these analyses since the {\it eROSITA} observations are of a comparable depth (actually $\sim \times 2$ times deeper) to those utilised in the 4MOST X-ray AGN survey and it has good-quality multi-wavelength follow-up observations and optical spectroscopic redshifts\footnote{We note that the eFEDS has much shallower X-ray observations than the {\it Chandra} and {\it XMM-Newton} data that we used in the deep field analyses, and does not have MIR and FIR photometric coverage. Hence, we cannot perform the same detailed UV-to-FIR SED fitting and X-ray analyses that we performed in the deep fields.}. We note that we do not use the publicly available 1st-pass all-sky survey \citep[eRASS1;][]{2024Merloni} because it does not have redshift estimates and is much shallower than the {\it eROSITA} observations used to define the 4MOST X-ray AGN survey.

To compare the properties of the 4MOST IR AGN and the 4MOST X-ray AGN surveys, we would ideally compare the obscured IR AGN candidates in the eFEDS footprint with the eFEDS X-ray AGN. However, the majority of the eFEDS area does not have extensive optical spectroscopic redshifts of non X-ray sources or multi-wavelength photometry, and therefore lacks the accurate redshifts and SED-fitting constraints we require for the obscured IR AGN analyses. Instead, we decided to use the sources that meet our 4MOST obscured IR AGN selection in the deep fields (see Section\,\ref{subsec:predictions}) to perform the redshift comparison with the eFEDS X-ray AGN. 

\begin{figure*}
\includegraphics[trim={0.5cm 0 2.5cm 2.5cm},clip, scale=0.5]{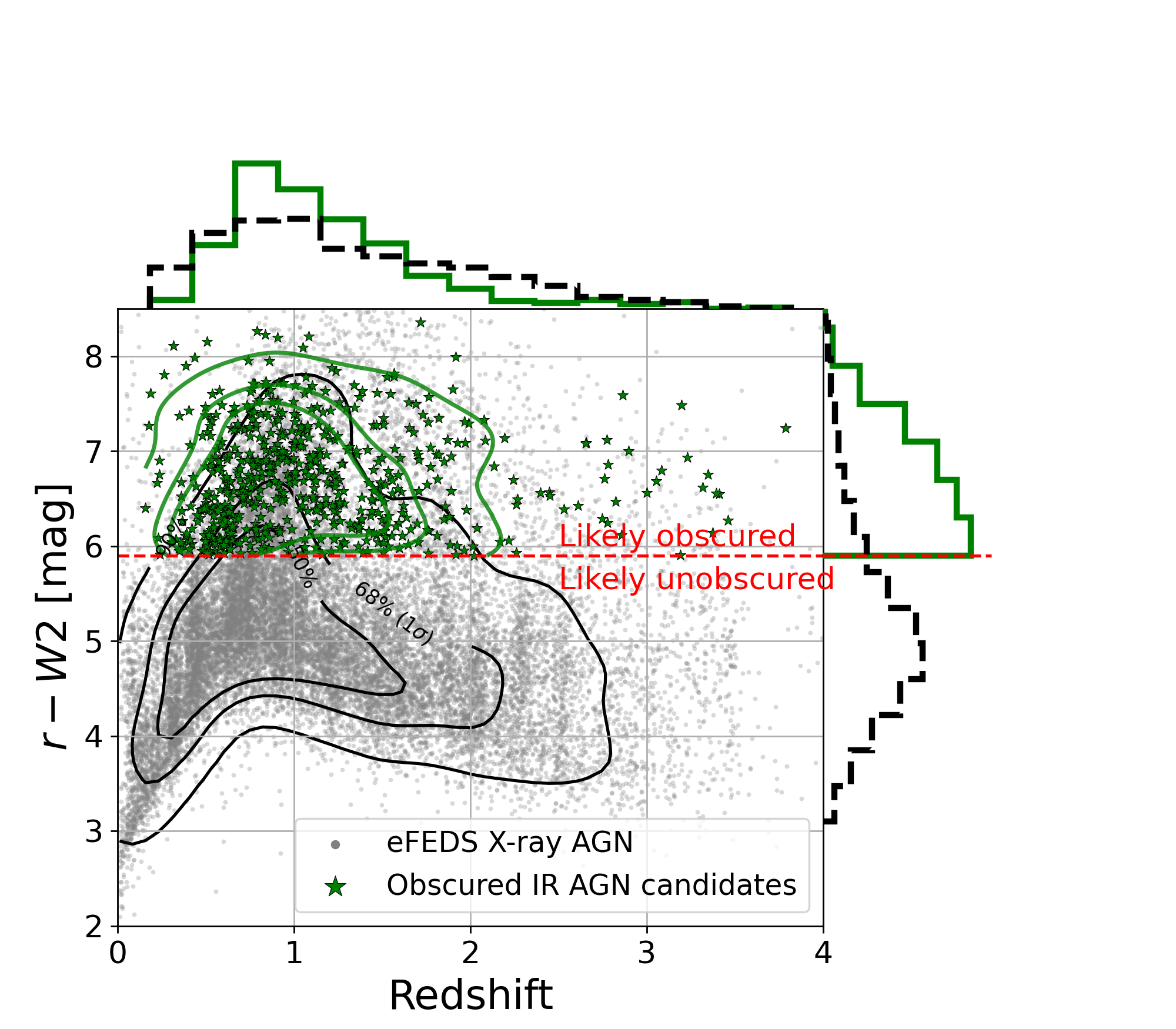}
\includegraphics[trim={0.5cm 0 2.5cm 2.5cm},clip, scale=0.5]{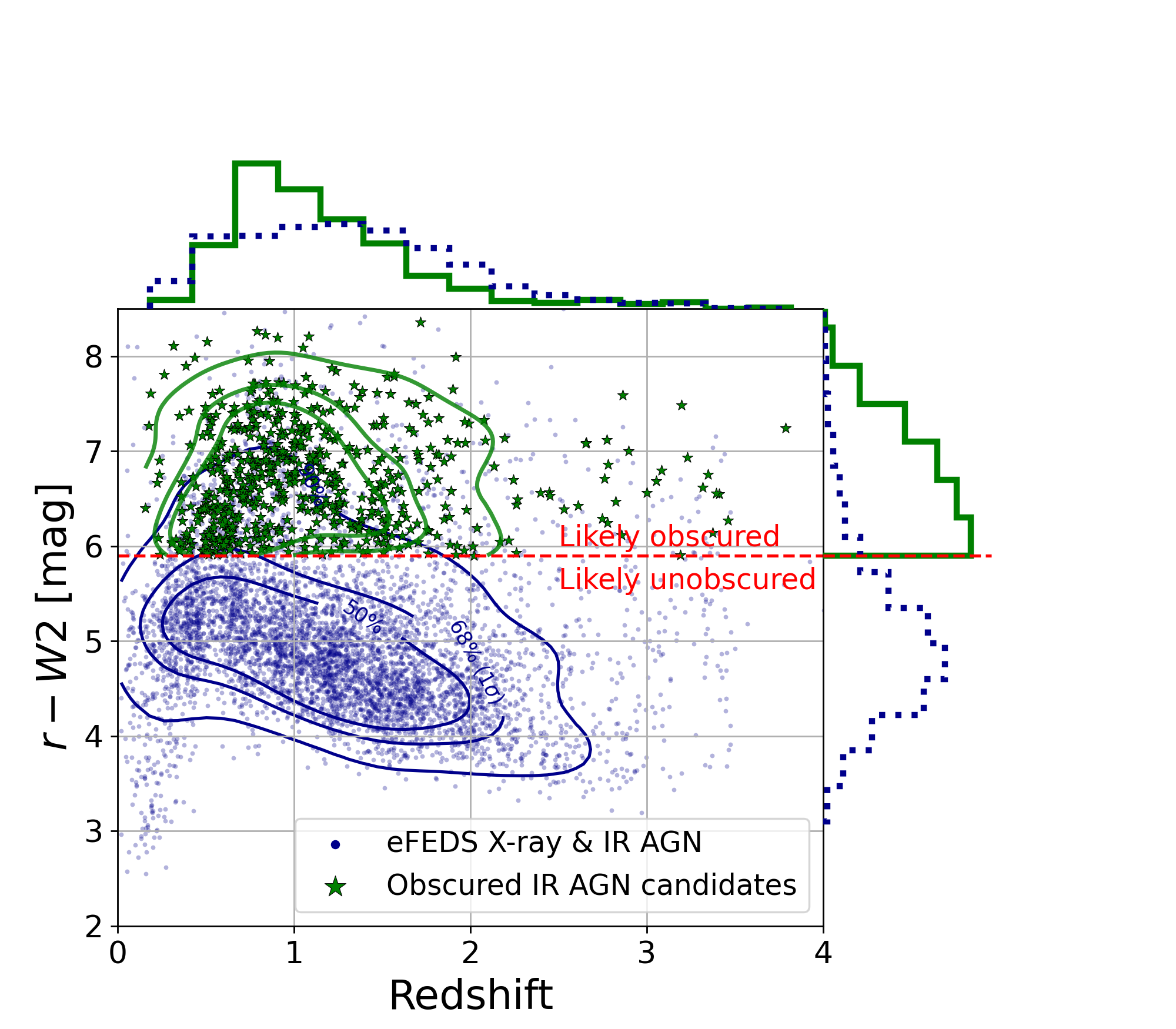}

\caption{ {\it Left panel:} $r-W2$ against redshift of the {\it eROSITA} AGN in the eFEDS (eFEDS X-ray AGN, grey dots), equivalent to the 4MOST X-ray AGN wide survey, and obscured IR AGN candidates in the deep fields down to $r_{\rm AB}=22.8 \, \rm mag$ (green stars, see Section\,\ref{subsec:predictions}). The contours show where the samples lie at 50\%, 68\%, and 90\% confidence levels. The figure also shows the distribution of the redshifts (top histograms) and $r-W2$ (right histograms) for the eFEDS X-ray AGN (grey dashed lines) and obscured IR AGN candidates (green solid lines). {\it Right panel:} Same as left panel, but only showing the eFEDS X-ray AGN that are also identified as IR AGN by the \citetalias{2018Assef} R90 selection, using unWISE photometry (eFEDS X-ray \& IR AGN, blue dots and blue dotted lines).} 
\label{fig:CompeRO}
\end{figure*}

Figure\,\ref{fig:CompeRO} depicts the $r-W2$ colour versus redshift for both samples, showing with contours where the samples lie at 50\%, 68\% ($\sim 1 \sigma$), and 90\% confidence level. The left panel plots the eFEDS X-ray AGN with a $W2$ S/N$>3$ in the LSDR8, equivalent to 21,320 sources, which clearly shows that the eFEDS X-ray AGN and the obscured IR AGN candidates have different $r-W2$ colours. By design, all the obscured IR AGN candidates have $r-W2\geq 5.9 \rm \: mag$, while only $\sim 21\% \rm (4425/21320)$ of the eFEDS X-ray AGN have $r-W2\geq 5.9 \rm \: mag$. We perform a Kolmogorov-Smirnov (KS) test and we find a $\rm p-value \ll 0.001$, confirming that both distributions are different. This plot also shows that the eFEDS X-ray AGN and the obscured IR AGN candidates have different redshift distributions ($\rm p-value \ll 0.001 $); the eFEDS X-ray AGN have more sources at $z\lesssim0.5$ and $z\gtrsim 1.5$. Indeed,  we find that $\sim 19\%$ and $\sim 34\%$ of the eFEDS X-ray AGN have $z<0.5$ and $z>1.5$, respectively, while only $\sim 9\%$ and $\sim 20\%$ of the obscured IR AGN candidates have $z<0.5$ and $z>1.5$, respectively. These differences are likely due to the X-ray data identifying lower-luminosity AGN at lower redshifts (where the host-galaxy will dilute the AGN signal in the IR waveband), and unobscured AGN at higher redshifts which are bright in the optical waveband due to a direct view of the nucleus.

\begin{figure}
\includegraphics[scale=0.45]{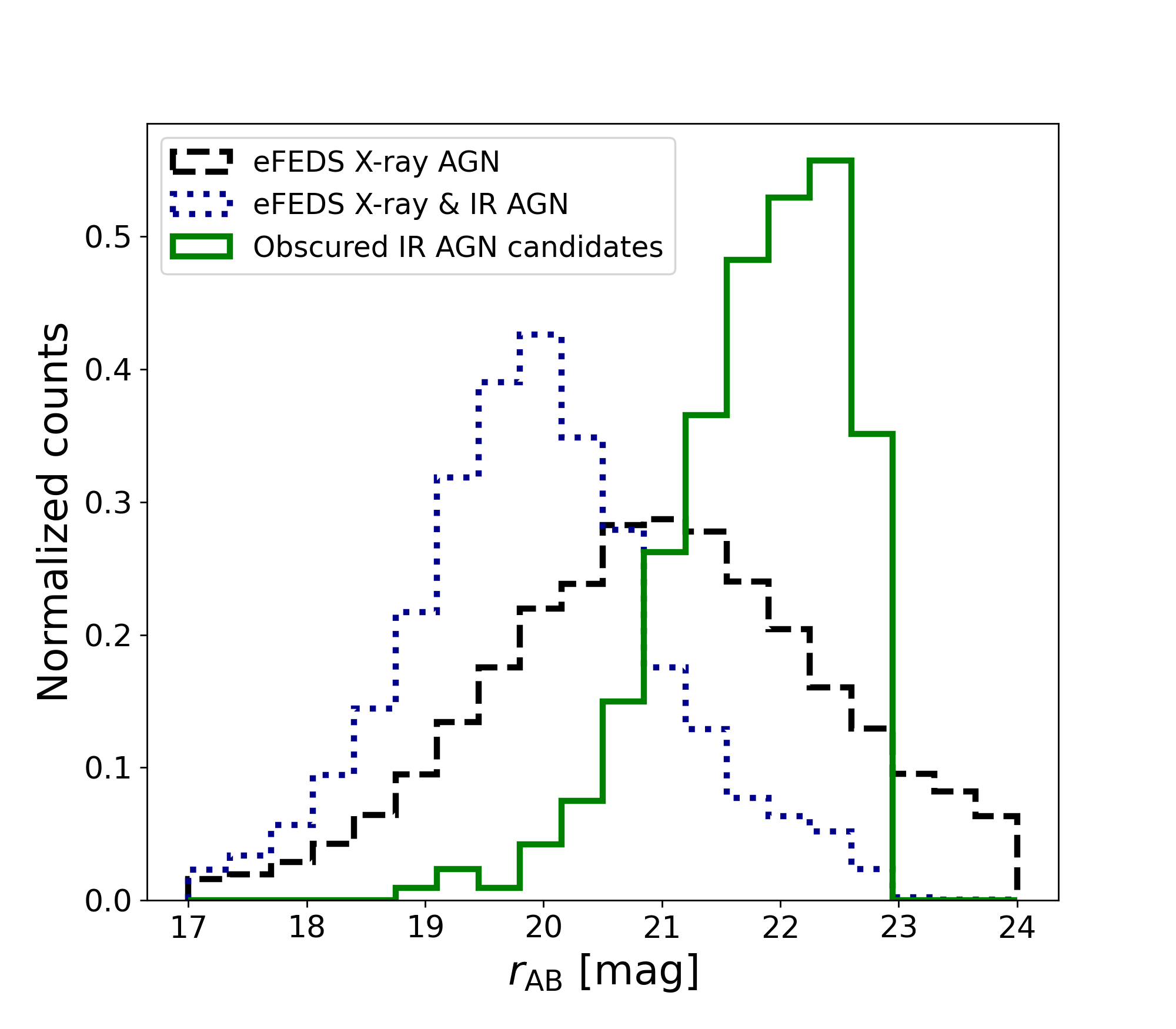}

\caption{Distributions of $r_{\rm AB}$ of the {\it eROSITA} AGN in the eFEDS (black dashed lines), the {\it eROSITA} AGN in the eFEDS that are also identified as IR AGN by the \citetalias{2018Assef} R90 selection using unWISE photometry (blue dotted lines), and the obscured IR AGN candidates in the deep fields down to $r_{\rm AB}=22.8 \, \rm mag$ (green continuous line).} 
\label{fig:CompeRO_R}
\end{figure}

\begin{table*}
\centering

 \begin{tabular}{lccccl} 
 \hline
 \hline
\noalign{\smallskip}
 Sample  & N &Redshift & $r_{\rm AB}$ &$f_{\rm eRO}$ & $f_{N_{\rm H}>10^{22}}$  \\ 
\noalign{\smallskip}
(1) & (2) &(3) & (4)&(5) & (6)  \\ 
\noalign{\smallskip}

 \hline
\noalign{\smallskip}

eFEDS X-ray AGN & $22,079$ &$1.2\pm 0.8$ & $21.1\pm 1.7$ &1 & $\sim 0.08$ \\ 
\noalign{\smallskip}
eFEDS X-ray \& IR AGN & $5,912$ &$1.3 \pm 0.9$ & $19.9\pm 1.2 $ &1 &  $\sim 0.07$  \\
\noalign{\smallskip}

4MOST IR AGN deep & $4,676$&$0.8 \pm 0.6$ &$22\pm 0.6$& 0.18 ($856/4676$) &$\sim 0.68$\\ 
\noalign{\smallskip}
4MOST IR AGN wide & $2,594$ &$1 \pm 0.6$ & $21.6\pm 0.5$ & 0.24 ($633/2594$)&$\sim 0.68$  \\ 

 \noalign{\smallskip}
\hline
 \noalign{\smallskip}
 \end{tabular}
 \caption{Sample statistics for the {\it eROSITA} AGN and the 4MOST IR AGN deep ($r_{\rm AB}<22.8$) and 4MOST IR AGN wide ($r_{\rm AB}<22.1$) surveys in the {\it eROSITA} Final Equatorial-Depth Survey (eFEDS) footprint. Column descriptions: (1) the sample; (2) sample size; (3) median and standard deviation of the redshift. For the 4MOST samples, the redshift statistics are calculated using the predictions in the deep fields (see Section\,\ref{subsec:predictionsAGN}); (4) median and standard deviation of the $r$-band values in AB units; (5) the fraction of the sample detected by {\it eROSITA} in eFEDS; (6) the fraction of the sample expected to have $N_{\rm H}>10^{22} \: \rm cm ^{-2}$.} \label{t:IRAGN_eRO}

\end{table*} 

Figure\,\ref{fig:CompeRO_R} depicts the $r$-band distributions of the eFEDS X-ray AGN and the obscured IR AGN candidates, showing that they also differ ($\rm p-value \ll 0.001$). The eFEDS X-ray AGN have an $r$-band distribution which resembles a Gaussian centred around $r_{\rm AB}\approx 21.1$; i.e., half of the sources have $r_{\rm AB}< 21.1$. On the other hand, the obscured IR AGN have an $r$-band distribution which resembles an inverse exponential function, where the r-band rapidly rises from $r_{\rm AB}\approx 21$ to a peak at the limit of our 4MOST IR AGN survey. Indeed, we find, that $\sim 80\%$ of the obscured IR AGN have $r_{\rm AB}>21.1$. The fainter optical magnitudes for the obscured IR AGN are expected since the obscuration will extinguish most of the bright accretion-disc emission in the optical band (see Figure\,\ref{fig:NH_obsfrac_rW2}c). We note that the eFEDS X-ray AGN that are also IR AGN are optically brighter than the overall eFEDS X-ray AGN population. This apparently surprising result is likely due to the soft X-ray sensitivity of {\it eROSITA} meaning that only the brightest obscured AGN can be detected, demonstrating that {\it eROSITA} only selects an extreme end of the obscured AGN population.

To provide a more direct prediction of the {\it eROSITA} properties we also investigate the 4MOST IR AGN survey sources in the eFEDS footprint: overall, $\sim 4,476$ and $\sim 2,594$ obscured IR AGN candidates reside in eFEDs with $r_{\rm AB}<22.8$ and $r_{\rm AB}<22.1$, respectively. We find that only $\sim 18\%$ (856 out of 4,476) and $\sim 24\%$ (633 out of 2,594) 4MOST IR AGN sources are detected by {\it eROSITA}, down to $r_{\rm AB}<22.8$ and $r_{\rm AB}<22.1$, respectively; see Table\,\ref{t:IRAGN_eRO}. The remaining $\sim 76-82\%$ are undetected, probably due to the soft X-ray emission being absorbed by large column densities, as predicted by our analyses in Section\,\ref{subsec:predictionsAGN}. We can directly test this scenario using the X-ray spectral constraints in eFEDS from \citet{2022Liu}. Interestingly, we find that the {\it eROSITA}-detected 4MOST IR AGNs have a median $\log N_{\rm H}/(\rm cm^{-2})\approx 21.2 \rm \: cm^{-2}$, while the {\it eROSITA} sample in eFEDs have a median $\log N_{\rm H}/(\rm cm^{-2})\approx 20.8 \rm \: cm^{-2}$. Additionally, the obscured ($N_{\rm H}>10^{22} \rm \: cm^{-2}$) fraction of the {\it eROSITA}-detected 4MOST IR AGNs is $\sim 22\%$, which is $\sim 3 \times$ larger than for the entire eFEDS sample \citep[][]{2022Liu}. We note that the {\it eROSITA}-detected 4MOST IR AGNs are significantly less obscured than the obscured IR AGN candidates in Section\,\ref{subsec:predictionsAGN} and the X-ray AGN in Section\,\ref{subsec:rW2} because the {\it eROSITA} observations in eFEDs are significantly shallower than {\it Chandra} and {\it XMM-Newton} in the deep fields, missing the most obscured AGNs. Similarly, \citet{2022Toba} analysed the properties of {\it WISE} $22 \rm \: \mu m$ ($W4$) detected IR AGN candidates in eFEDS and found that only $\sim 10\%$ are detected by {\it eROSITA}. They concluded, in agreement with our results here, that the wide majority of the {\it WISE} $22 \rm \: \mu m$ sources are undetected by {\it eROSITA} due to extreme absorption.

Finally, we apply a self-consistent IR AGN selection to the eFEDS {\it eROSITA} X-ray AGN, identifying the \citetalias{2018Assef} R90 IR AGN using the unWISE magnitudes\footnote{To find the unWISE counterparts of the eFEDS sources, we match the optical positions of the {\it eROSITA} AGN in eFEDs with the unWISE catalogue, adopting the same approach as that outlined in Section\,\ref{subsec:wise}}. This analysis will inform us of the total sample of IR AGNs that will be targeted by the 4MOST X-ray and IR AGN surveys combined. The expected $r-W2$ and redshift distributions are shown in the right panel of Figure\,\ref{fig:CompeRO}. We find that $\sim 26\%$ (5785 out of 22079) and $\sim 25\%$ (5574 out of 22079) of the eFEDS X-ray AGN meet the \citetalias{2018Assef} R90 IR AGN selection down to $r_{\rm AB}<22.8$ and $r_{\rm AB}<22.1$, respectively. The numbers are almost the same for $r_{\rm AB}<22.8$ and $r_{\rm AB}<22.1$ since most eFEDS X-ray AGN are optically bright. We find that the majority ($\sim 86\%$: 5001 out of 5785) of the IR-eFEDS X-ray AGN have $r-W2<5.9 \rm \, mag$ colours, as expected given the majority of the {\it eROSITA} sources are X-ray unobscured. Extrapolating to the rest of the 4MOST IR AGN footprint, we expect that the 4MOST X-ray AGN survey will identify $\approx 300,000$ IR AGNs with $r-W2<5.9 \rm \: mag$ colours, directly complementing our $\approx 200,000$ obscured IR AGN with $r-W2\geq 5.9 \rm \: mag$. Hence, the combination of the 4MOST IR AGN and the 4MOST X-ray AGN surveys will produce a sample of $\sim 500,000$ IR-selected AGN, the largest optical spectroscopic survey targeting IR AGNs to date by $>2$ orders of magnitude. Table\,\ref{t:IRAGN_eRO} summarizes the statistics reported in this Section.

\subsection{Scientific goals }\label{sec:goals}

The 4MOST IR AGN survey will provide the first large-scale optical spectroscopic sample of IR AGN to investigate the accretion and host galaxy properties of obscured IR AGN. When combined with the IR-detected sources in the 4MOST X-ray AGN survey, it will provide the most comprehensive and complete census of the IR AGN population to date, allowing us to determine what (sub) populations have been missed in previous studies, test whether obscured IR AGN are fundamentally different to unobscured IR AGN, and to understand whether obscured IR AGN preferentially reside in different environments to unobscured IR AGN. This will be achieved from the combination of the 4MOST optical spectra along with the rich multi-wavelength coverage overlapping the 4MOST IR AGN survey footprint (see Figure\,\ref{fig:coveragemap}). Here, we describe some of the specific analyses that can be performed with this comprehensive dataset to characterize the AGN and host galaxy properties of the IR AGN population.

\begin{enumerate}
    \item Spectroscopic characterization: the 4MOST optical spectra will allow, for the first time, the large-scale exploration of the optical spectral properties of the obscured IR AGN population. We will use the spectra to measure redshifts and constrain the accretion properties of our sources. Due to the 4MOST spectral coverage ($3700-9000$\AA), we expect to detect many important AGN emission lines such as $\rm [NeV] \rm \lambda 3346, 3426$\AA, $\rm [HeII] \rm \lambda 4687$\AA, and $\rm [OIII] \rm \lambda 5008$\AA ; see Figure\,\ref{fig:opt_spec}.  These AGN emission lines will probe AGN activity up to $z\sim 1.5$, where most of the 4MOST IR AGN sources lie, and will allow us to perform optical classification of sources. For example, we can use different emission line diagnostic to identify optical AGN signatures in our sample such as ``BPT" diagnostics out to $z\sim 0.35$ \citep[e.g.,][]{1981Baldwin, 2003Kauffmann}, or $\rm [OIII]/H\beta$ and $\rm [NeIII]/[OII]$ vs $\rm [OIII]/[OII]$ emission line ratio analyses at higher redshifts \citep[e.g.,][]{2011Yan,2011Juneau,2024Feuillet}. The optical spectra will also facilitate other analyses, depending on the detected emission lines and S/N (e.g., energetic outflows from broadened forbidden lines; black-hole masses from broad permitted lines; dust obscuration from continuum and emission-line fitting; \citealp[e.g.,][]{2002McLure,2006Vestergaard,2011Shen,fawcett2022,2020Rakshit}). In the case of the systems without clear optical, X-ray, or IR SED AGN signatures but reliable redshift, we can use optical spectral stacking to study different subsets of the IR AGN population (e.g., extremely red quasars and hot dust obscured galaxies; \citealp[e.g.,][]{2014Stern,2015Assef,2015Ross,2017Hamann}) and search for faint AGN features \citep[e.g.,][]{2015Zhu, 2018Rigby, 2024Arnaudova}. Little is known about the overall obscured IR AGN population and it is also quite probable that we will discover new sub populations of obscured AGN.

    \item Dust-obscured SFR: the 4MOST footprint overlaps with ten well-observed fields of the sky with {\it Herschel} coverage, which together cover an area of $\sim 480\rm \: deg^2$; see Figure\,\ref{fig:coveragemap}. In those fields, we can use UV-to-FIR SED fitting to reliably calculate the SFR of the IR AGN sample. Combined with the stellar masses, we can also explore where the obscured IR AGN lie on the main sequence of star formation \citep[e.g.,][]{2007Noeske,2012Whitaker,2015Schreiber}, and investigate the incidence and properties of extreme starbursts in IR quasars, extending the results found by \citet{2022Andonie, 2024Andonie}.

    \item X-ray analyses: for the $\sim 20\%$ of the obscured IR AGN that will be detected by {\it eROSITA}, we can perform X-ray spectral fitting to constrain key AGN properties such as X-ray luminosity and $N_{\rm H}$ to compare with the 4MOST X-ray AGN survey. In addition, we can perform more detailed X-ray spectral analyses in the fields with deeper X-ray data from {\it Chandra} and {\it XMM-Newton} (which typically align with the fields also with {\it Herschel} far-IR coverage). For the $\sim 80\%$ of the sample that will be X-ray undetected, we can use a similar approach to that adopted in Section\,\ref{subsec:predictionsAGN}, and perform X-ray stacking analyses to assess the level of obscuration for different AGN sub populations \citep[e.g.,][]{2021Carroll,2022Andonie, 2023Carroll}. 
    
    \item Spectral energy distribution analysis: all our sources have information in the optical bands $g$, $r$, $i$, and $z$ and the {\it WISE} bands $W1$ and $W2$. We also find that $\sim 60\%$ and $\sim 40\%$ of our sources have $>2\sigma$ detections in $W3$ and $W4$, respectively. This combination of photometry already allows us to characterize the UV-to-MIR SED of the entire 4MOST IR AGN survey. In addition, the X-ray analysis will provide the X-ray luminosity and obscuring gas column ($N_{\rm H}$) for individual sources, which can be used to place priors and improve the constraints on the contribution of the AD emission to the rest-UV SED, as adopted in Section\,\ref{sec:sedfitting}. With this information, we can perform a simple UV-to-MIR SED fitting, and retrieve the AGN luminosity and stellar masses of the entire sample and make comparisons between different sub populations. To demonstrate the viability of this analysis, we have developed a SED fitting approach that uses the physical constraints from the deep-field analysis (see Section\,\ref{subsec:predictions}) as priors to help guide the SED fitting in the rest of the sky with less photometric coverage. We find that this approach will be particularly useful to constrain the average AGN luminosity and stellar masses of AGN samples, but the values of individual sources are more uncertain. Appendix\,\ref{ap:sed} describes this approach and shows its performance.

    \item Placing obscured IR AGN within the overall AGN population: the 4MOST IR and X-ray AGN surveys will provide a rich and large sample that can be used to compare the AGN, host galaxy properties, and large-scale environments of IR and X-ray selected AGN as a function of different physical parameters to  determine the relationship between obscured and unobscured AGN and test the standard AGN orientation model. These comparisons can be used to study the incidence of obscuration in AGN, the distribution of the AGN obscuring material \citep[e.g.,][]{2014Merloni}, the galaxy clustering of different AGN populations \citep[e.g.,][]{2022Petter,2023Petter}, and to test SMBH-galaxy co-evolution scenarios \citep[][]{2022Andonie,2024Andonie}.
\end{enumerate}

\section{Summary} \label{sec:conc}

We present the motivation and design of the 4MOST IR AGN survey, the first large-scale optical spectroscopic survey targeting MIR-selected obscured AGN. Our parent samples are the Legacy Surveys DR10 and the unWISE catalogues. To select the obscured AGN, we first identify the IR AGN using the \citet{2018Assef} 90\% AGN reliability colour selection using unWISE magnitudes, and then identify the obscured IR AGN selecting the objects with $r-W2\geq 5.9 \rm \: mag$. The final 4MOST IR AGN survey will target $\approx 212,000$ sources and has two components: a deep component which will target $\approx 30,000$ sources down to $r_{\rm AB}<22.8$ over a well-observed region of $\sim 850 \rm \: deg^2$, and a wide component which will target $\approx 182,000$ sources down to $r_{\rm AB}<22.1$ over a larger area of $\sim 9,000 \rm \: deg^2$. Our survey aims to complement the 4MOST X-ray AGN survey, which will target $\sim 1\rm M$ {\it eROSITA} AGN, by identifying the obscured AGN that {\it eROSITA} will miss across the same spatial footprint.

To investigate the expected properties of the 4MOST IR AGN survey selection function, we analyze a sample of $\sim 650$ obscured IR AGN candidates in four deep, well-observed fields in the sky using detailed UV-to-FIR SED fitting and X-ray spectral constraints from {\it Chandra} or {\it XMM-Newton}. The deep fields analysis provides information on the AGN reliability of our adopted approach and helps to predict the AGN and host galaxy properties of the final 4MOST IR AGN survey sample. Our findings can be summarized as follows:

\begin{enumerate}

    \item Based on UV-to-FIR SED fitting constraints, we find that the AGN reliability of our obscured IR AGN selection approach is $\sim 80\%$ and $\sim 87\%$ for $r_{\rm AB}<22.8$ and $r_{\rm AB}<22.1$, respectively. Our analysis finds that the AGN reliability of the \citetalias{2018Assef} R90 when applied to unWISE photometry is $\sim 70\%$; however, the majority of the contamination is coming from sources with $r-W2<4 \rm \: mag$. Hence, after removing those sources, the reliability of the \citet{2018Assef} selection increases to $\sim 80-90\%$ (see Section\,\ref{sec:finalselection} and Appendix\,\ref{ap:reliability}).

    \item We find that the optical-to-MIR colour cut $r-[4.5] \geq 5.9  \: \rm mag$, adapted from \citet{2007Hickox}, is a good proxy of AGN obscuration. The \citet{2007Hickox} selection was defined using the {\it Spitzer} IRAC 2 channel at $4.5 \rm \: \mu m$, but we have demonstrated it can also be used with {\it WISE} $W2$ at $4.6 \rm \: \mu m$. We tested the colour cut using X-ray spectral constraints in the COSMOS field, and find that $N_{\rm H}$ systematically increases with $r-[4.5]/W2$ (see Figure\,\ref{fig:NH_obsfrac_rW2}). We find that sources with $r-[4.5]/W2\geq 5.9 \rm \: mag$ have a median $N_{\rm H, r-4.5/W2\geq 5.9} =3.5\times 10^{22}\rm \: cm^{-2}$, and $\sim 65\%$ of them have a $N_{\rm H}>10^{22}\rm \: cm^{-2}$. On the other hand, sources with $r-[4.5]/W2<5.9 \rm \: mag$ have a median $N_{\rm H,r-4.5/W2<5.9} =3\times 10^{21}\rm \: cm^{-2}$ and only $\sim 30-35\%$ of them have a $N_{\rm H}>10^{22}\rm \: cm^{-2}$ (see Section\,\ref{subsec:rW2}).

    \item Based on the results of our optical spectral simulations, we predict that the observations from our survey will allow us to successfully calculate the redshifts (the overall recovery rate is $\approx 72\%$ but it increases to $\approx 82\%$ when excluding sources in the ``redshift desert") and identify prominent AGN emission lines in our targets. We find that the average exposure times of the 4MOST IR AGNs targets are $\sim 14.5$ and $\sim 29.4$ minutes for the wide and deep surveys, respectively; see Section\,\ref{subsec:spec_predictions}.     

    \item On the basis of the UV-to-FIR SED fitting results and X-ray analyses in the deep fields, we predict the AGN and host galaxy properties of the 4MOST IR AGN wide ($r_{\rm AB}<22.1$) and deep ($r_{\rm AB}<22.8$) surveys. We find that
    
    \begin{itemize}
        \item The obscuration properties of the obscured IR AGN candidates of the wide and deep surveys are similar with $\sim 45-50\%$ detected by {\it Chandra} or {\it XMM-Newton}. The X-ray constraints indicate that $\sim 70\%$ of the X-ray detected sources have $N_{\rm H}>10^{22}\rm \: cm^{-2}$. In addition, based on X-ray stacking analyses of the X-ray undetected obscured IR AGN candidates, we conclude that the majority of the X-ray undetected sources are also obscured by $N_{\rm H}>10^{22}\rm \: cm^{-2}$ (see Section\,\ref{subsec:predictionsAGN}).

        \item Our sources have redshifts ranging $\approx 0.5-3.5$, with median values of $z\approx 0.8$ and $z\approx 1$ for the wide and deep surveys, respectively. In addition, the AGN luminosities of both samples are consistent, with median values of $\log L_{\rm AGN, IR}/({\rm erg \: s^{-1}}) \approx 45.1$, where $\approx 55\%$ of sample are powerful IR quasars with $L_{\rm AGN, IR} > 10^{45} \rm \: erg \: s^{-1}$  (see Section\,\ref{subsec:predictionsAGN}).  

        \item The host galaxy properties of obscured IR AGN candidates are similar for the wide and deep surveys. Our sources are hosted by massive galaxies with a median stellar mass of $\log M_{\star}/\rm M_{\odot} \approx 10.8$. The star formation properties of the obscured IR AGN candidates span a wide range, with SFR ranging $\approx \rm 2-3300 \rm \: M_{\odot} \: yr^{-1}$, and a median value of $\sim 27  \rm \: M_{\odot} \: yr^{-1}$. In addition, we find that $\approx 33\%$ of the obscured IR AGN candidates are hosted by galaxies lying within the starburst region above the main sequence of star-forming galaxies (see Section\,\ref{subsec:hostprop}).  
        
    \end{itemize}

\item We compare the properties of the 4MOST IR AGN and 4MOST X-ray AGN surveys in the eFEDS area, using the {\it eROSITA} observations in eFEDS as an equivalent to the 4MOST X-ray AGN survey (see Section\,\ref{sec:comparison_ero}). We find that $\sim 17\%$ and $\sim 22\%$ of the obscured IR AGN are detected by {\it eROSITA} in eFEDS down to $r_{\rm AB}=22.8$ and $r_{\rm AB}=22.1$, respectively. We show that both surveys are complementary and target samples with different optical, MIR, and X-ray properties. We find (1) only $\sim 21\%$ of the eFEDS X-ray AGN meet the obscured 4MOST IR AGN selection of $r-W2\geq 5.9 \rm \: mag$; (2) the eFEDS X-ray AGN are optically bright with $\sim 50\%$ of them having a $r_{\rm AB}\lesssim 21.1$, while the 4MOST IR AGN sources are optically faint with $\sim 80\%$ of them having $r_{\rm AB}> 21.1$; (3) the 4MOST X-ray AGN survey covers a wider redshift range than the IR AGN survey with $\sim 2.1$ and $\sim 1.7$ times more sources at $z<0.5$ and $z>1.5$, respectively, and (4) the majority ($\sim 92\%$) of the eFEDS X-ray AGN are X-ray unobscured with $N_{\rm H}<10^{22} \rm \: cm^{-2}$, while the majority ($\sim 70\%$) of the 4MOST IR AGN are expected to be obscured by $N_{\rm H}>10^{22} \rm \: cm^{-2}$.

\end{enumerate}

On the basis of our analysis, we show that the 4MOST IR AGN survey will target a key population of $\sim 212,000$ IR AGN, where $\sim 140,000$ sources will be obscured by $N_{\rm H}>10^{22} \rm \: cm^{-2}$, $\sim 121,000$ sources will be powerful IR quasars with $L_{\rm AGN, IR} > 10^{45} \rm \: erg \: s^{-1}$, and only $\sim 40,000$ sources will be detected by {\it eROSITA}. In addition, we predict that the 4MOST X-ray AGN survey will identify $\sim 300,000$ IR AGN with $r-W2<5.9\rm \: mag$ colours giving, when combined with the 4MOST IR AGN survey, $\sim 500,000$ IR-selected AGN, which is $\sim 100 \times$ larger than any existing spectroscopic MIR-selected AGN sample allowing for the first large-scale spectroscopic studies of this previously poorly studied population. Using the X-ray, optical/UV, and IR data overlapping with the 4MOST footprint, we will reliably characterise (both in the physical and empirical properties) the obscured IR AGN population for the first time and identify previously missed AGN sub populations, extending our taxonomy of the overall AGN and quasar population. This rich dataset will allow us to test whether (and how) obscured IR AGN are fundamentally different to unobscured IR AGN, distinguishing between the standard orientation-dependent unified model and evolutionary based models, understand whether these differences are universal or restricted to key sub populations, and to determine whether obscured IR AGN preferentially reside in different environments to unobscured IR AGN (see Section\,\ref{sec:goals}).

\section*{Acknowledgements}

We thank the referee for their positive and constructive comments, which substantially improved the paper. This work has been supported by the EU H2020-MSCA-ITN-2019 Project 860744 “BiD4BESt: Big Data applications for black hole Evolution STudies”. CA is supported by the Alexander von Humboldt Foundation through a Humboldt Research Fellowship for postdoctoral researchers. DMA and CG thank the Science Technology Facilities Council (STFC) for support from the Durham consolidated grant (ST/T000244/1). DJR acknowledges the support of STFC grant NU-012097.

This research used data obtained with the Dark Energy Spectroscopic Instrument (DESI). DESI construction and operations is managed by the Lawrence Berkeley National Laboratory. This material is based upon work supported by the U.S. Department of Energy, Office of Science, Office of High-Energy Physics, under Contract No. DE–AC02–05CH11231, and by the National Energy Research Scientific Computing Center, a DOE Office of Science User Facility under the same contract. Additional support for DESI was provided by the U.S. National Science Foundation (NSF), Division of Astronomical Sciences under Contract No. AST-0950945 to the NSF’s National Optical-Infrared Astronomy Research Laboratory; the Science and Technology Facilities Council of the United Kingdom; the Gordon and Betty Moore Foundation; the Heising-Simons Foundation; the French Alternative Energies and Atomic Energy Commission (CEA); the National Council of Science and Technology of Mexico (CONACYT); the Ministry of Science and Innovation of Spain (MICINN), and by the DESI Member Institutions: www.desi.lbl.gov/collaborating-institutions. The DESI collaboration is honored to be permitted to conduct scientific research on Iolkam Du’ag (Kitt Peak), a mountain with particular significance to the Tohono O’odham Nation. Any opinions, findings, and conclusions or recommendations expressed in this material are those of the author(s) and do not necessarily reflect the views of the U.S. National Science Foundation, the U.S. Department of Energy, or any of the listed funding agencies.

\section*{Data Availability}

The datasets generated and/or analysed in this study are available from the corresponding author on reasonable request.



\bibliographystyle{mnras}
\bibliography{references} 




\appendix




\section{{\it WISE} colour selection performance for different {\it WISE} catalogues } \label{ap:reliability}

This appendix extends the analysis of Section\,\ref{sec:finalselection} and explores the efficiency of the \citetalias{2018Assef} R90 selection for the AllWISE, unWISE and LS DR10 {\it WISE} catalogues. In addition, here we also investigate the origin and nature of the contamination of the AGN candidates. Section\,\ref{ap:a:wise_desc} provides a more detailed description of each {\it WISE} catalogue, and Section\,\ref{ap:a:wise_rel} presents the properties of the IR AGN candidates selected by \citetalias{2018Assef} R90. 

\subsection{{\it WISE} catalogues description } \label{ap:a:wise_desc}

This section describes the datasets used to construct the three {\it WISE} catalogues used in this work. The AllWISE catalogue contains photometric information for $\sim 700$ million sources with $5\sigma$ point source limits at $W1=16.83$ and $W2=15.6$ mag in the Vega system, identified in the first 6 months of {\it WISE} observations. The aperture photometry in AllWISE is optimized for point-source characterization, underestimating the flux in extended sources relative to the {\it WISE} PSF. The unWISE catalogue uses the deep 5-year image unblurred adds \citep{2019Meisner}, which combines five years of {\it WISE} and {\it NEOWISE} \citep{2014Mainzer} data, and adopts a more careful treatment to model the {\it WISE} photometry for blended sources, detecting roughly three times as many sources ($\sim 2$ billion sources) as in AllWISE. Lastly, the LS DR10 catalogue uses the position of the $griz$ detected sources to perform forced deblended photometry in the unWISE coadds using the {\it The Tractor} algorithm \citep{2016Lang}, providing even deeper {\it WISE} photometry, reaching $5\sigma$ limiting magnitudes at $W1=20$ and $W2=19.3$ mag for $\sim 2.8$ billion sources.

\subsection{AGN reliability and contamination of our IR AGN selection} \label{ap:a:wise_rel}

\begin{figure*}
\centering
\includegraphics[scale=0.32]{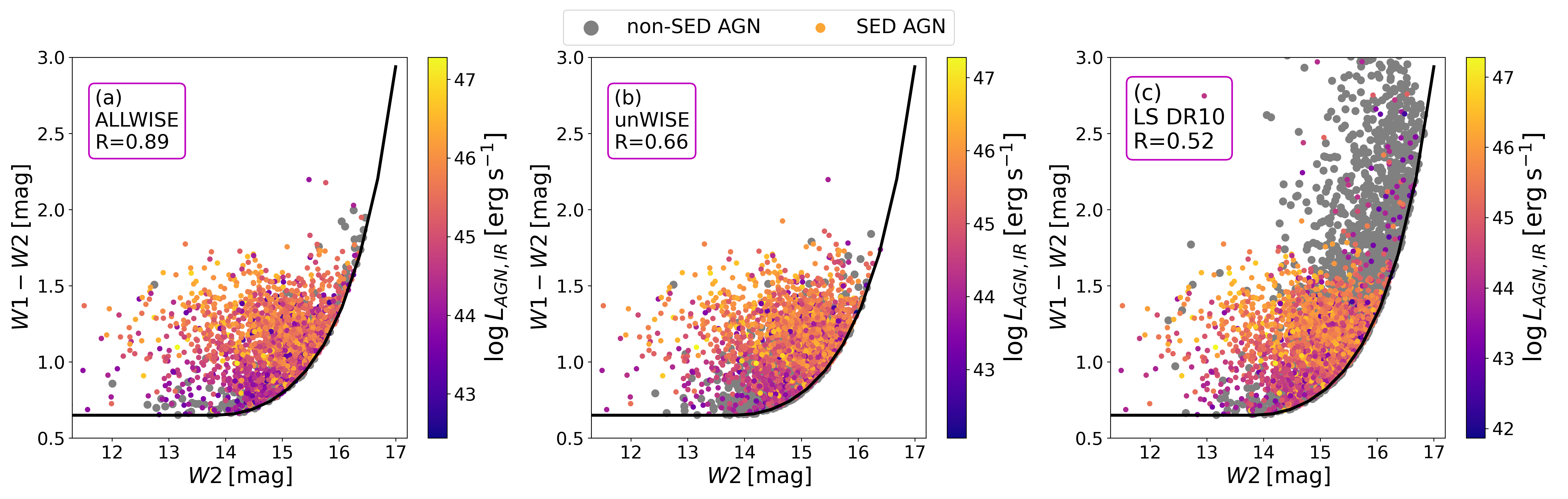}

\caption{IR AGN candidates identified with the \citet{2018Assef} R90 reliability criterion across the $W1-W2$ vs $W2$ plane for $r_{\rm AB }<22.8 \, \rm mag$, coloured by the logarithmic $8-1000\rm \: \mu m$ AGN luminosity obtained from the SED fitting. Grey points are sources without an AGN component in their IR SEDs. We show the selection using the {\it WISE} photometry from the AllWISE (panel a), unWISE (panel b), and Legacy Survey DR10 (panel c) catalogues. Each panel reports the AGN reliability of the \citet{2018Assef}  selection based on our IR SED fitting. } 
\label{fig:A90_cat}
\end{figure*}

This section extends the analysis presented in Section\,\ref{sec:finalselection} to two additional {\it WISE} catalogues: AllWISE and LS DR10. Likewise, we use UV-to-FIR SED fitting in four deep fields (see Section\,\ref{sec:data:photo}) to calculate the AGN reliability and the AGN and host galaxy properties of the \citetalias{2018Assef} R90 AGN candidates identified using the three {\it WISE} catalogues. To identify the IR AGN candidates, we follow the \citetalias{2018Assef} R90 AGN wedge (see Equation\,\ref{eq:A18R90}). Figure\,\ref{fig:A90_cat} shows IR AGN candidates identified using the \citetalias{2018Assef} R90 criteria across the $W1-W2$ vs $W2$ plane, and highlights the sources that are also identified as IR AGN by our SED fitting. We find that the \citetalias{2018Assef} selection only reaches 89\% reliability when using the AllWISE photometry, and decreases to 66\% when using unWISE photometry, and to 52\% when using LS DR10 photometry. It is clear from the figure that the LS DR10 catalogue contains more sources with high S/N, and faint $W1$ and $W2$ photometry than AllWISE and unWISE. These faint, red sources meet the \citetalias{2018Assef} R90 selection but are not identified as AGN by our UV-to-FIR SED fitting; i.e., they are contaminant galaxies.

\begin{figure*}
\centering
\includegraphics[scale=0.4]{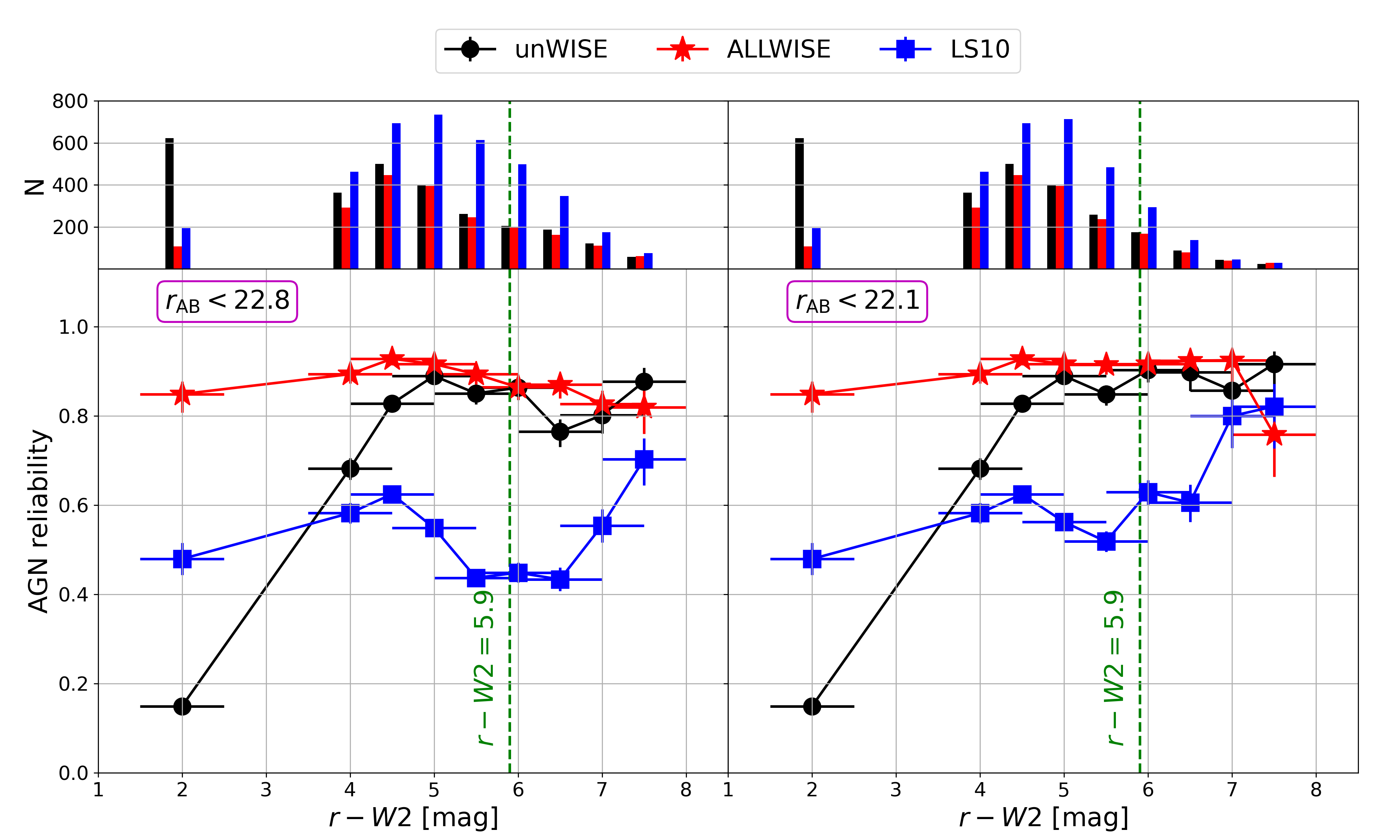}

\caption{AGN reliability as a function of the $r-W2$ colour for the AGN candidates selected by the \citet{2018Assef} R90 AGN colour selection as in Figure\,\ref{fig:rel_unwise}, but for three {\it WISE} catalogues: unWISE (black circle), AllWISE (red star), and LS DR10 (blue square). The left and right panels show the AGN reliabilities for sources with $r_{\rm AB}<22.1 \, \rm mag$ and $r_{\rm AB}<22.8 \, \rm mag$, respectively. The upper panels show the number of sources per $r-W2$ bin for the unWISE (black bar),  AllWISE (red bar), and LS DR10 (blue bar) catalogues. } 
\label{fig:ap:rel}
\end{figure*}

Figure\,\ref{fig:ap:rel} further explores the AGN reliability as a function of $r-W2$, similar to Figure\,\ref{fig:rel_unwise}, but this time for the AllWISE, unWISE, and LS DR10 {\it WISE} catalogues\footnote{We note that \citetalias{2018Assef} used UV-to-MIR SED fitting to arrive at their 90\% AGN reliability threshold using AllWISE photometry (see \citet[][]{2013Assef} for details on their SED fitting approach). In our work, we also independently assess the reliability of the approach using other deeper {\it WISE} photometry, such as unWISE and the LSDR10, and we use a more sophisticated SED modelling which also includes the FIR band.}, for $r_{\rm AB}<22.1$ and $r_{\rm AB}<22.8$. Figure\,\ref{fig:ap:rel} shows that when using AllWISE photometry, the AGN reliability does not have a strong dependence on $r-W2$. However, when using the unWISE photometry, the \citetalias{2018Assef} R90 colour-selection approach provides a highly contaminated sample at $r-W2<4 \rm \: mag$ colours, with an AGN reliability of $\approx 20\%$, but it drastically improves to $>80\%$ for $r-W2\gtrsim 4.5 \rm \: mag$.  At $r-W2\geq 5.9 \rm \: mag$, which is our obscured AGN selection, the \citetalias{2018Assef} R90 selection provides a clean AGN sample with an AGN reliability of $87\%$ for $r_{\rm AB}<22.1$ and $80\%$ for $r_{\rm AB}<22.8$. Finally, when using the {\it WISE} photometry from the LS DR10, the AGN reliability is $<60\%$ for most $r-W2$ colour bins, including $r-W2\geq 5.9 \rm \: mag$, indicating that the \citetalias{2018Assef} AGN selection does not perform well when using LS {\it WISE} photometry.

\begin{table*}
\centering

\begin{tabular}{c|c|l|l|l|l|l|}
 \hline
 \hline
\noalign{\smallskip}
& &  & AllWISE& unWISE & LS10 \\
\noalign{\smallskip}

\hline
\hline
\noalign{\smallskip}
\citetalias{2018Assef} R90 AGN & $r_{\rm AB}<22.8$ & AGN candidates density & 97&140 & 219& \\
\noalign{\smallskip}
& & AGN reliability  & 89\% & 68\% & 56\% & \\
\noalign{\medskip}
 & $r_{\rm AB}<22.1$ & AGN candidates density &84 &125 & 165& \\
\noalign{\smallskip}
& & AGN reliability & 91\% & 69\% & 62\% & \\
\noalign{\medskip}
\citetalias{2018Assef} R90 AGN + $r-W2\geq 5.9$ & $r_{\rm AB}<22.8$ & AGN candidates density & 29& 32&73 & \\
\noalign{\smallskip}
& & AGN reliability & 87\% & 80\% & 48\% & \\
\noalign{\medskip}
& $r_{\rm AB}<22.1$ & AGN candidates density & 17& 18&31 & \\
\noalign{\smallskip}
& & AGN reliability & 92\% & 87\% & 67\% & \\
\hline
\hline

\end{tabular}
\caption{AGN reliability and AGN number density (per deg$^2$) of the AGN candidates identified by the \citetalias{2018Assef} R90 selection using the AllWISE, unWISE, and LS DR10 {\it WISE} photometry (top rows). The AGN reliability is calculated based on the UV-to-FIR SED fitting results. We also report the values for the obscured \citet{2018Assef} AGN with $r-W2\geq 5.9 \rm \: mag$ colours (bottom rows). }
\label{t:reliability}
\end{table*}

Table\,\ref{t:reliability} provides a summary of the AGN reliability and number density of the IR AGN candidates identified by \citetalias{2018Assef} R90 for different {\it WISE} catalogues. We note that the AGN density is not always the same: when using the {\it WISE} magnitudes from the unWISE and LS10 catalogues, the \citetalias{2018Assef} R90 selection identifies $\sim 1.4 \times$ and $\sim 2.3 \times$ more AGN than when using the AllWISE catalogue, respectively. However, we find that most of the additional IR AGN candidates are galaxies misidentified as IR AGN. After the $r-W2\geq 5.9 \rm \: mag$ colour cut is applied, unWISE finds only a slightly higher AGN number density than AllWISE, probably due to the fact the colour cut removed most of the non-AGN contamination from the unWISE sample.

\begin{figure*}
\centering
\includegraphics[scale=0.32]{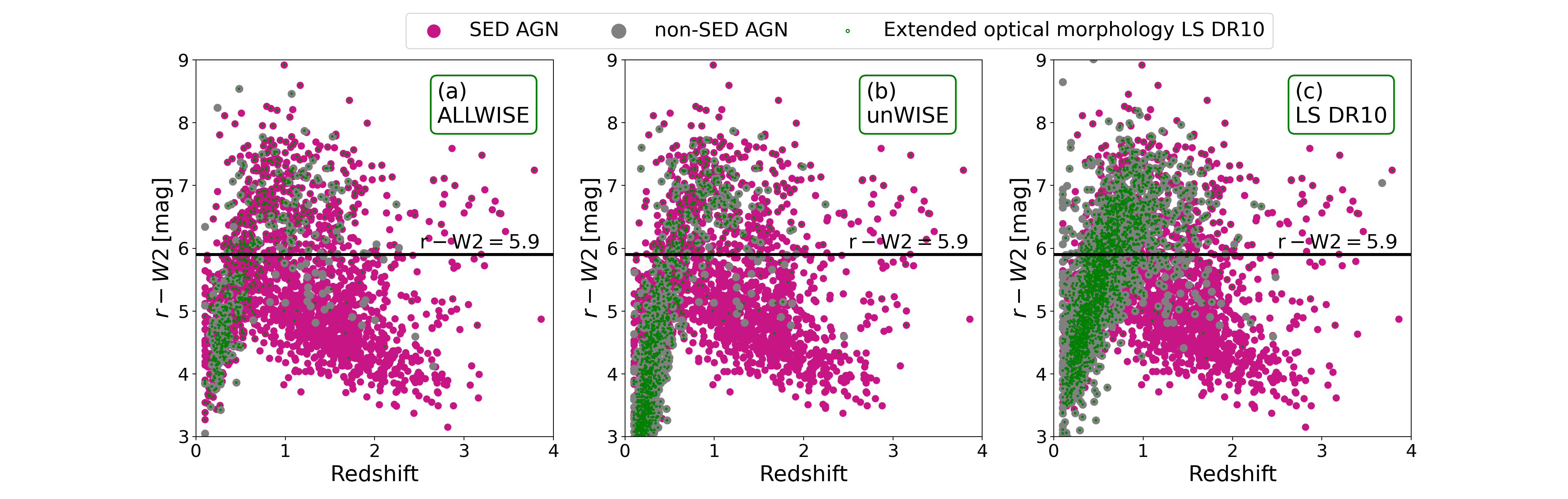}

\caption{$r-W2$ against redshift for the IR AGN candidates identified with the \citet{2018Assef} R90 criterion for $r_{\rm AB }<22.8 \, \rm mag$, using AllWISE (panel a), unWISE (panel b), and LS DR10 (panel c) photometry. Grey points are sources without an AGN component in their IR SEDs, and small empty green circles are sources classified as optically extended in the LS DR10 catalogue. The figures show that the majority of the non-SED AGN also have an extended optical morphology.} 
\label{fig:ap:rW2_extmorph}
\end{figure*}

Figure\,\ref{fig:ap:rW2_extmorph} plots the \citetalias{2018Assef} R90 IR AGN candidates across the $r-W2$ vs redshift plane, highlighting the sources that are identified as AGN by our SED fitting and also the sources that have an extended optical morphology in the LS DR10. We find that for the three catalogues, the contaminant sources are typically morphologically classified as optically extended in the LS DR10 catalogue. When using {\it WISE} photometry from the AllWISE catalogue, $\approx 76\%$ of the non-AGN contaminants are optically-extended. When using {\it WISE} photometry from the unWISE and LS DR10 catalogues, $>90\%$ of the non-AGN contaminants are optically extended, and in consequence, the AGN reliability of the \citetalias{2018Assef} R90 significantly drops to $\approx 30-40\%$. However, for optically point-like sources, the AGN reliability for the unWISE and LS DR10 catalogues is about 90\%, as shown by \citet{2018Assef}.  Table\,\ref{t:reliability_subsamples} summarizes the AGN reliability for each aforementioned subsample. 

\begin{figure*}
\centering
\includegraphics[scale=0.32]{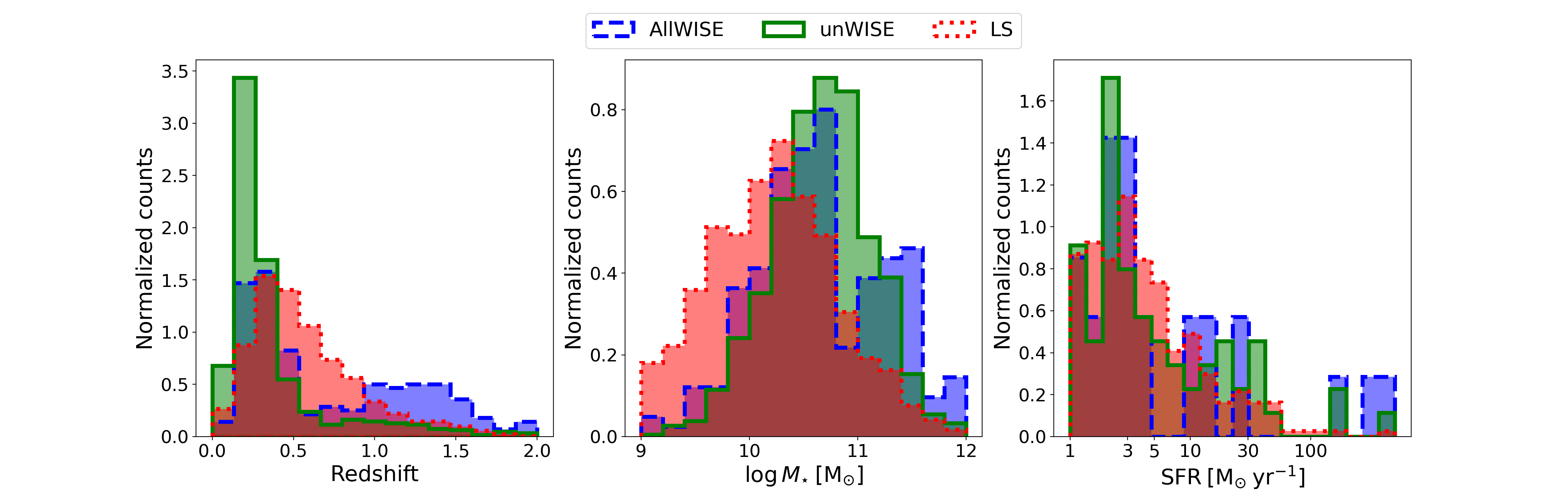}

\caption{Redshift (left panel), stellar masses (central panel), and star-formation rates (right panel) of the non-AGN contaminants identified from our UV-to-FIR SED fitting of the IR AGN candidates selected with the \citet[][]{2018Assef} R90 criterion using AllWISE (dashed blue), unWISE (solid green), and LS DR10 (dotted red) catalogues. } 
\label{fig:ap:contamination}
\end{figure*}

Figure\,\ref{fig:ap:contamination} examines the host galaxy properties of the non-AGN contaminants. For the three catalogues, the majority ($>50\%$) of the sources have $z<0.5$. The stellar masses of the non-AGN contaminants are not so different from the SED AGN, ranging $\approx 10^{10}-10^{12}\rm \: M_{\odot}$. However, the non-AGN contaminants are systems with little ongoing star-formation, with more than $80\%$ of the sources having $\rm SFR<10 \rm \: M_{\odot} \: yr^{-1}$. These results indicate that the non-AGN contaminants are typically low redshift, quiescent galaxies, with red optical colours.

In order to understand the higher number of ``apparent" AGN in the unWISE and LS DR10 catalogues, we visually inspect a random sample of optically bright, extended IR AGN selected by \citetalias{2018Assef} R90. We find that the unWISE and LS DR10 catalogues split the extended sources into one central source plus several fainter components. This is a result of the deblending approach used to extract the photometry, which is acknowledged in \citet{2019Schlafly}. Hence, a significant part of the AGN contamination for the unWISE and LS DR10 catalogues is due to the misidentification of faint and red extended components as AGN. A solution to increase the purity of the \citetalias{2018Assef} R90 AGN samples identified using unWISE and LS DR10 photometry would be only to apply the \citetalias{2018Assef} R90 selection to point-like sources in the LS DR10 catalogue. However, that would significantly decrease the completeness of the AGN samples as only $\sim 46\%$ and $\sim 38\%$ of the \citetalias{2018Assef} R90 AGN have point-like morphologies in the unWISE and LS DR10 catalogues, respectively.

\begin{table}
\centering

\begin{tabular}{c c c c c c}
 \hline
 \hline
\noalign{\smallskip}
 \citetalias{2018Assef} R90 AGN subsample & AllWISE& unWISE & LS10 \\
\noalign{\smallskip}

\hline
\hline
\noalign{\smallskip}
Entire sample & 89\% & 68\% & 52\% \\
\noalign{\smallskip}

$z<0.5$ & 79\% & 35\% & 30\% \\
\noalign{\smallskip}
Extended optical morphology &79\% & 41\% & 31\% \\
\noalign{\smallskip}
Point-like source & 96\% &96\% &88\% \\

\hline
\hline

\end{tabular}
\caption{AGN reliability of the \citetalias{2018Assef} R90 selection for sources with $z<0.5$, sources that are classified as optically-extended by the LS DR10 catalogue, and sources that are classified as optically-point-like by the LS DR10 catalogue. We show the results for the three different {\it WISE} catalogues explored in this work.}
\label{t:reliability_subsamples}
\end{table}

\begin{figure*}
\centering
\includegraphics[scale=0.43]{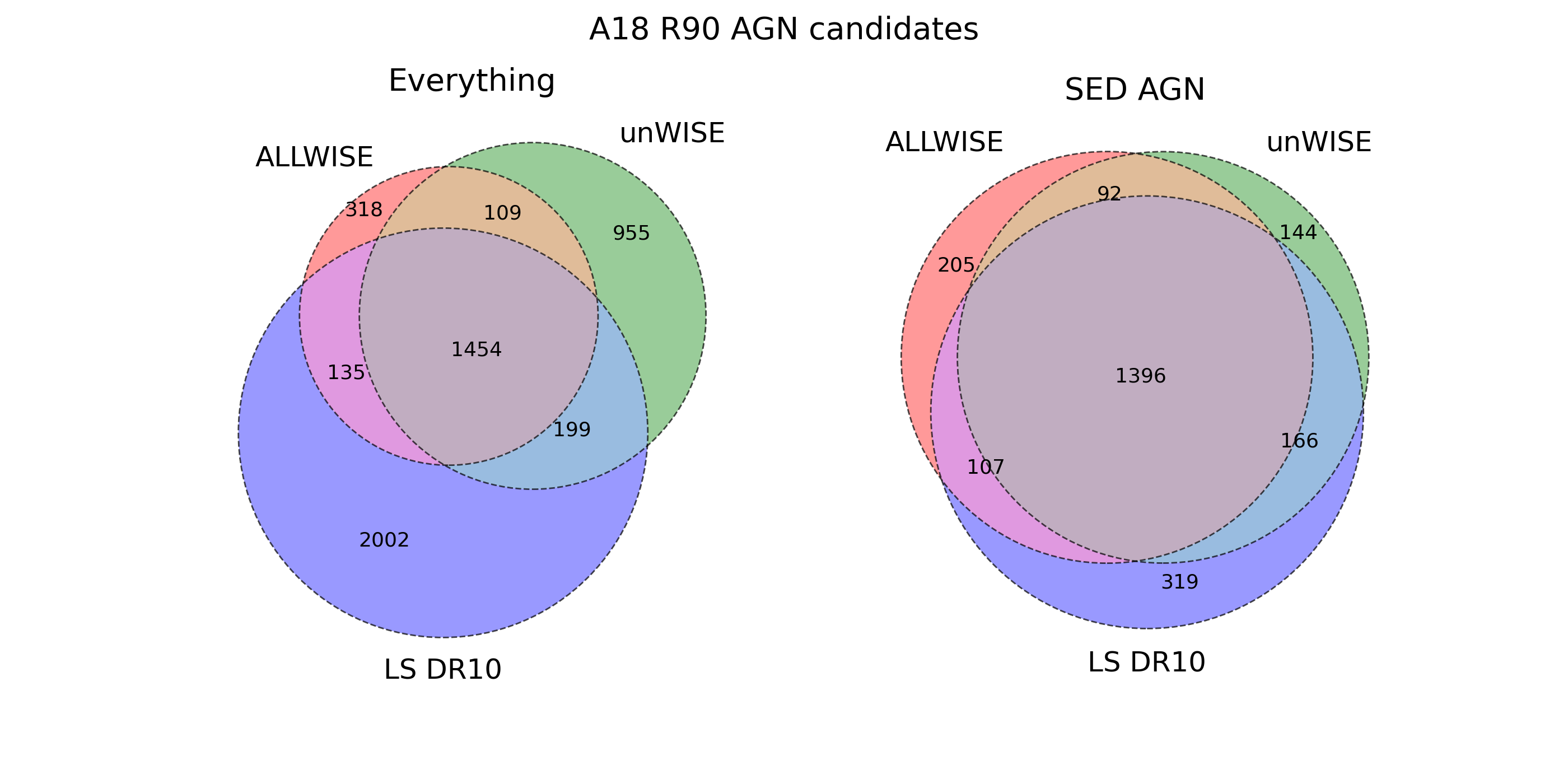}
\includegraphics[scale=0.43]{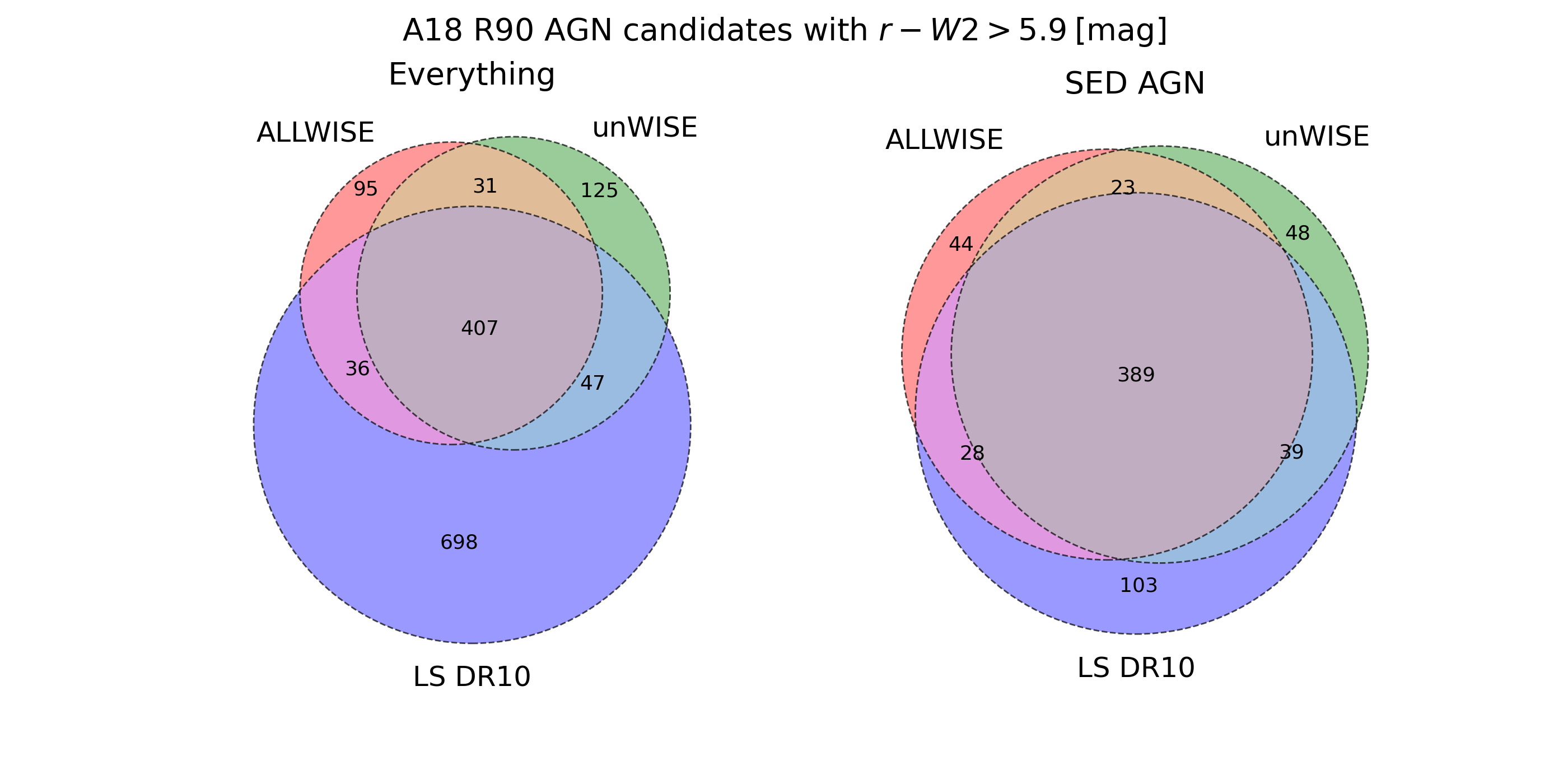}

\caption{Venn diagrams showing the overlap of \citetalias{2018Assef} R90 AGN candidates (labelled "Everything") using AllWISE, unWISE, and LS DR10 photometry (left plots) and the same overlap only for the sources that are also classified as IR AGN by our UV-to-FIR SED fitting (labelled "SED AGN", right plots). The bottom plots show the overlap only for sources with $r-[4.5]\geq 5.9 \rm \: mag$ colours. In the four cases, the majority ($\sim 78\%$) of the AGN identified using the AllWISE photometry are also identified when using the unWISE and LS DR10 photometry.} 
\label{fig:venn}
\end{figure*}

We also investigate the overlap of the different {\it WISE} catalogues. Figure\,\ref{fig:venn} shows Venn diagrams of the \citetalias{2018Assef} R90 selected AGN using the AllWISE, unWISE, and LS DR10 catalogues for all the sources, the sources that are also identified as AGN by the UV-to-FIR SED fitting, and the sources with $r-W2\geq 5.9 \rm \: mag$. We find that $\sim 78\%$ of the AGN selected using the AllWISE photometry are also selected when using the unWISE and LS DR10 photometry. The overlap increases to 86\% when we only consider the sources that are also identified as AGN by UV-to-FIR SED fitting. The numbers are the same when considering only the sources with $r-W2\geq 5.9 \rm \: mag$ colours. This analysis shows that the unWISE and the LS DR10 {\it WISE} photometry find more IR AGN candidates than AllWISE, but the three catalogues find almost the same real AGN. 

Overall, since we are interested in the obscured IR AGN candidates with $r-W2\geq 5.9 \rm \, mag$, we conclude that the unWISE and AllWISE photometry are both good options for selecting IR AGN when using the \citetalias{2018Assef} R90 approach, as both provide high AGN reliabilities (i.e., $>80-90\%$; see Figure\,\ref{fig:ap:rel} and Table\,\ref{t:reliability}) and similar AGN number densities. The LS DR10 also provides a clean AGN sample when only considering the sources with point-like optical morphologies but at the expense of significantly reducing the completeness of the AGN sample. Based on these results, we decided to use the unWISE catalogue to perform the target selection of the 4MOST IR AGN survey for three reasons: 1) it provides a clean AGN sample, finding AGN with 80\% and 87\% reliability down to $r_{\rm AB}=22.8$ and $r_{\rm AB}=22.1$, respectively; 2) its slightly higher AGN number density (it finds 1-2 more real AGN per $\rm deg^2$) provides a more complete AGN sample, adding $\sim 11,000$ sources in the final survey, and 3) it is more closely aligned with the LS DR10 catalogue, which is our parent sample, since it uses the same unWISE image co-adds and a consistent {\it WISE} photometric extraction approach (see Appendix\,\ref{ap:a:wise_desc} for details). Hence, the unWISE photometry provides the best balance between reliability and completeness


\section{SED fitting of the 4MOST IR AGN survey sources with UV-to-MIR photometry} \label{ap:sed}

\begin{table}
\centering

 \begin{tabular}{lccc} 
 \hline
 \hline
\noalign{\smallskip}
 Parameter &   Redshift &   Gaussian prior parameters\\
\noalign{\smallskip}
\hline
\noalign{\smallskip}

\multirow{3}{*}{$\log L_{\rm AGN, IR}$} &  $z<1$ & $\mu=44.9 \rm \: erg\: s^{-1}$, $\sigma = 0.41 $ \\
 & $z=1-2$ & $\mu=45.5 \rm \: erg\: s^{-1}$, $\sigma = 0.44 $\\ 
 & $z>2$ & $\mu=46.5 \rm \: erg\: s^{-1}$, $\sigma = 0.42 $\\ 

 \hline
\noalign{\smallskip}
\multirow{3}{*}{$\log M_{\star}$} &  $z<0.5$ & $\mu=\rm 10.2\: M_{\odot}$, $\sigma = 0.5 $ \\
 & $z=0.5-1$ & $\mu=\rm 10.8 \: M_{\odot}$, $\sigma = 0.4 $\\ 
 & $z=1-1.5$ & $\mu=\rm  11.0 \: M_{\odot}$, $\sigma = 0.4 $\\ 
& $z=1.5-2$ & $\mu=\rm  11.3 \: M_{\odot}$, $\sigma = 0.4 $\\ 
& $z=>2$ & $\mu=\rm  11.9 \: M_{\odot}$, $\sigma = 0.9 $\\ 

\noalign{\smallskip}
\hline
\noalign{\smallskip}

\multirow{3}{*}{$\log L_{\rm SF, IR}$} &  $z<0.5$ & $\mu=45.0 \rm \: erg\: s^{-1}$, $\sigma = 0.5 $ \\
 & $z=0.5-1$ & $\mu=45.3 \rm \: erg\: s^{-1}$, $\sigma = 0.5 $\\ 
 & $z=1-1.5$ & $\mu=45.9 \rm \: erg\: s^{-1}$, $\sigma = 0.6 $\\
 & $z>2$ & $\mu=46.0 \rm \: erg\: s^{-1}$, $\sigma = 0.4 $\\ 
 
 \noalign{\smallskip}

 \hline
\hline
 \noalign{\smallskip}
 \end{tabular}
 \caption{ Gaussian priors placed on the simple LS DR10 plus {\it WISE} SED fitting. The values reported in this table correspond to the best-fitting mean ($\mu$) and standard deviation ($\sigma$) for the predicted AGN luminosity, stellar masses, and star formation luminosities for different redshifts bins of the 4MOST IR AGN sources, calculated from the detailed UV-to-FIR SED fitting in the deep fields (see Sections\,\ref{sec:sedfitting} and\,\ref{subsec:predictions}). } \label{t:priors}
\end{table}

In this paper, we have used SED fitting in well-observed fields in the sky with X-ray and UV-to-FIR photometry to predict the AGN and host galaxy properties of the 4MOST IR AGN survey sample. However, for the majority of the sources in the 4MOST IR AGN survey, we will be restricted to just 6 to 8 photometric measurements (i.e., the LS bands $griz$, and the $WISE$ bands $W1$, $W2$, $W3$ and $W4$) in the UV-to-MIR band to perform SED fitting. This all-sky available photometry is limited, but, as we demonstrate below in this section, it can still reliably constrain the AGN torus emission and the stellar mass for obscured AGN.

To fit the all-sky photometry, we use a more simplified SED fitting model to the one described in Section\,\ref{sec:sedfitting}. This time, we reduce the degree of free parameters for the stellar population and torus models. For the stellar population emission, we consider the simple stellar population model from \citet{2003BC}, which assumes one single instantaneous starburst at $\tau=0$. For the torus emission, we only allow the short wavelength slope to be free to vary while fixing the long wavelength slope to $\Gamma=-0.9$, the best-fitting value found in our more-detailed SED template fitting in the deep fields (see Section\,\ref{sec:data:sed}). 

To help guide the simplified SED fitting modelling, we use the knowledge gained from the AGN and host galaxy property constraints from the deep fields. For this, we take advantage of the Bayesian nature of \textsc{fortesfit}, and place priors on the AGN luminosity and stellar mass fitting parameters. To calculate the priors, we use the deep-field results and fit Gaussian functions to the distributions of both parameters for different redshift ranges; see table\,\ref{t:priors} for the best-fitting parameters (i.e., mean $\mu$ and standard deviation $\sigma$) of the AGN luminosity and stellar masses. We note that the standard deviation is typically $\sigma \approx 0.4 \rm \: dex$, which gives the SED fitting process freedom to explore a wide range of values around the central position of the Gaussian. 

\begin{figure*}
\centering
\includegraphics[scale=0.35]{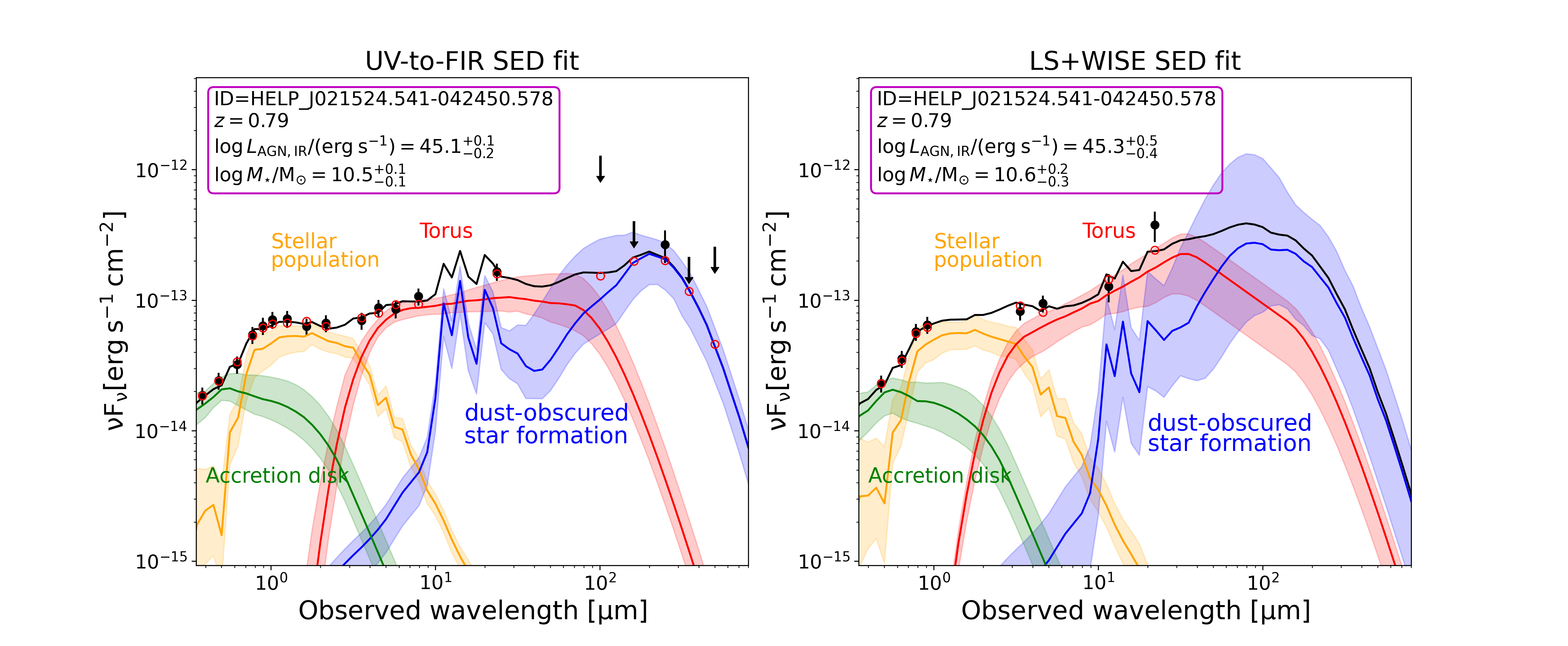}
\includegraphics[scale=0.35]{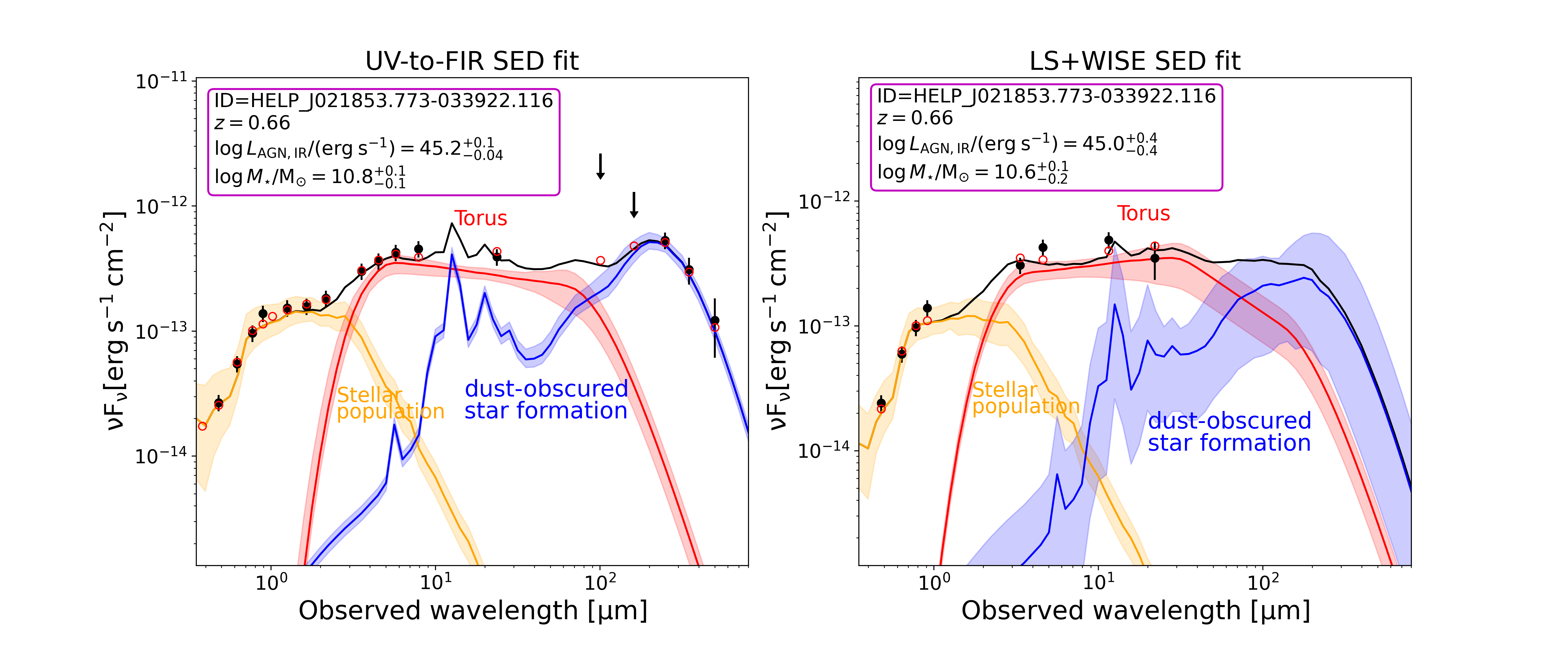}

\caption{Example best-fitting SED models for an X-ray detected (top panels) and an X-ray undetected (bottom panels) obscured IR AGN candidates in the XMM-LSS field. The left figures show the best-fitting SED using all the available UV-to-FIR photometry, and the left figures show the best-fitting SED only using LSDR10 and {\it WISE} photometry. Black points with error bars are the photometric data with their $1\sigma$ uncertainties, and red empty circles are the best-fitting model photometry. The blue curves represent the dust-obscured SF component, the red curves are the torus components, the orange curves are the stellar population emission, and the green curves are the accretion disc emission. The spread in the curves represents the approximate $1\sigma$ scatter in the SED components as constrained by \textsc{fortesfit}. The black line is the sum of all the best-fitting components. The magenta outline box shows the best-fitting AGN luminosity and stellar masses and their $1\sigma$ uncertainties. The ID and redshifts are as in the HELP catalogue \citet{2019Shirley,2021Shirley}.
These figures show samples of the full SED and are only meant to guide the reader; hence, they should not be used to measure the merit of the fit.}. 
\label{fig:SEDcomp_ex}
\end{figure*}

\begin{figure*}
\centering
\includegraphics[scale=0.38]{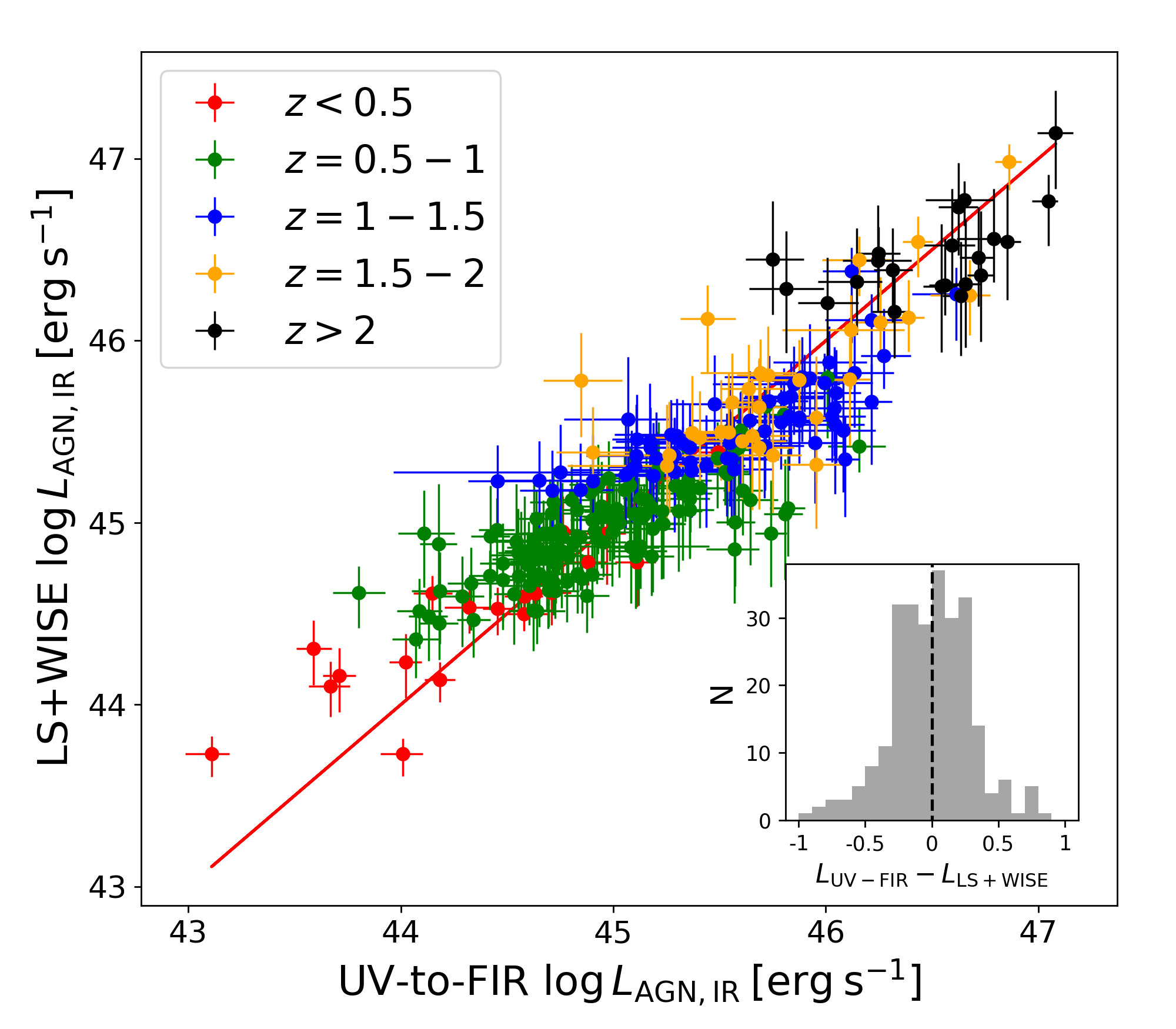}
\includegraphics[scale=0.38]{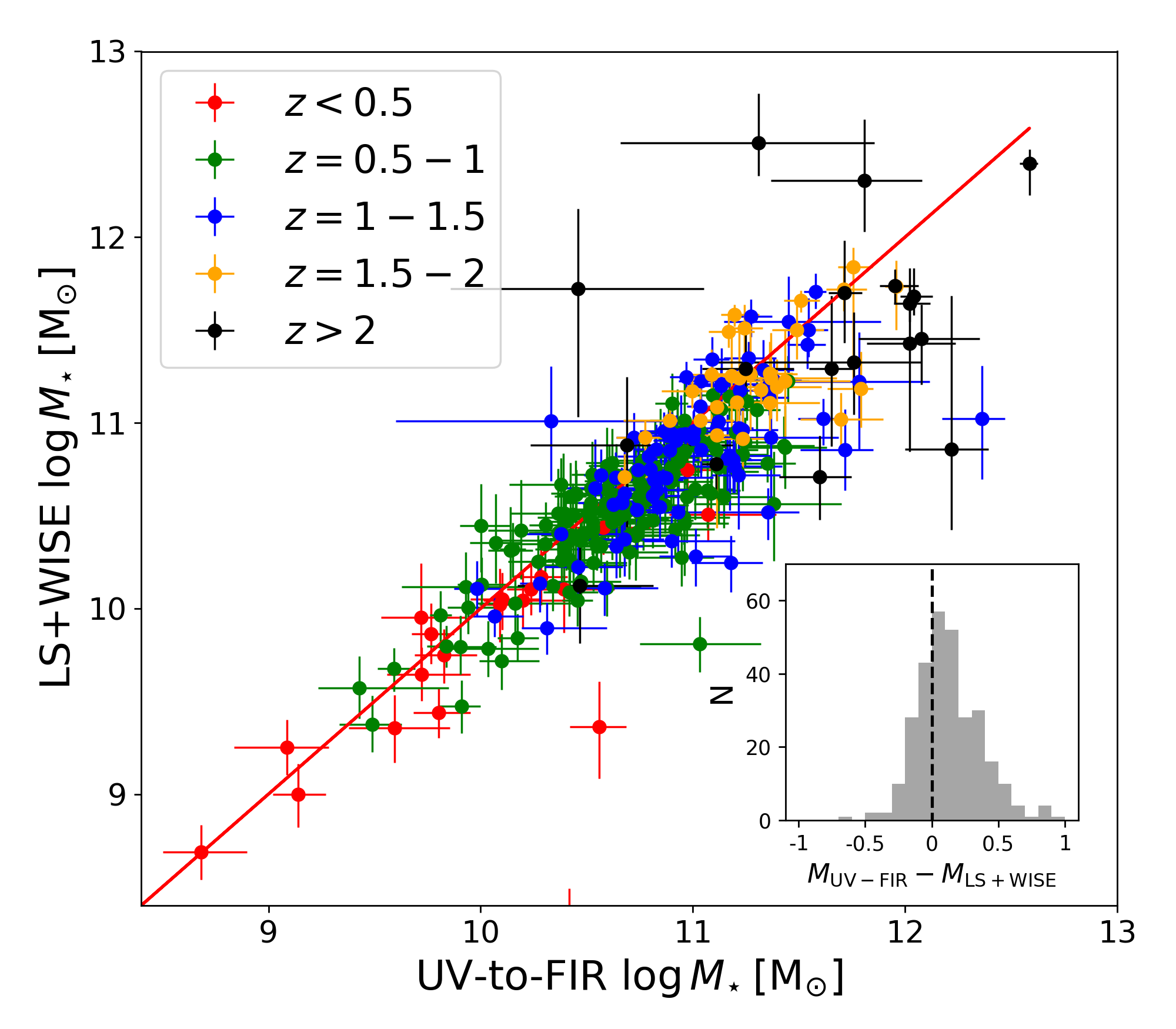}
\caption{Comparison of SED fitting constraints when only using the UV-to-MIR LS+{\it WISE} photometry (y-axis), and when using all the available UV-to-FIR photometry (x-axis). The left panel compares the $8-1000\rm \: \mu m$ AGN luminosity ($L_{\rm AGN, IR)}$) and the right panel compares the stellar mass ($M_{\star}$). Red circles are sources with $z<0.5$, green circles sources with $z=[0.5,1]$, blue circles sources with $z=[1,1.5]$, orange circles sources with $z=[1.5,2]$, and black circles sources with $z>2$. The small panel on the right bottom corner of each plot shows the distribution of the difference between the parameter values obtained using UV-to-FIR photometry and only UV-to-MIR photometry. }. 
\label{fig:SEDcomp}
\end{figure*}

We tested this approach in the XMM-LSS and ELAIS-S1 fields. We excluded the  Bo\"{o}tes field from the analysis because it does not have $i$-band photometry, which is not the case for the 4MOST IR AGN survey sources. Figure\,\ref{fig:SEDcomp_ex} shows two examples of best-fitting SED models for an X-ray undetected (top) and detected (bottom) obscured IR AGN candidates in the XMM-LSS field. The figure shows the best SED fitting results when using all the available UV-to-FIR photometry (left) and only the LSDR10 and {\it WISE} photometry. The AGN luminosity and the stellar masses are consistent within $1\sigma$ between the two SED fitting implementations. We note that with the current UV-to-MIR available photometry, it is not possible to constrain the contribution from the accretion disk and dust-obscured star formation. Both components are well-constrained in the UV-to-MIR SED modelling only because we set priors in the accretion disk and star formation luminosities, which the SED fitting is assuming as truthful given the lack of photometry in the UV and FIR wavelengths. When available, we can use an X-ray luminosity prior to constrain the AD luminosity. However, constraining the dust-obscured SFRs will be challenging as the majority of the IR AGN wide survey footprint does not have FIR photometric coverage. Figure\,\ref{fig:SEDcomp} shows the comparison of the AGN luminosity and stellar masses when using only LSDR10 and {\it WISE} photometry (y-axis), and when using all the available UV-to-FIR photometry (x-axis). Overall, we find a good agreement between both SED fitting values; however, we still find many sources where both values are inconsistent. Hence, this approach will be more useful to constrain the average properties of a given sample. We note that when having additional accretion disk or stellar masses constraints provided by the 4MOST optical spectra and/or X-ray constraints, we will be able to use that information to improve even more the SED fitting priors, producing more accurate results.

\bsp	
\label{lastpage}
\end{document}